\documentclass[11pt,a4paper]{article}
\usepackage{jheppub}
\usepackage{amsmath,amstext,amssymb}
\usepackage{graphicx}
\usepackage{feynmp}
\usepackage{epstopdf}
\usepackage{caption}
\usepackage{float}

\def\be{\begin{equation}}
\def\ee{\end{equation}}
\def\ba{\begin{eqnarray}}
\def\ea{\end{eqnarray}}

\newcommand{\chidag}{\chi^{\dag}}
\newcommand{\Adag}{A^{\dag}}
\newcommand{\adag}{a^{\dag}}

\newcommand{\bq}{ \textbf{q}}

\newcommand{\bk}{ \textbf{k}}
\newcommand{\bl}{ \textbf{l}}
\newcommand{\bpartial}{ {\bf \partial}}

\usepackage{graphics}
\usepackage{graphicx}% Include figure files
\usepackage{dcolumn}% Align table columns on decimal point
\usepackage{bm}% bold math
\usepackage{epsfig}
\usepackage{graphicx}
\usepackage{multirow}
\usepackage{dcolumn}% Align table columns on decimal point
\usepackage{graphicx,epsfig}

\begin{document}

%\preprint{preprint USM-TH-298}
\today

\title{\center A functional RG approach for the BFKL Pomeron}

\author[a]{Jochen Bartels,}%,\note{Corresponding author.}}
\author[b]{Carlos Contreras,}
\author[c]{Gian Paolo Vacca}%\note{Also at Some University.}}

\affiliation[a]{II. Institut f\"{u}r Theoretische Physik, Universit\"{a}t Hamburg, Luruper Chaussee 149,\\
D-22761 Hamburg, Germany}
\affiliation[b]{Departamento de Fisica, Universidad Tecnica Federico Santa Maria, Avda. Espana 1680, Casilla 110-V, Valparaiso, Chile }
\affiliation[c]{INFN Sezione di Bologna, DIFA, via Irnerio 46, I-40126 Bologna, Italy}

% e-mail addresses: one for each author, in the same order as the authors
\emailAdd{jochen.bartels@desy.de}
\emailAdd{carlos.contreras@usm.cl}
\emailAdd{vacca@bo.infn.it}

%%%%%%%%%%%%%%%%%%%%%%
\abstract{In this paper we encode the perturbative BFKL leading logarithmic resummation, relevant for the Regge limit behavior of QCD scattering amplitudes, in the IR-regulated effective action which satisfies exact functional renormalization group equations. This is obtained using a truncation with a specific infinite set of non local vertices  describing the multi-Regge kinematics (MRK). The goal is to use this framework to study, in the high energy limit and at larger transverse distances
the transition to a much simpler effective local reggeon field theory, whose critical properties were recently investigated in the same framework.
We perform a numerical analysis of the spectrum of the BFKL Pomeron deformed by the introduction of a Wilsonian infrared regulator to understand
the properties of the leading poles (states) contributing to the high energy scattering.
}      
\date{today}
%%%%%%%%%%%%%%%%%%%%%%

%\pacs{12.38.Cy, 12.38.Aw,12.40.Vv} momentujm 

\maketitle
 
%%%%%%%%%%%%%%%%%%%%%%%%%%%%%%%%
\section{Introduction}
%%%%%%%%%%%%%%%%%%%%%%%%%%%%%%%%

Understanding the physics of strong interactions (QCD) through the study of scattering processes in different kinematical regimes is still a challenging task. Particularly interesting is the kinematic region of very high energies and finite momentum transfers (Regge limit): in contrast to those high energy limits where collinear factorization holds and provides a clean separation between perturbative and nonperturbative QCD, the Regge limit has no such strict separation. It is therefore tempting  to 
investigate, in the Regge limit, the transition from the perturbative QCD regime (controlled by small distances) to the region of large distances, for which still very little can be derived from a fundamental QCD description. Also, with the start of LHC elastic and multiparticle processes in the Regge limit have become an important issue; a deeper theoretical understanding therefore is not a simple  academic question. 

Some time ago  \cite{Bartels:2015gou,Bartels:2016ecw} we have started a program which aims at finding an interpolation between the perturbative QCD Pomeron (BFKL Pomeron) and the nonperturbative soft Pomeron which describes elastic proton-proton scattering at high energies. Such a study must also include the Odderon. 
In transverse space, the BFKL Pomeron and the soft Pomeron, are probing different regions:
the BFKL Pomeron describes scattering processes which are dominated by small transverse  
distances (large transverse momenta), whereas the soft Pomeron is sensitive to transverse distances 
of the order of or larger than the proton radius.   
As a theoretical framework, which can be used both on the perturbative (UV) region and in the nonperturbative (IR) region, we employ the Wilsonian exact renormalization group techniques. In particular we  have found it convenient to study the flow of an effective average action (IR regulated generator of the one particle irreducible (1PI) vertices) whose RG-flow is described by the Wetterich equation~\cite{Wetterich,Morris:1994ie}.

In two recent papers \cite{Bartels:2015gou,Bartels:2016ecw} we have started with the long distance region. We have studied the flow of the so called Reggeon Field Theory (RFT) for a single Pomeron field and for a Pomeron coupled to an Odderon field, in particular the critical universal properties of these two QFTs and some features of the flow associated to the scale change. 
These theories were designed and formulated before the QCD era, expecting that they could be relevant for the description of the non perturbative large distance behavior of the strong interations.
Pioneering work in RFT has been done already more that 40 years ago \cite{Gribov:1968fg,Gribov:1968uy,AB,Migdal:1973gz}. In particular, a critical solution has been found, and scalings laws for the asymptotic scattering amplitude have been derived.
We also find this as one possible solution, but in our recent analysis, we went beyond this analysis. New results include:\\
(i) we have performed a more general search for (IR) fixed points (i.e. in the presence of all possible Pomeron self couplings) \\ 
(ii) we have studied the behavior also in the vicinity of the fixed points, i.e. the high energy behavior of scattering processes at large but finite energies, and we made a preliminary analysis of global flows within a RFT description\\    
(iii) we have included the Odderon.\\
We mention that this formalism is also well suited to pose questions about more general multi-particle interactions in the Regge limit~\cite{Abramovsky:1973fm,Bartels:2005wa}: in the context of LHC experiments this has become an important issue.

The present paper addresses the UV region: we start from high energy QCD, as formulated in Lipatov's effective action~\cite{Lipatov:1995pn}, generically describing the BFKL Physics regime. 
Here the most important degree of freedom is the reggeized gluon, built from gluon exchanges at high energies. The perturbative BFKL Pomeron \cite{BFKL} then appears as a composite state of two reggeized gluons in color singlet. In leading order it has a continuous spectrum for the variable angular momentum which is dual to the rapidity. The presence of vertices, local in rapidity, but nonlocal in the transverse space, maps QCD in the Regge limit into an effective $2+1$ dimensional QFT of reggeized gluons. Considering their composite states which couple directly to external asymptotic states (i.e. hadrons) through impact factors, it is tempting to switch, starting from the reggeized gluons, to another effective description in terms of colorless composite states of reggeized gluons. As an  example, with such a 'switch', the $2\to 4$ reggeized gluon vertex~\cite{Bartels:1994jj,Bartels:1995kf,Braun:1997nu,Bartels:2002au,Bartels:2004ef}
induces a perturbative triple Pomeron vertex, so that the BFKL Pomeron interacts with itself. Similarly, other self-interaction vertices will appear, including vertices which 
describe the interactions of the BFKL Pomeron with the perturbative Odderon \cite{Bartels:1999yt,Janik:1998xj}. 

It is known that in NLO \cite{Lipatov:1985uk}, when the QCD coupling becomes momentum dependent, some infrared cutoff has to be introduced which, strictly speaking, already goes beyond perturbation theory. The BFKL Pomeron spectrum then becomes discrete, even at relatively short distances. This corresponds to the appearance of bound state Regge poles (Pomeron states), and it needs to be investigated which of them and in what way they will contribute to the scattering processes at large distances and large rapidities. In order to study the interactions of these Regge poles and in order to be able to move to larger distances it will be convenient to switch to bound state fields, to consider the local approximation and to make use of the well-known formalism of RFT. This step will make it much easier to include pomeron interactions which in the full QCD description are described by non local vertex functions. In the field theory based upon reggeized gluons these vertex functions would lead to an extremely hard problem to solve, including many details which we expect to be irrelevant and washed out at large distances.

We stress that both frameworks, BFKL physics and RFT, live in $2+1$-dimensions, transverse momenta (transverse distances) and rapidity (angular momentum). In order to study the transition from the UV to the IR region it is necessary to have a description which allows to trace the transmutation of the degrees of freedom at the different relevant scales. 
This can be controlled by introducing an infrared cutoff for the transverse momentum: initially (in the UV region) the momentum cutoff has to be large to justify the applicability of perturbation theory for QCD in the Regge limit. Lowering this cutoff brings us into the infrared, nonperturbative region: first still for the QCD description, then for the local RFT (with an IR cutoff).    
The main task then is to bring these two last pictures together: the bound states of the BFKL Pomeron define Pomeron fields which serve as an input into RFT. 
As a theoretical framework, we choose the IR regulated effective action, satisfying exact renormalization group equations, as the way to encode the different dynamical regimes.              

In this paper we perform a careful analysis of the BFKL Pomeron and its spectrum which, in the next (future) step, will be used to define the input to the exact RG (ERG) flow equations. To this end it is necessary to find a formulation of the BFKL Pomeron which is compatible with the ERG flow equations. We begin with the BFKL equation and introduce an infrared regulator as suggested by the RG equations. This regulator is different from all others that have been used before (e.g. Higgs mechanism \cite{Braun:1996tc,Levin:2016enb,Levin:2015noa,Levin:2014bwa}, infrared boundary values \cite{Lipatov:1985uk,Kowalski:2017umu,Kowalski:2015paa,Kowalski:2014iqa}). With this regulator we derive a differential equation for the BFKL Pomeron Green's function with respect to the IR cutoff parameter $k$. This equation is nonlinear and characterized by a two loop structure, i.e. has the form of the infrared evolution equations in QED and QCD, derived earlier by Lipatov and Kirschner 
\cite{Kirschner:1982xw,Kirschner:1982qf,Kirschner:1983di}.

In the next step we find an effective field theory which can be used to derive the BFKL Pomeron. We start from Lipatov's effective action and formulate an effective field theory for the 1PI vertices associated to composite states of two reggeized gluons. This field theory contains separate fields for the reggeized gluon and for elementary gluons. 
We also show how impact factors to external asymptotic particle states permit to construct, at the level of the effective action, a link to another description based upon Pomeron bound state fields
derived from the the BFKL equation.

Introducing the infrared regulator we derive, for the effective field theory of the reggeized gluon fields, exact functional RG flow equations. 
The flow equation for the 1PI 4-point function of the process: two reggeized gluons $\to$ two reggeized (i.e. the BFKL Green's function) appears  as part of an infinite set of coupled flow equations which also contain higher order processes, e. g. two reggeized gluons plus elementary gluon $\to$ two reggeized plus elementary gluon. Making use of special features of this field theory, in particular constraints derived from the MRK and features of Lipatov's effective action, we then construct explicitly the truncation needed to derive, from the functional RG flow equations, the same nonlinear differential equation for the BFKL Pomeron Green's function which was obtained before. In other words, we show, for the BFKL Green's function of two incoming and two outgoing reggeized gluons, that the infinite set of functional RG flow equations set is equivalent to a single nonlinear differential equation.  By construction the solution of the flow equation can be obtained from the knowledge of the IR regulated BFKL kernel and Green's function. 

In the second part of our paper we perform a numerical analysis of the BFKL Pomeron with our infrared regulator. In this part of our analysis we will not yet make use of the functional RG equations derived before, but, instead, find explicit solutions of the BFKL equation for a few fixed values of the infrared cutoff.  
To make contact with a previous analysis \cite{Levin:2016enb,Levin:2015noa,Levin:2014bwa} of the BFKL spectrum we begin with a numerical study of  
the BFKL Pomeron with a gluon mass regulator.\footnote{In our paper we will refer to  this mass 
regulator as "Higg's" mass which should not be confused with the physical Higgs boson.}  We then 
compute the BFKL spectrum using our infrared regulator which, in a future part of our analysis, will allow to use the BFKL Pomeron 
as an input to the RG flow. We compute energy eigenvalues (i.e. poles in the angular momentum plane), in particular intercepts and $q^2$ slopes of the trajectory functions and eigenfunctions of the BFKL kernel: these parameters define a set of Pomeron states. Their numerical values exhibit important features which will play a crucial role in the next step of our analysis: \\
(i) the intercepts and the $\bq^2$ slopes of the bound state poles decrease and go to zero, once we move from the leading eigenvalues down towards the accumulation point at $\omega=0$.\\
(ii) Only the eigenfunction of the leading trajectory is typically 'soft', i.e. it has support in the small-momentum region. The other nonleading eigenfunctions extend more and more into the UV region, i.e. they become 'hard'.\\
As we have already said, these results, in the next step of our program, can be used to define the (local) Pomeron fields which  serve as input to the RFT. Its infrared behavior will have to be matched to the results of our previous papers devoted to the IR behavior.    

The paper is organized as follows. We begin with the leading order BFKL equation (section 2), introduce an infrared regulator (section 3) and derive a nonlinear differential equation for the  dependence upon the  infrared regulator. In section 4 we define an effective field theory of reggeized gluons which describes the leading order BFKL Green's function  and discuss how to make a transition to bound state Pomeron fields. In section 5 we formulate the flow equations for this field theory: we obtain an infinite set of coupled equations. We then make use of a special feature of this field theory and derive, for the BFKL Green's function, a nonlinear equation which coincides with the equation derived in section 3. In sections 7 and 8 we perform our numerical studies.
In the concluding section we summarize and give an outlook on the following steps of our program.                
An appendix follows with a prescription of how to deal with the gluon fields which do not propagate in rapidity.

%%%%%%%%%%%%%%%%%%%%%%%%%%%%%%%%
\section{The setup: the LO BFKL equation}
%%%%%%%%%%%%%%%%%%%%%%%%%%%%%%%%

\begin{figure}[H]
\epsfig{file=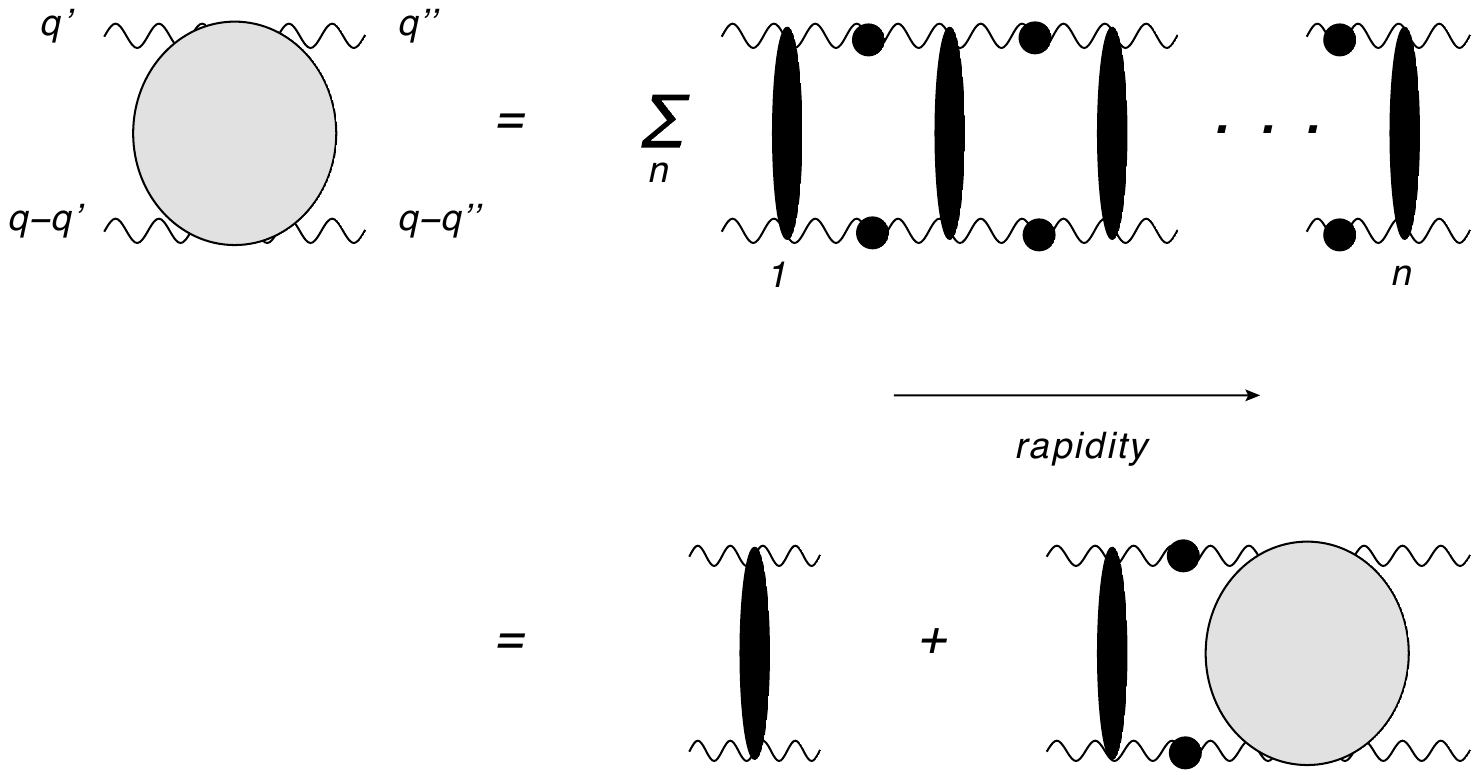,width=12cm,height=5cm}
\caption{the BFKL ladder} 
\label{Fig1}
\end{figure}

We begin by recalling the massless color singlet BFKL equation~\cite{BFKL} in the leading approximation  (MRK). 
Let us start with the amputated BFKL Green's function $G(\bq',\bq-\bq';\bq'',\bq-\bq''|\omega)$. It is obtained as an infinite sum of ladder diagrams and satisfies the integral (Bethe-Salpeter like) equation:
\ba
\label{BFKL-eq1}
 &&G_{\text{BFKL}}(\bq',\bq-\bq';\bq'',\bq-\bq''|\omega)=   K_{\text{BFKL}}(\bq,\bq-\bq';\bq'',\bq-\bq'') \\&&+ \int d^2\bk  \,K_{\text{BFKL}}(\bq,\bq-\bq';\bk,\bq-\bk)  \frac{1}{{\bq'}^2 (\bq-\bq')^2}
\frac{1}{\omega- \omega_g(\bk) -\omega_g(\bq-\bk)}\nonumber\\&& G_{\text{BFKL}}(\bk,\bq-\bk;\bq'',\bq-\bq''|\omega),\nonumber
\ea
where $\omega$ is the angular momentum variable dual to the rapidity in the scattering process and the $\bq$s are two 
dimensional transverse momentum variables.
The same equation can be illustrated as in Fig.~\ref{Fig1}.

The analytic expression of the LO BFKL kernel (the so called real part, induced by rapidity separated real gluon emissions) has the form:
\ba
\label{amp-BFKL-kernel}
K_{\text{BFKL}}(\bq',\bq-\bq';\bq'',\bq-\bq'') 
= \frac{\bar{\alpha}_s}{2\pi} \left(-\bq^2 +\frac{{\bq''}^2 (\bq-\bq')^2}{(\bq'-\bq'')^2} +  \frac{{\bq'}^2 (\bq-\bq'')^2}{(\bq'-\bq'')^2}\right),
\ea  
where 
\be
\bar{\alpha}_s=\frac{N_c \alpha_s}{\pi}.
\label{alphasb}
\ee
This kernel is illustrated  in Fig.~\ref{Fig2}a:
\begin{figure}[H]
\epsfig{file=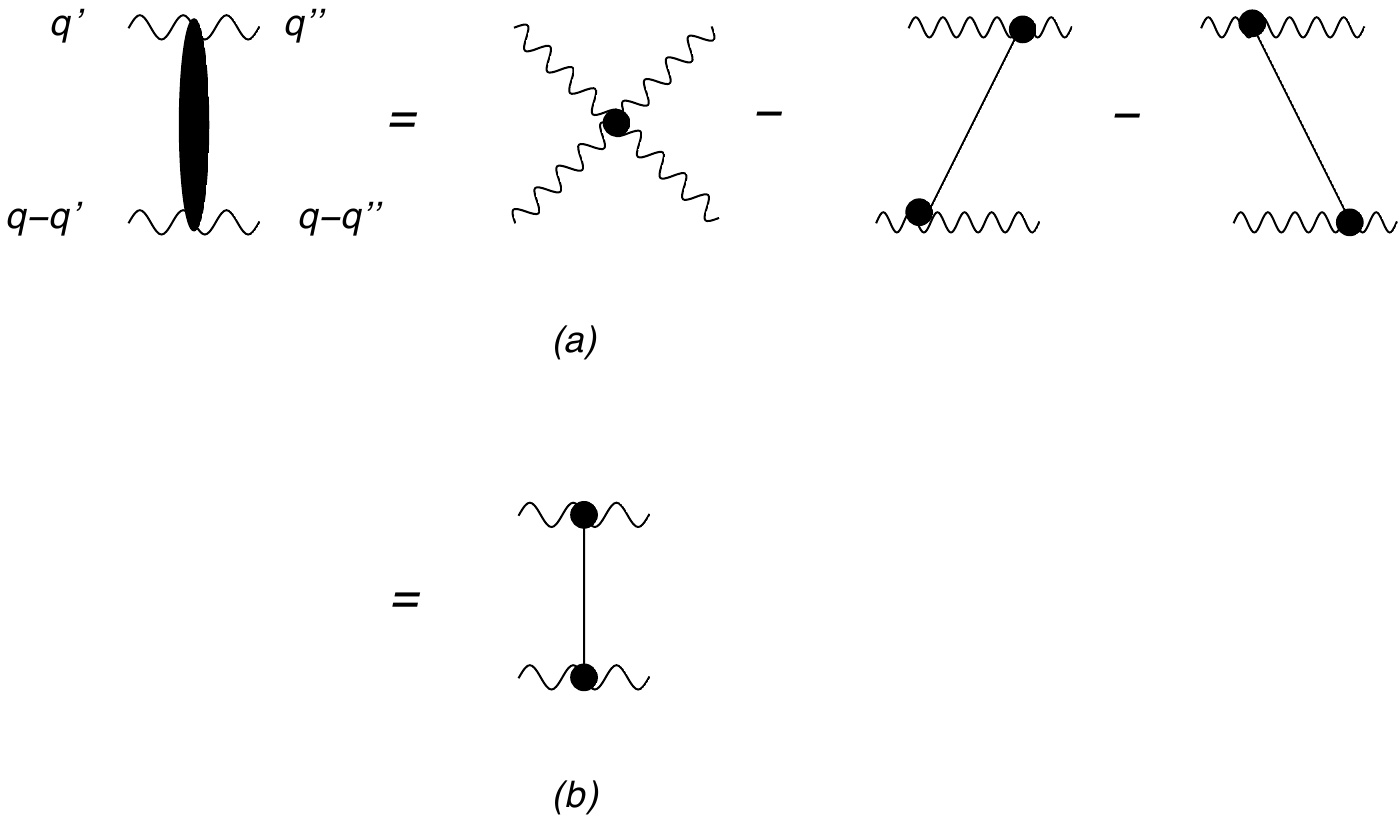,width=14cm,height=5cm}\\
\caption{Structure of the BFKL kernel with (a) real momenta and with (b) complex momenta.
The dots remind of the momentum factors of the vertices which are specified in (\ref{BFKL-kernel}).} 
\label{Fig2}
\end{figure}
The gluon trajectory function has the form:
\be
\alpha_g(\bq^2)= 1+\omega_g(\bq^2)
\ee
with 
\be
\omega_g(\bq^2)= -q^2 \frac{\bar{\alpha}_s}{4\pi}  \int d^2\bk \,\frac{1}{\bk^2 (\bq-\bk)^2}.
\ee 
It is this form of the BFKL equation which is best suited for implementing the ERG equations.   

It is well known that the BFKL equation can be reformulated to arrive at an IR safe version.
After modifying the kernel by including gluon propagators and symmetrizing 
\ba
\label{BFKL-kernel}
&&K(\bq,\bq-\bq';\bq'',\bq-\bq'') \nonumber\\ 
&&=\frac{1}{\sqrt{{\bq'}^2 (\bq-\bq')^2}} K_{\text{BFKL}}(\bq',\bq-\bq';\bq'',\bq-\bq'') 
\frac{1}{\sqrt{{\bq''}^2 (\bq-\bq'')^2}}\nonumber\\
&&= \frac{\bar{\alpha}_s}{2\pi} \frac{1}{\sqrt{{\bq'}^2 (\bq-\bq')^2}} \left(-\bq^2 +\frac{{\bq''}^2 (\bq-\bq')^2}{(\bq'-\bq'')^2} +  \frac{{\bq'}^2 (\bq-\bq'')^2}{(\bq'-\bq'')^2}\right) \frac{1}{\sqrt{{\bq''}^2 (\bq-\bq'')^2}}
\ea  
we introduce another Green's function $\tilde{G}$ which satisfies the equation 
\ba
\label{BFKL-eq2}
&&\tilde{G}(\bq',\bq-\bq';\bq'',\bq-\bq''|\omega)= \frac{1}{\omega}  K(\bq,\bq-\bq';\bq'',\bq-\bq'') \nonumber \\
&&+ \frac{1}{\omega} \int d^2k  \tilde{K}(\bq,\bq-\bq';\bk,\bq-\bk) \,\,
 \tilde{G}(\bk,\bq-\bk;\bq'',\bq-\bq''|\omega),
\ea 
where $\tilde{K}$ now contains a real part (gluon emission) and a virtual part (gluon trajectory):
\be
 \tilde{K}(\bq,\bq-\bq'',\bq-\bq'')= K(\bq,\bq-\bq';\bq'',\bq-\bq'')+ \delta^{(2)}(\bq'-\bq'') \left( \omega_g(q')+\omega_g(q-q')\right).
\ee

In this form, the IR singularities of the BFKL kernel can explicitly be seen to cancel. The connection between the two Green's function is given by:
\ba
&&\tilde{G}(\bq',\bq-\bq';\bq'',\bq-\bq''|\omega)\\
&&=\frac{1}{\Omega(q')}\frac{1}{\sqrt{{\bq'}^2 (\bq-\bq')^2}}G_{\text{BFKL}}(\bq',\bq-\bq';\bq'',\bq-\bq''|\omega) \frac{1}{\sqrt{{\bq''}^2 (\bq-\bq'')^2}} , \nonumber
\ea
where we have put 
\be
\Omega(\bq')=\omega-\omega_g(\bq')-\omega_g(\bq-\bq').
\ee

So far we have limited ourselves to the leading order (LL) BFKL equation with a fixed coupling $\alpha_s$. To get closer to realistic physics we need to include the momentum dependent coupling, $\alpha_s(\bq^2)$ which appears in NLO. In this paper, as a first step, we will adopt the approximation of using, for the BFKL kernel, the leading terms only and introduce the running coupling (with some prescription to be discussed later). For the fixed coupling $\alpha_s$ we simply substitute the leading order running coupling $\alpha_s(\bq^2)$, which at a fixed low momentum scale will be approximated by a constant value. The infrared cutoff which will be introduced in the following 
section has to lie above this 'freezing' scale.

One can express the BFKL Green's function in terms of eigenfunctions of the 
kernel $\tilde{K}$. In the functional RG approach we shall work with coarse-grained quantities controlled by an IR cutoff $k$ which will be specified below. We shall therefore add this $k$ dependence in the kernel, the trajectory function $\Omega$ and Greens's functions. In the presence of this cutoff the eigenvalue spectrum of the BFKL kernel then consists of a discrete and a continuum part (in the Regge limit we expect that the continuous part  gives a small correction to the high energy behavior). We write:   
\ba
{\cal{G}}_k&=&\frac{1}{\omega- \tilde{K}}\nonumber\\
&=&\sum_n \frac{\psi_{n,k}(\bq',\bq-\bq') \psi_{n,k}^*(\bq'',\bq-\bq'')}{\omega -\omega_{n.k}}+{\rm continuous \,part},
\label{Gtilde}
\ea
where the $\psi_n$ are the eigenfuctions of the  (symmetrized) BFKL kernel  $\tilde{K}$ which we will compute numerically in the second part of this paper.  The sum of the discrete set and
the continuous part of eigenfunctions defines a complete set. This Green's function ${\cal{G}}_k$ satisfies the equation 
\be
{\cal{G}}_k= \frac{1}{\omega} + \frac{1}{\omega} \tilde{K}_k {\cal{G}}_k,
\ee
where the inhomogeneous term contains the delta functions $\delta^{(2)}(\bq'-\bq'')$.
We also can write:
\be
\label{Gtilde1}
{\cal{G}}_k= \frac{1}{\Omega_k(\bq')}+\frac{1}{\Omega_k(\bq')}K_{k} \frac{1}{\Omega_k(\bq'')}+...\,.
\ee
This Green's function ${\cal G}_k$ can be named as the (nonamputed) Green's function of four reggeized gluons.  

It is easy to see the connection of ${\cal{G}}_k$ with our previous Green's function $\tilde{G}_k$: 
\be
\tilde G_k={\cal {G}}_k K_{k},
\label{relation1}
\ee
which equivalently can be written as 
\be
\tilde{G}_k +1 = \,\,{\cal{G}}_k  \times \Omega_k(\bq'') . 
\ee

In the flow equations, we will deal with amputated 1PI vertex functions generator $\Gamma_k$. The most interesting one, $\Gamma_k^{(4)}$, is  
obtained from $G_{\text{BFKL}}$  by subtracting the BFKL kernel:
\ba
&&\Gamma_{k}^{(4)}(\omega;\bq',\bq-\bq';\bq'',\bq-\bq'')
\nonumber\\
&&=G_{k} (\bq',\bq-\bq';\bq'',\bq-\bq''|\omega)-K_{\text{BFKL};k}(\bq',\bq-\bq';\bq''-\bq-\bq'').
\ea
It can also be written in the following form:
\ba
&&\Gamma_{k}^{(4)}(\bq',\bq-\bq';\bq'',\bq-\bq''|\omega)
=\int d^2 \bl  \,K_{\text{BFKL};k}(\bq',\bq-\bq';\bl,\bq-\bl)\nonumber\\
&&\frac{1}{\Omega_k(\bl)}
 \frac{1}{\bl^2 (\bq-\bl)^2} 
G_{\text{BFKL};k}(\bl,\bq-\bl;\bq'', \bq-\bq'';\omega)
\ea
or, alternatively,
\ba
&& \Gamma_{k}^{(4)}(\omega;\bq',\bq-\bq';\bq'',\bq-\bq'')
\nonumber\\
&&=\int d^2 \bl \,d^2\bl' \,K_{\text{BFKL};k}(\bq',\bq-\bq';\bl,\bq-\bl)
 \frac{1}{\sqrt{\bl^2 (\bq-\bl)^2}} \nonumber\\
&&\hspace{2cm} \cdot {\cal{G}}_k (\bl,\bq-\bl;\bl',\bq-\bl'|\omega)\nonumber\\ && 
\cdot  \frac{1}{\sqrt{{\bl'}^2 (\bq-\bl')^2}} K_{\text{BFKL};k}(\bl',\bq-\bl';\bq'',\bq-\bq'').
\ea
In short,
\be
\Gamma_k^{(4)}= K_{\text{BFKL};k} \otimes {\cal{G}}_k \otimes K_{\text{BFKL};k},
\ee
where the symbol $\otimes$ includes the square roots of the propagators.

Finally, in an elastic scattering process (e.g. $\gamma^*\gamma^*$ scattering) the BFKL Green's function couples to impact factors $\Phi(\bq',\bq-\bq')$ (which go to zero when $\bq' \to 0$ or $\bq-\bq'\to 0$) and leads to an infrared finite scattering amplitude:
\ba
\label{T_elastic}
T_{el}(s,t)&&= i s \int \frac{d\omega\, d^2\bq' d^2 \bq''}{2\pi i} (-s)^{\omega} \Phi(\bq',\bq-\bq') 
\frac{1}{\sqrt{{\bq'}^2(\bq-\bq')^2}} \nonumber\\
&&\hspace{1cm}\times  {\cal{G}}_k  \frac{1}{\sqrt{{\bq''}^2(\bq-\bq'')^2}}\Phi (\bq'',\bq-\bq'')\nonumber\\
&&=is \int \frac{d \omega}{2 \pi i}  (-s)^{\omega}\Phi \otimes  {\cal{G}}_k \otimes \Phi.
\ea     
With (\ref{Gtilde}) one easily derives expressions for the coupling of a single BFKL  pole to the 
external particle.

Eq. (\ref{T_elastic}) can also be expressed in terms of the 1PI vertex function  $\Gamma_{k}^{(4)}$. Starting from (\ref{Gtilde1}) we write 
\ba
\label{Gtilde2}
{\cal{G}}_k&=&\frac{1}{\Omega_k} + \frac{1}{\Omega_k} K_k \frac{1}{\Omega_k} + \frac{1}{\Omega_k} K_k {\cal{G}}_k  K_k \frac{1}{\Omega_k}\nonumber\\ 
&=&\frac{1}{\Omega_k} + \frac{1}{\Omega_k} K_k \frac{1}{\Omega_k} + \frac{1}{\Omega_k} \otimes 
\Gamma_k^{(4)} \otimes \frac{1}{\Omega_k}.
\ea
The  last line of (\ref{T_elastic} then becomes:
\ba
\hspace{-0.7cm}T_{el}(s,t)&&=is \int \frac{d \omega}{2 \pi i} (-s)^{\omega} \Phi \otimes \Big[
\frac{1}{\Omega_k} + \frac{1}{\Omega_k} \otimes K_{\text{BFKL} k} \otimes \frac{1}{\Omega_k} + 
\frac{1}{\Omega_k} \otimes \Gamma^{(4)} \otimes \frac{1}{\Omega_k}\Big]
\otimes \Phi.
\ea
Here the symbols $\otimes$ are written in order take care of the momentum propagators.  
%%%%%%%%%%%%%%%%%%%%%%%%%%%%%%%%
\section{Regulators  and the $\tau$-derivative of the  BFKL Geen's function}
\label{sect3}
%%%%%%%%%%%%%%%%%%%%%%%%%%%%%%%%

In order to investigate the implementation in a Functional RG setup of the perturbative BFKL Pomeron
it will be necessary to introduce an infrared cutoff.
An obvious candidate is the Higgs mass for the gluon propagator. This has been used to study the energy spectrum of the Higgs mass regulated BFKL Pomeron~\cite{Levin:2016enb,Levin:2015noa,Levin:2014bwa}: for positive $\omega$ values there is a discrete set of infinitely many Regge poles (with accumulation point at $\omega=0$), accompanied by a cut along the negative $\omega$ axis. This spectrum has been studied numerically, both for fixed and for running coupling $\alpha_s$. An earlier attempt~\cite{Braun:1996tc} has started from the assumption of the bootstrap equation for gluon reggeizzation: in this model the discrete spectrum was analysed, both for intercept and slope. The model represents a deformation of the BFKL Pomeron equation, including most of NLO corrections and, in particular, the running coupling with an IR-regulator. In the UV region it matches the  DGLAP results in the double logarithmic limit (large $Q^2$ and $1/x$, using standard Deep Inelastic Scattering variables). Most recently another approach has been pursued in \cite{Kowalski:2017umu,Kowalski:2015paa,Kowalski:2014iqa}: using the massless BFKL equation in the forward direction, infrared boundary conditions have been imposed at a fixed momentum scale $k_0^2$. The resulting BFKL equation has then been used to fit the small-x and low-$Q^2$ HERA data, which allows to fix the infrared boundary values. 

All these approaches only use the BFKL Pomeron, and so far no attempt has been made to introduce the triple Pomeron vertex and to study the Regge cut corrections. 

In our approach we aim at embedding the BFKL Pomeron into a reggeon field theory which includes corrections to the 
BFKL Pomeron based upon interaction vertices, in particular the triple Pomeron vertex. A necessary first step is the introduction of an infrared regulator which later on will allow to make use of the exact renormalization group equations.
In the following we describe this regulator for the BFKL Pomeron and derive an equation for the derivative of the BFKL Pomeron Green's function with respect to the IR cutoff parameter $k$. We obtain a nonlinear equation which is of the same form as the infrared evolution equations introduced by Lipatov and Kirschner \cite{Kirschner:1982xw,Kirschner:1982qf,Kirschner:1983di}. Later on we will show that, with this regulator, the BFKL Pomeron can be formulated within the exact renormalization group approach using a truncation for the effective average action compatible with the MRK.

To implement the coarse-graining characterizing the Wilsonian RG approach we introduce the following momentum regulator for the elementary gluon propagator 
inside the BFKL kernel and the gluon trajectory function:
\be
\label{glu-prop}
\frac{1}{\bq^2} \rightarrow \frac{1}{\bq^2+R_k(\bq^2)}
\ee
with the optimized form~\cite{Litim}
\be
\label{reg1}
R_k(\bq^2)  = (k^2-\bq^2) \theta(k^2-\bq^2).      
\ee
Alternative choices can be taken, e.g. such as
\be
\label{reg2}
R_k(\bq^2)=\frac{\bq^2}{e^{\frac{\bq^2}{k^2}}-1}.
\ee

These regulators are such that the propagation of quantum fluctuations is inhibited for $\bq^2\ll k^2$, but is unaltered in the UV region.
This fact is implemented in the functional integral which defines the IR-regulated effective action.
In this paper we will use the regulator (\ref{reg1}). For the reggeized gluon with the bare propagator 
\be 
 \frac{1}{\bq^2} \frac{1}{\omega-\omega_g(\bq^2)}
 \label{reg-gluon_prop_nok}
\ee
we will use 
\be
\label{reg-gluon-prop}
\frac{1}{\omega (\bq^2+R_k(\bq^2))-\bq^2 \omega_{g,k}(\bq^2) },
\ee
where the regulated trajectory function is given by
\be
\label{reg-traj}
\omega_{g, k}(\bq^2)=-\bq^2 \frac{\bar{\alpha}_s}{4\pi}  \int d^2\bl \frac{1}{\big[\bl^2+R_k(\bl^2)\big] 
\big[(\bq-\bl)^2+R_k((\bq-\bl)^2)\big]}.
\ee
For the kernel $K_{\text{BFKL}}$ we find:
\ba
\label{reg-amp-BFKL-kernel}
&&K_{\text{BFKL};k}(\bq',\bq-\bq';\bq'',\bq-\bq'')  \nonumber\\
&&= \frac{\bar{\alpha}_s}{2\pi} \left(-\bq^2 
\frac{(\bq'-\bq'')^2}{(\bq'-\bq'')^2+R_k((\bq'-\bq'')^2)} 
 +\frac{{\bq''}^2 (\bq-\bq')^2+{\bq'}^2(\bq-\bq'')^2}{(\bq'-\bq'')^2+R_k((\bq'-\bq'')^2)}\right)
\ea  
with the corresponding Green's function $G_{\text{BFKL};k}$:
\ba
\label{reg-BFKL-amp}
 &&G_{\text{BFKL};k}(\bq',\bq-\bq';\bq'',\bq-\bq''|\omega)=   K_{\text{BFKL};k} (\bq,\bq-\bq';\bq'',\bq-\bq'') \\&&+ \int d^2\bk \, K_{\text{BFKL};k}(\bq,\bq-\bq';\bk,\bq-\bk)  \frac{1}{({\bq'}^2+R_k({\bq'}^2) ((\bq-\bq')^2+R_k((\bq-\bq')^2))}\nonumber\\
&&\cdot \frac{1}{\omega- \omega_{g,k}(\bk^2) -\omega_{g,k}((\bq-\bk)^2)} G_{\text{BFKL};k}(\bk,\bq-\bk;\bq'',\bq-\bq''|\omega)\,.\nonumber
\ea

With these regulators we take the derivative of the BFKL Green's function with respect to $\tau=\ln k/k_0$. 
\begin{figure}[H]
\epsfig{file=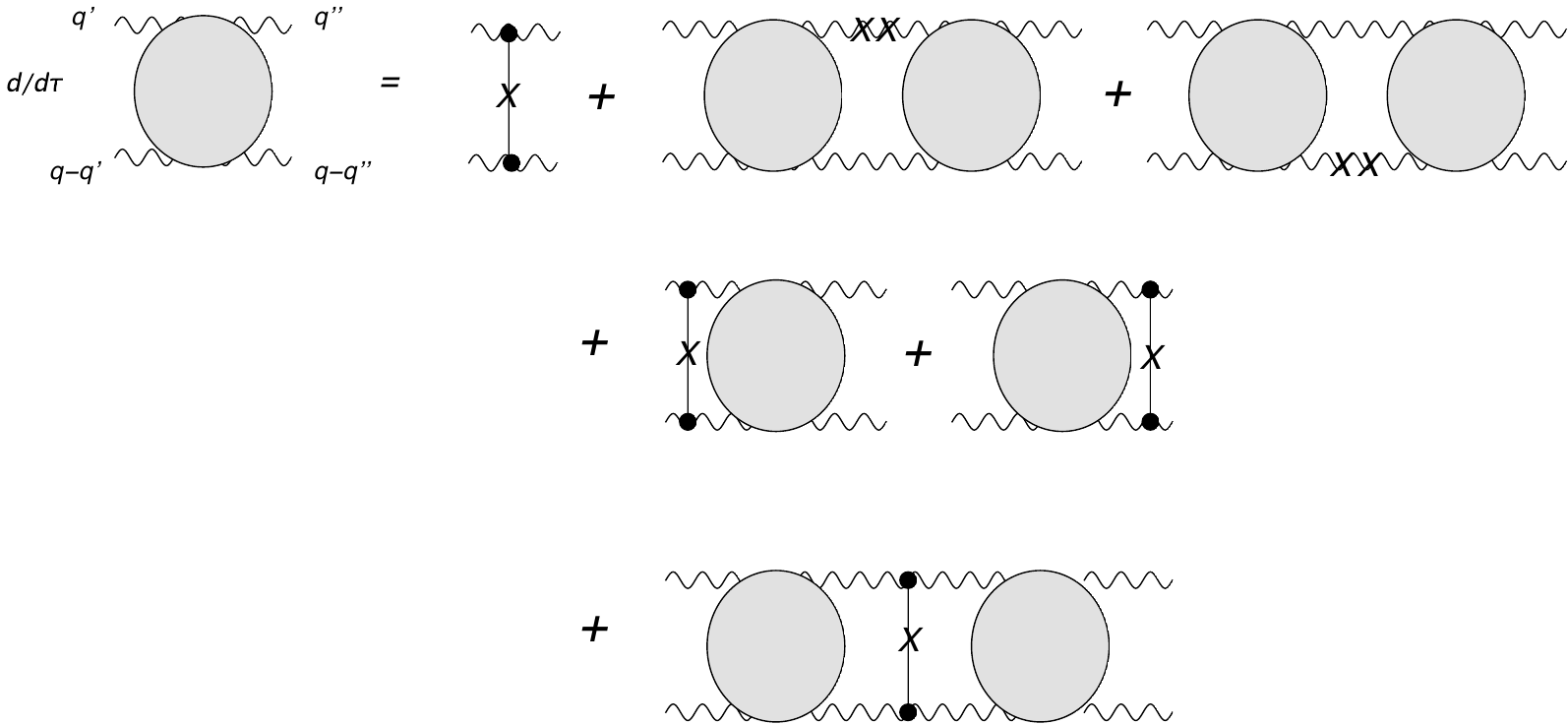,width=14cm,height=6cm}
\caption{k-derivative of the BFKL ladder} 
\label{Fig3}
\end{figure} 
We start from the integral equation in Fig.~\ref{Fig1} and obtain the result illustrated in Fig.~\ref{Fig3},
where the internal elementary gluon (straight line) has the propagator given in (\ref{glu-prop})
whereas the reggeized gluon (wavy line) has the form described  in (\ref{reg-gluon-prop}).
The crosses on internal lines denote the derivatives with respect to $\tau=\ln k/k_0$. For 
the elementary gluon inside the kernel we find:
\be
-\frac{1}{\bq^2+R_k(\bq^2)} \dot{R}_k(\bq^2) \frac{1}{q^2+R_k(\bq^2)}
\ee
with 
\be
\dot{R}_k(\bq^2)=k\frac{d}{dk} R_k(\bq^2)=\frac{d}{d\tau} R_k(\bq^2)=2 k^2 \theta(k^2-\bq^2),
\ee
whereas for the reggeized gluon propagator
this derivative consists of two contributions, marked by a double cross:
\be
\frac{d }{d \tau} G_{regge}(\bq^2) = -G_{regge}(\bq^2) \omega \dot{R}_k(\bq^2) G_{regge}(\bq^2) + G_{regge}(\bq^2) \bq^2 \frac{d  \omega_{g,k}(\bq^2) }{d \tau} G_{regge}(\bq^2)
\ee
with 
\be
\frac{d \omega_{g,k}(\bq^2) }{d \tau}=2 \bq^2 \frac{\bar{\alpha}_s}{4\pi}  \int d^2\bl \left( \frac{1}{\bl^2+R_k(\bl^2)}  \dot{R}_k(\bl^2) \frac{1}{\bl^2+R_k(\bl^2)}\right) \frac{1} {(\bq-\bl)^2+R_k((\bq-\bl)^2)}.
\ee
These two terms are illustrated in Fig.~\ref{Fig4}:
\begin{figure}[H]
\begin{center}
\epsfig{file=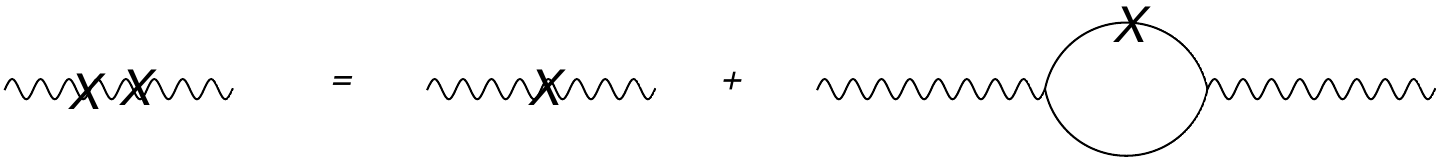,width=8cm,height=1cm}
\end{center}
\caption{k-derivative of the reggeized gluon }
\label{Fig4}
\end{figure} 

As already mentioned, the nonlinear structure of this equation is the same as in the infrared evolution equation of Lipatov and Kirschner  \cite{Kirschner:1982xw,Kirschner:1982qf,Kirschner:1983di}, derived for summing double logs. A more detailed analysis of this equation will be performed in a future paper. 

%%%%%%%%%%%%%%%%%%%%%%%%%%%%%%%%
\section{An effective  field theory for deriving the BFKL Green's function}
%%%%%%%%%%%%%%%%%%%%%%%%%%%%%%%%

Before we perform a numerical analysis of the BFKL equation with this infrared regulator in order to be able to select the leading contributions, 
we would like to demonstrate that the same  equation for the $\tau$-derivative can be obtained from the RG flow equations for a suitable truncation of the associated effective average action in the MRK. 
To this end we will not start from the full QCD Lagrangian but from the effective field theory derived by 
L.~Lipatov~\cite{Lipatov:1995pn}: this effective action has been designed to describe the Regge limit of QCD, 
and in contrast to the QCD Lagrangian it introduces a new degree of freedom, the field of the reggeized gluon. 
We therefore start from this effective action and formulate an effective field theory for which 
we shall then derive the consistent flow equations.
 In order to switch later on to the RFT description at larger distances one needs to rewrite the flowing dynamics of reggeized gluons in terms of Pomeron fields (associated to composite states of reggeized gluons at the same rapidity). This will be  illustrated in the second part of this section.

%%%%%%%%%%%%%%%%%%%%%%%%%%%%%%%%
\subsection{Elements of the field theory}
%%%%%%%%%%%%%%%%%%%%%%%%%%%%%%%%

\begin{figure}[H]
\epsfig{file=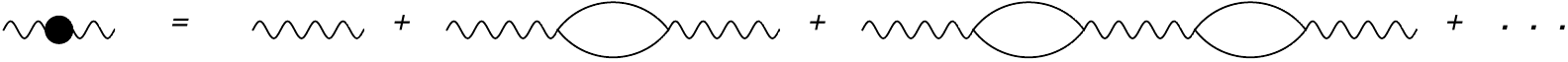,width=14cm,height=0.8cm}
\caption{the propagator of a reggeized gluon (see text)}
\label{Fig5}
\end{figure}

There are a few basic ingredients of the effective action which we summarize as follows. First, when calculating in perturbative QCD the high energy behavior of, say, a 2 by 2 scattering process, the t-channel is divided into intervals with small and large rapidity separations. Those with small rapidities are described by the normal QCD degrees of freedom (gluons, quarks) whereas the large rapidity intervals are described by the exchange of reggeized glouns. As described in detail in \cite{hentschinski} the propagator of a reggeized gluon (given in Eq.(\ref{reg-gluon_prop_nok})) is built up successively. We illustrate this in Fig.~\ref{Fig5},          
\noindent
where the wavy line stands for the (bare) reggeized gluon propagator,
\be
\label{bareprop}
\frac{1}{\omega} \frac{1}{\bq^2}
\ee
and the bubbles on the rhs  denote the expression
\be
\bq^2 \omega_g(\bq^2)\,.
\ee
The sum of all diagrams equals the (dressed) reggeized gluon propagator in (\ref{reg-gluon_prop_nok}). In terms of rapidity, in Fig.~\ref{Fig5} each wavy lines 
belongs to a rapidity gap and generates one power of $\ln s$ (being $s$ the squared center of mass energy in the scattering process), whereas the bubble has no propagation in rapidity. Formally we therefore introduce a gluon propagator which propagates in transverse momentum but not in rapidity (for details see Appendix A). 
The (real) kernel of the BFKL equation in (\ref{BFKL-kernel}) belongs to small rapidity interval: in LO the extension can be taken to zero. Therefore,
the kernel also has to be described by fields which to not propagate in rapidity. It is well known, the kernel simplifies if we use complex momenta $q=q_1+iq_2$, $q^*=q_1-iq_2$, $\partial=\partial_1+i\partial_2$, $\partial^*=\partial_1-i\partial_2$ (cf. Fig.~\ref{Fig2}b).
With the identity
\be
\label{compl-vertex1}
q'(q-q')^* {q''}^*(q-q'') +cc=-\bq^2(\bq'-\bq'')^2 +{\bq'}^2 (\bq-\bq'')^2+(\bq-\bq')^2 {\bq''}^2
\ee
we obtain
 \be
\label{compl-vertex2}
K_{\text{BFKL}}= \frac{\bar{\alpha}_s}{2\pi}\Big[ q' (q'')^* \frac{1}{( \bq'-\bq")^2} (q-q')^*(q-q'') + cc\Big].
\ee 
This suggests to introduce, for the exchanged gluon with momentum $(q-q')$, another gluon field which has no evolution in rapidity. 
We therefore introduce the following (color octet) fields (the subscript 'a' denotes the color):\\
(1) $A_a,\Adag_a$ for the (bare) reggeized gluon with the propagator $\frac{1}{\omega} \frac{1}{\bq^2}$. 
In order stay with the planar BFKL ladders we distinguish (in a horizontal ladder) between the upper and lower reggeized gluons and introduce two different fields:
 $A_{1a},\Adag_{1a}$ and  $A_{2a},\Adag_{2a}$.\\
(2) To generate the trajectory functions we need, for each reggeized gluon, a scalar gluon: $a_{1a},\adag_{1a}$ and $a_{2a},\adag_{2a}$ which has no propagation in rapidity. Their propagators are simply $1/\bq^2$. 
\\
3) For the produced gluon inside the BFKL kernel we need another field, $\chi_a,\chidag_a$  without propagation in rapidity.  
\begin{figure}[H]
\begin{center}
\epsfig{file=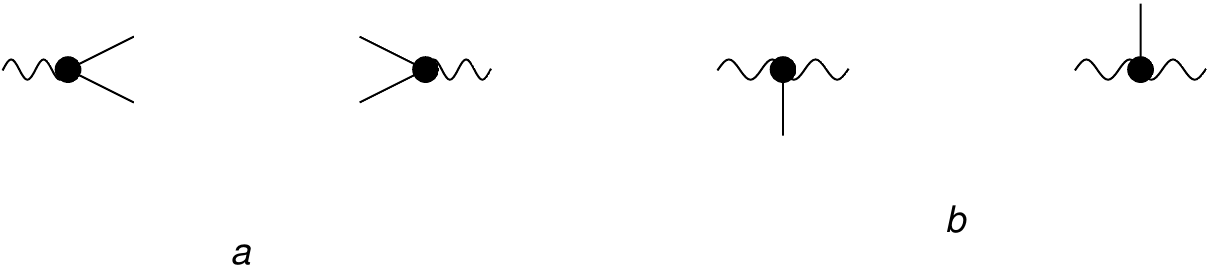,width=12cm,height=2cm}
\end{center}
\caption{the interaction vertices.}
\label{Fig6}  
\end{figure}

The interaction vertices are illustrated in Fig.~\ref{Fig6}.
In the Lagrangian they correspond to:
\be
g f_{abc} \left( (\bpartial^2\Adag_{ia}) a_{ib} a_{ic}+ \adag_{ia}\adag_{ib}  (\bpartial^2A_{ic })\right) \,\,\,\,(i=1,2),
\ee
\be
g f_{abc} \Big[(\partial \Adag_{ia}) (\partial^* A_{ib)} \chi_c + (\partial^* \Adag_{ia}) (\partial A_{ib}) \chidag_c\Big]\,\,(i=1,2),
\ee
for Fig.~\ref{Fig6}a and Fig.~\ref{Fig6}b respectively.

It is not difficult to construct, using these ingredients, all-order amputated vertex functions. In particular, after summation over the self-energies the bare propagator of the reggeized gluon becomes:
\ba
&&\frac{1}{\omega \bq^2} \Big[1+ \bq^2 \omega_g(\bq^2) \frac{1}{\omega \bq^2} +  \left( \bq^2 \omega_g(\bq^2) \frac{1}{\omega \bq^2} \right)^2 +... \Big] \nonumber\\
&&= \frac{1}{\bq^2} \frac{1}{\omega-\omega_g(\bq^2)}.
\label{reg-gluon}
\ea

\begin{figure}[H]
\begin{center}
\epsfig{file=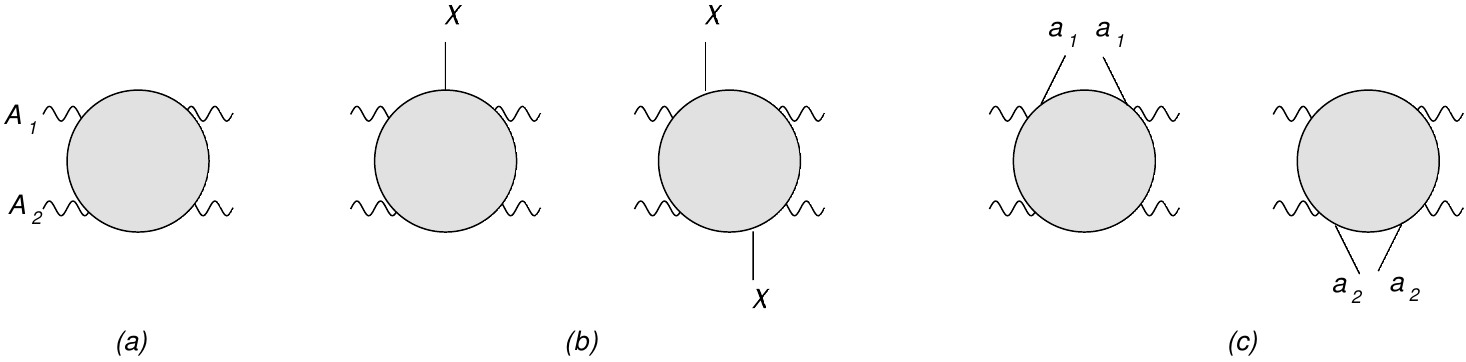,width=12cm,height=3cm}
\caption{vertex functions.
(a) elastic process (b) emission of one or two $\chi$ fields (c) emission of $a_1$ or $a_2$)}.
\label{Fig7}
\end{center}
%\label{Fig7}
\end{figure}

A few observations maybe useful. By construction we implement the interactions in the MRK and therefore:\\
(i) for any vertex function, in the strict MRK, the number of reggeized gluons is conserved;\\
(ii) there is no renormalization of vertices. In particular, there are no vertex corrections to the local vertex $A \Adag \chi$, since the $\chi$ field does not propagate in rapidity (see Appendix A).
 \\
With these remarks it is easy to see that the all order four point vertex function non local in rapidity and transverse space for the elastic process (reggeized gluon 1 + reggeized gluon 2 $\rightarrow$  reggeized gluon 1 + reggeized gluon 2) as in Fig.~\ref{Fig7}a  coincides with (\ref{BFKL-eq1}).  

Eventually one may think of relaxing the condition (i) by working in the generalized leading log approximation MRK where $n\to m$ reggeized gluon vertices (with $n-m$ even by signature conservation) are allowed. In such a case the vertices with reggeized gluons will be renormalized by loops. Since the related truncation will be very difficult to deal with at the level of reggeized gluon fields, we shall study effects related to nonconservation of the number of reggeized gluons
only after switching to composite fields and using the local approximation. Moreover, we expect that several details of the reggeized gluon dynamics will be mostly washed out by interactions of the composite fields, and the main global features will be captured by the much more simple local RFT description.

In addition to the BFKL Green's function, this field theory describes an infinite number of other inelastic non local vertex functions. As we will see below, the flow equation for the BFKL 4-point function includes non local vertex functions with 4 reggeized gluons plus elementary gluons. There can be additional $\chi$-fields or $a$-fields, produced at the same rapidites. Examples are shown in Fig.~\ref{Fig7}b (right) and c. Although in some of these vertex functions the two reggeized gluons can no longer be in color singlet states, it is important to stress that, in the context of the flow equations for the BFKL Green's function, color nonsinglet states never appear: 
in all cases of interest pairs of external $a$ or $\chi$ fields will be contracted.  As a result, states of two reggeized gluons will always remain in color singlet states.   

All these inelastic vertex functions in (b) and (c), however,  can be expressed in terms of the elastic
vertex function and $3$-point vertices. For the cases illustrated in Fig.~\ref{Fig7}b we find a structure illustrated in Fig.~\ref{Fig8} and Fig.~\ref{Fig9}:
\begin{figure}[H]
\begin{center}
\epsfig{file=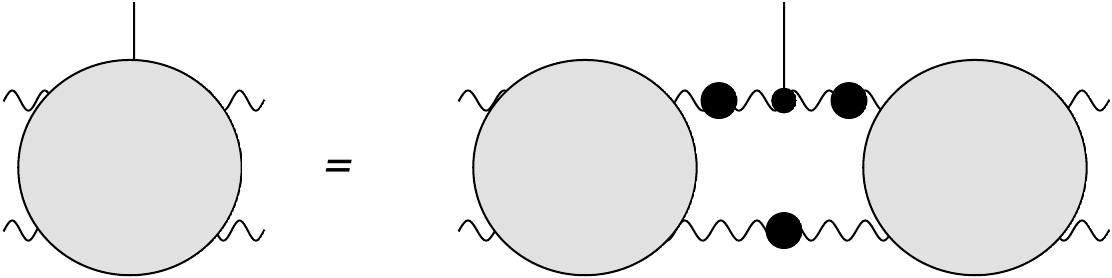,width=8cm,height=2cm}
\caption{5-point vertex function.
\label{Fig8}}  
\end{center}  
%\label{Fig8}
\end{figure}
\begin{figure}[H]
\begin{center}
\epsfig{file=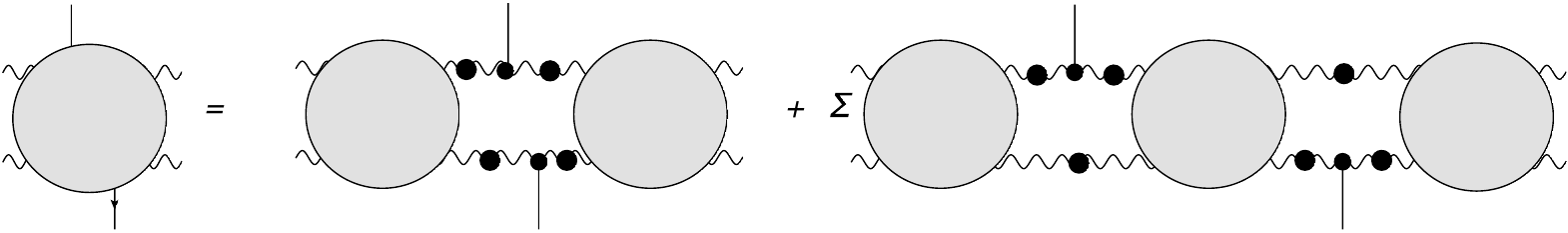,width=14cm,height=2.5cm}
\caption{A 6-point-vertex function.
\label{Fig9}}
\end{center}
%\label{Fig9}
\end{figure}
For this second example it is important to note the order in rapidity:
the left and right gluons are emitted (or absorbed) at rapidities $y_1$, $y_2$, resp., and 
the ordering means that $y_1 \le y_2$. Similarly, for Fig.~\ref{Fig7}c we find:
\begin{figure}[H]
\begin{center}
\epsfig{file=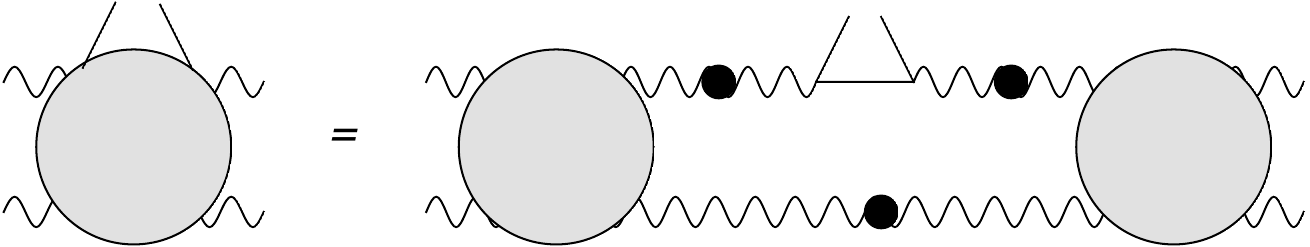,width=11cm,height=2cm}
\caption{Another 6-point-vertex function.
\label{Fig10}}
\end{center}
\end{figure}
As we will see later,  with these identities it will be possible to derive, from the flow equations, the same nonlinear equation as in section~\ref{sect3}.

%%%%%%%%%%%%%%%%%%%%%%%%%%%%%%%%
\subsection{From reggeized gluons to Pomeron fields}
%%%%%%%%%%%%%%%%%%%%%%%%%%%%%%%%

At large distances the  interactions in the high energy limit are naturally described in terms of colorless objects, such as Pomeron or Odderon Green's functions and their interacting vertices. 
In particular, for the BFKL Pomeron it is natural to shift from the description in terms of reggeized gluons to bound state fields, Pomeron fields. As we have already said before, in the Regge limit one has high energy factorization, that is a scattering amplitude is decomposed in impact factors (depending on the external particles such has hadrons) which couple to Pomeron states evolving 
with Green's functions. In the following we will give a short description of how this transition to 
bound state fields is carried out, keeping in mind that we work with the generator of 1PI proper vertices. In addition to the BFKL Pomeron  
 Green's function which we discussed in some detail
we shall be a bit more general by allowing also higher order reggeized gluon transition vertices, 
in particular the $2\to 4$ gluon vertex which leads to the triple Pomeron vertex.  

To begin with, it is useful to introduce the impact factors $\Phi$ and $\Phi^\dagger$, which depend on one rapidity variable and on two transverse momenta. This is done by introducing in the bare Wilsonian action source terms which couple to the composite state of two reggeized gluons in color singlet states at the same rapidity. We write a standard local term in the path integral:
\be
e^{ \int \left( \Phi^\dagger A A+ A^\dagger A^\dagger \Phi \right)}.
\ee
As we outlined above, the solution to the BFKL equation is encoded in the Green's function
${\cal{G}}$ which, by means of~(\ref{Gtilde}), can be expressed in terms of eigenfunctions and eigenvalues of the kernel $\tilde{K}$. 
This Green's function is related to the 1PI vertex functions involving reggeized gluons ( the explicit relation to $\Gamma_k^{(4)}$
is given in (\ref{Gtilde2})) of the effective action 
$\Gamma_k$ and is one element which appears in the dependence of the latter on the "sources" $\Phi$ and $\Phi^\dagger$, which were not undergoing the Legendre transform. 
Let us consider a truncation which contains the 1PI vertices $\Gamma^{(4)}_{A^\dagger A^\dagger A A}$.
Once we allow for the $2 \to 4$ reggeized gluon transitions we also have $\Gamma^{(6)}_{A^\dagger A^\dagger A^\dagger A^\dagger A A}$ and $\Gamma^{(6)}_{A^\dagger A^\dagger A A A A}$. Writing 
the effective action as 
\be
\Gamma[A,A^\dagger, \cdots; \Phi,\Phi^\dagger]=\sum_i \Gamma_i,
\ee
where in $\Gamma_i$ the subscript $i-$ denotes the power of the sources $\Phi$ and $\Phi^\dagger$, the lowest non trivial order terms are:
\ba
\Gamma_2 &=& \Phi^\dagger {\cal G} \Phi+ \cdots \nonumber\\
\Gamma_3 &=&\left( \Phi^\dagger {\cal G} V_{2\to 4} {\cal G} \Phi  {\cal G} \Phi  +{\rm h.c.} \right)+\cdots,
\label{truncation_source}
\ea
where the dots stand for more complicated terms with may include also external reggeized gluon fields, e.g.
\be
\Gamma_1 = \int \left( \Phi^\dagger {\cal G} A A + {\rm h.c.} \right)+\cdots \,.
\ee

These examples show why it is useful to introduce sources for the composite states of reggeized gluons at fixed rapidity. As we have shown in section 2, it is not straightforward to relate in a 1PI effective action the proper vertices $\Gamma^{(4)}$ or $\Gamma^{(6)}$ to the IR safe BFKL Green’s function associated to the Pomeron. Using sources gives us access to the IR safe dynamics of the reggeized gluons, in particular to the composite states of $2$, $4$, and $6$ reggeized gluons in the MRK and to the effective vertex $V_{2\to4}$ (related to $\Gamma^{(6)}$), which is local in rapidity and represents the key ingredient to the triple Pomeron vertex~\cite{Bartels:1994jj,Bartels:1995kf,Braun:1997nu,Bartels:2002au,Bartels:2004ef}.

In order to find an alternative formulation in terms of bound state (Pomeron) fields, $\Psi$ and $\Psi^\dagger$, we start from another field theory which must lead to the same Green's function
${\cal G}$. Again we start from the source terms:   
\be
\label{SourcePom}
 \int \left( \Phi^\dagger  \Psi + \Psi^\dagger \Phi \right)
\ee
and introduce the kinetic term:
\be
\int \Psi^\dagger {\cal G}^{-1} \Psi.
\label{KinPom}
\ee
For higher order interactions we include terms like
\be
\int \left( \Psi^\dagger V_{2\to 4} \Psi  \Psi  +{\rm h.c.} \right).
\label{IntPom}
\ee
To show the equivalence with the previous formulation in terms of reggeized giuon fields $A,\Adag$ one can use standard perturbation theory. For example in a functional approach we replace in the interaction terms the fields by the 
derivatives:
\be 
\psi \to \frac{\delta}{ \delta \Phi^\dagger},\,\, \psi^\dagger \to \frac{\delta}{ \delta \Phi},
\ee
and then integrate the quadratic form over the fields $\Psi$ and $\Psi^\dagger$, by completing the square. Then expanding the exponential of the interactions in powers and acting with the derivatives one can reproduce the previous results. For example 
\be
\int \left( \Psi^\dagger V_{2\to 4} \Psi  \Psi  +{\rm h.c.} \right) \to \int \left( \Phi^\dagger {\cal G} V_{2\to 4}  {\cal G}\Phi   {\cal G}\Phi  +{\rm h.c.} \right)
\ee
in agreement with $\Gamma_3$ in (\ref{truncation_source}).

We conclude with a few remarks on this bound state effective action. First, the propagator
of the $\Psi$ field, ${\cal G}$, can be written as a spectral decomposition. Namely, by inserting the complete set of eigenstates (cf.(\ref{Gtilde}))
\be
{\cal G}^{-1} = \omega - \tilde{K} = \sum_n |\psi_n> (\omega-\omega_n) <\psi_n|
\ee
the kinetic term (\ref{KinPom}) turns into a sum of different Pomeron fields
\be
\int \sum_n  \Psi_n^\dagger \left( \omega - \omega_n \right) \Psi_n
\ee
with 
\be
\Psi_n = <\psi_n| \Psi,\,\,\,\, \Psi^\dagger_n= \Psi^\dagger |\psi_n>.
\ee
Including the $\bq^2$ dependence of the eigenvalues $\omega_n$ we decompose into 
intercept and slope
\be
\omega_n(\bq^2)  = \omega_n(0) -\alpha_n' \bq^2
\ee
and arrive at the standard form of the Pomeron propagator of a local Pomeron field. Inserting this 
spectral decomposition into the the triple interaction term (\ref{IntPom}) we obtain a set of triple 
Pomeron couplings. 

We emphasize that the study of the RG evolution at large transverse distances and large rapidities of QCD in terms of 1PI reggeized gluon vertices, which are non local in rapidity and transverse space variables, is a very difficult task. Our approach seeks to overcome this difficulty by shifting the problem to the RG study of a local RFT with non trivial interactions: this RFT uses
the leading Pomeron states selected by with the spectral analysis. In our previous paper~\cite{Bartels:2015gou} we have shown that the investigation of such a local RFT is a tractable problem.

Next we note that, in this bound state theory, the $A, \Adag$ are no longer dynamical fields which propagate in rapidity, since only composite states of  two reggeized gluons in color singlet are effectively considered. Nevertheless, we could indeed introduce states given by a pair of two reggeized gluons at fixed rapidity:
\be
\int \left( \Psi^\dagger AA + {\rm h.c.} \right).
\ee
From this we can re-derive the Green's function, ${\cal G}$, of four reggeized gluons with the rapidity evolution given by the propagator of the $\Psi$ field.
Finally, the triple Pomeron vertex should contain a factor $i$, since the two Pomeron cut contribution to the Pomeron self energy comes with a negative sign.

It would be interesting to derive this field theory of bound state Pomeron fields directly from our 
effective field theory, in analogy to the familiar Hubbard-Stratonovich transformation.  For the rest of this paper we make use only of the effective field theory of reggeized gluons,   

%%%%%%%%%%%%%%%%%%%%%%%%%%%%%%%%
\section{RG flow equations}
%%%%%%%%%%%%%%%%%%%%%%%%%%%%%%%%

Let us now discuss the form of the IR regulated effective action, satisfying the Wetterich equation~\cite{Wetterich,Morris:1994ie}
\be
k\frac{\partial}{\partial k}\Gamma_k=\frac{1}{2}{\rm Tr}\left[ \left(\Gamma_k^{(2)}+R_k
\right)^{-1}
k\frac{\partial}{\partial k} R_k\right]
\label{eq:exactflow1}
\ee
for this field theory which is strongly constrained in the MRK for its expansion in the 1PI non local vertices. 
We are interested in the scattering of two reggeized gluon in the color singlet 
state, i.e. the BFKL Pomeron channel.  To this end we find it convenient to examine the flow of each vertex. The self-energies of the basic fields such as the the elementary real gluons as well as the reggeized ones have already been described. We shall first discuss a bit more the reggeized gluon case. For the 3-point vertices in the MRK, as already said, there is no renormalization and it is sufficient to have the elementary ones. Then we shall consider the family of vertices with 4 reggeized gluons and an arbitrary number of real (inelastic) gluon emissions.

Turning now to the effective average action, we shall deal with a truncation which, in the pure MRK,  is consistent and can be schematically written as
\ba
\Gamma_k [A,\Adag, a, a^\dag, \chi,\chi^\dag] &=& 
\frac{1}{2} \left( \int  \Adag \Gamma^{(2)}_{k,\Adag A}  A \nonumber+a^\dag \Gamma^{(2)}_{k,a^\dag a}  a + \chi^\dag \Gamma^{(2)}_{k,\chi^\dag \chi}  \chi \right) \\
&{}&\hspace{-3.5cm}+\frac{1}{2}\int \left( \Adag \Gamma^{(3)}_{k,\Adag a a}  a a + a^\dag a^\dag \Gamma^{(3)}_{k,a^\dag a^\dag A} A \right) +\int \Adag \Gamma^{(3)}_{k,\Adag A \chi} A \chi
+\int \Adag \chi^\dag \Gamma^{(3)}_{k,\Adag \chi^\dag A} A\nonumber\\
&{}&\hspace{-3.5cm}+ \sum_{I} \int \Adag \Adag  a^\dag \cdots a^\dag \chi^\dag \cdots \chi^\dag
 \Gamma^{(4+|I|)}_{k,\Adag \Adag  a^\dag \cdots a^\dag \chi^\dag \cdots \chi^\dag  A A a \cdots a \chi \cdots \chi} 
A A a \cdots a \chi \cdots \chi.
\label{truncation}
\ea
In the first line, the k-dependent 2-point function of the A field can be inferred from  (\ref{reg-gluon});
those of the $a$ and $\chi$ fields contain the IR regulators. In the second line the vertices are not $k$ dependent, and in the last line all 1PI vertices with 4 reggeized gluons have to be included. The 
4-point vertex with no external gluons $a_i$ or $\chi$, $\Gamma^{(4)}_{k,\Adag \Adag AA}$, is associated to elastic processes, while all the others vertices with an arbitrary number of $a$ or $\chi$ lines, i.e. for $|I|>0$, are associated to inelastic amplitudes and, thanks to the MRK, are built from $\Gamma_k^{(2)}$, $\Gamma_k^{(3)}$ and $\Gamma_k^{(4)}$. 
This means that the system of flow equations for the infinite set of 1PI vertices closes in the MRK.

The transition to the Pomeron bound state field description can be done along the lines presented in the previous section.

%%%%%%%%%%%%%%%%%%%%%%%%%%%%%%%%
\subsection{Two-point function}
%%%%%%%%%%%%%%%%%%%%%%%%%%%%%%%%

We begin with the propagator for the reggeized gluon. For the kinetic term we make the 
ansatz
\be
\Adag \Gamma^{(2)}_{k,\Adag A}  A=\Adag_a \partial_{\tau} A_b - \Adag_a \Sigma_{kab}(\partial_x^2) A_b
\ee
with the flow equation illustrated in Fig.~\ref{Fig11}:

 \begin{figure}[H]
 \begin{center}
\epsfig{file=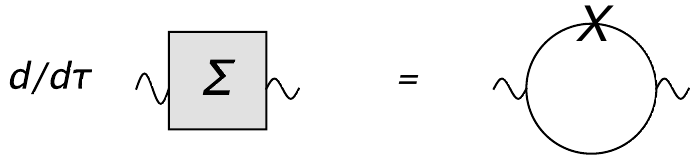,width=8cm,height=1.5cm}
\caption{flow equation for the selfenergy of the reggeized gluon   .
\label{Fig11} }
\end{center}  
\end{figure}
\noindent
Comparison with (\ref{reg-traj}) shows that the $\tau$-derivative of the selfenergy, $\Sigma_k(\bq^2)$, satisfies the same equation 
as $\tau$-derivative of the regularized trajectory function. In the following we therefore simply write $\omega_{g,k}(\bq^2)$  instead of  $\Sigma_k(\bq^2)$.

%%%%%%%%%%%%%%%%%%%%%%%%%%%%%%%%
\subsection{Vertex functions} 
%%%%%%%%%%%%%%%%%%%%%%%%%%%%%%%%
   
Let us first define 1PI vertex functions with two incoming and two outgoing reggeized gluons and an arbitrary number of elementary gluons, 
$\Gamma^{(4)m,n}_{\hspace{0.9cm}m',n'}$. Here $m(m')$ refers to the number of $\chi$ fields, $n(n')$ to the number of $a$ fields.

\begin{figure}[H]
\begin{center}
\epsfig{file=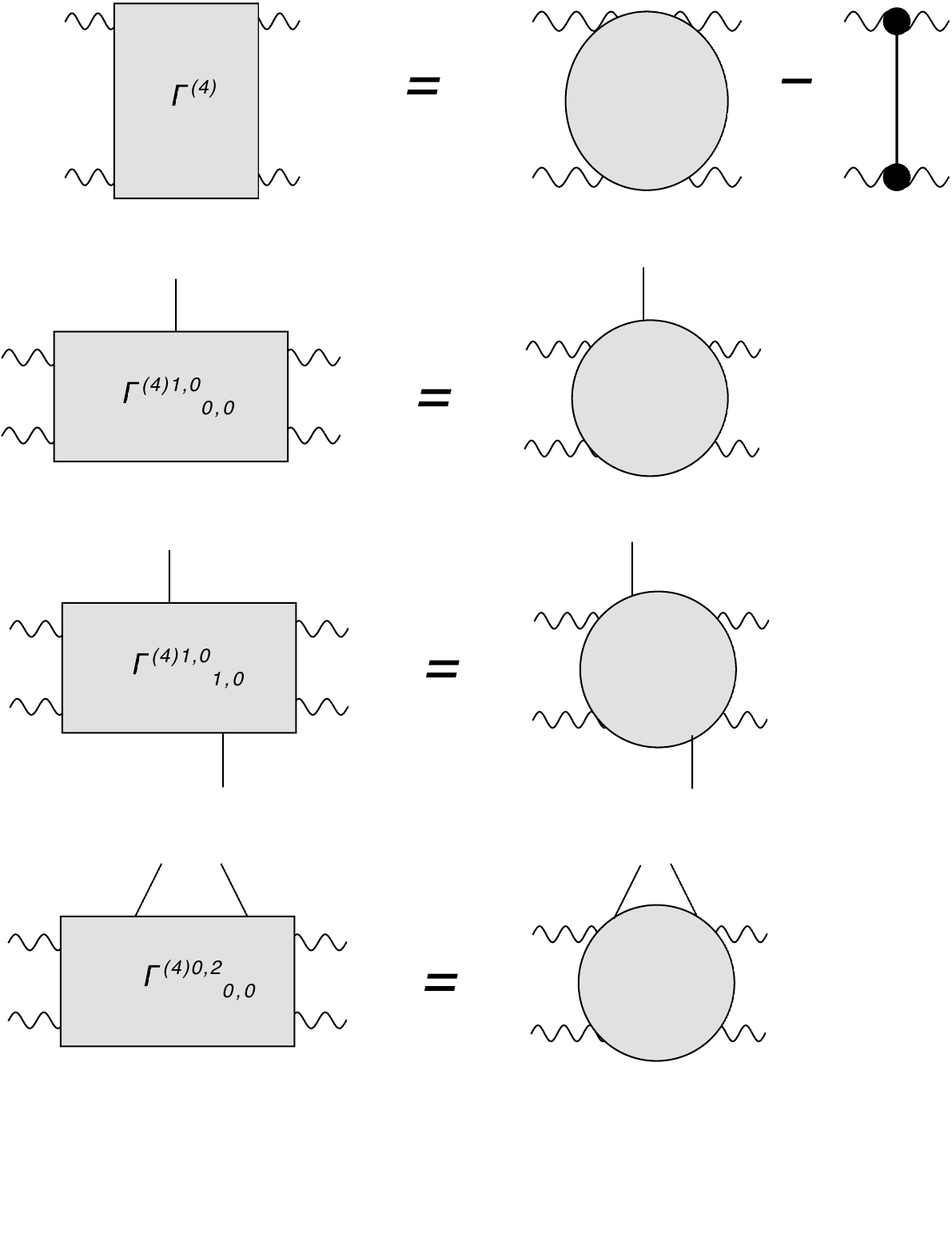,width=8cm,height=8cm}
\caption{Definition of one-particle irreducible vertex functions.  
\label{Fig12}}
\end{center}  
\end{figure} 

The upper (lower) numbers refer to the emission from the upper (lower) reggeized gluons in a horizontal ladder diagram.  
For the four point function of reggeized gluons without any elementary gluons, we define the one particle irreducible function
$\Gamma^{ (4)0,0}_{\hspace{0.9cm}0,0}=\Gamma^{(4)}$ by subtracting, from the sum of the BFKL ladder diagrams, the lowest order gluon exchange diagram (Fig.~\ref{Fig12}a). 
For the higher order vertex functions $\Gamma^{(4)m,n}_{\hspace{0.9cm}m',n'}$  one easily derives  from Figs.~\ref{Fig8} and~\ref{Fig9} that these vertex functions are generically already one-particle irreducible. 
Diagrammatically, we will denote 1PI vertex functions by boxes in the figures.

%%%%%%%%%%%%%%%%%%%%%%%%%%%%%%%%
\subsection{Flow equation for the 4-point function}
%%%%%%%%%%%%%%%%%%%%%%%%%%%%%%%%

\begin{figure}[H]
\begin{center}
\epsfig{file=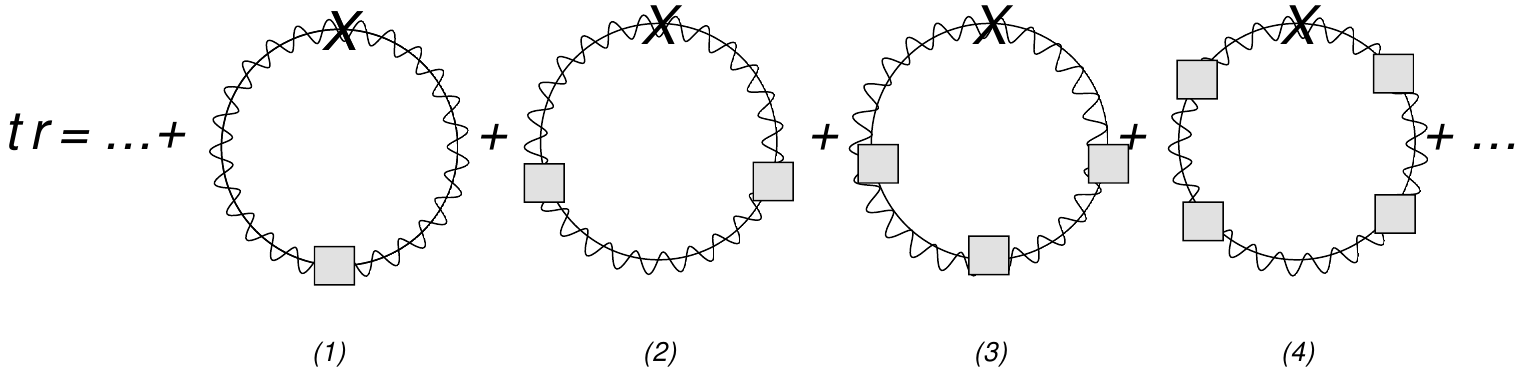,width=14cm,height=3cm}
\caption{expansion of the trace.
\label{Fig13}} 
\end{center}  
\end{figure}

\begin{figure}[H]
\begin{center}
\epsfig{file=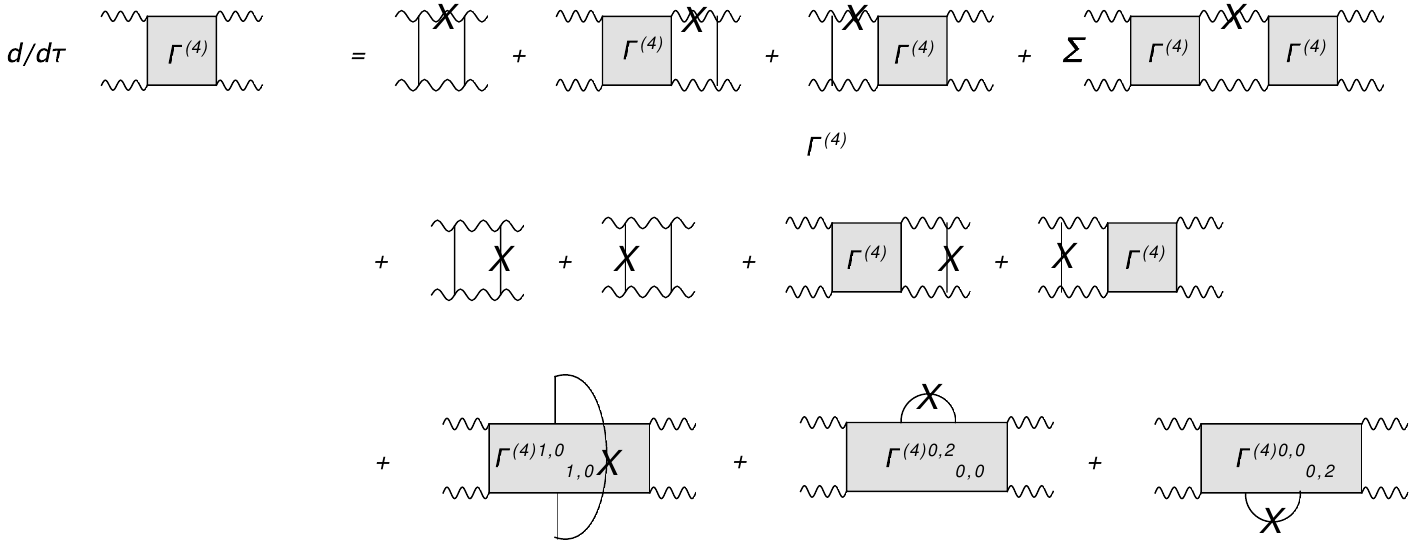,width=16cm,height=5cm}
\caption{flow equation for the 1PI 4-point function.
\label{Fig14}} 
\end{center} 
\end{figure}

Let us now write the flow equation for the 1PI 4-point function, $\Gamma^{(4)}$. Its flow equation can be obtained taking 4 functional derivatives in the reggeized gluon fields of the flow of the 1PI vertex generator given in Eq.~\eqref{eq:exactflow1}. 
We expand the trace on the rhs of the flow equations as depicted in Fig.~\ref{Fig13}, where
the crosses mark the derivative of the momentum regulator with respect to the RG "time":
\be
\dot{R}_k= k\frac{d}{dk} R_k.
\ee
As to the circle on the rhs of Fig.~\ref{Fig13}, we have to sum over elementary gluons with the regulated propagators
\be
\frac{1}{\bq^2+R_k(\bq^2)}
\ee
and the reggeized gluon with the regulated propagator (\ref{reg-gluon-prop}):
\be
\frac{1}{\omega \, (\bq^2+R_k(\bq^2)) - \omega_{g ,k}(\bq^2)}.
\ee  
Next we evaluate the rhs of Fig.~\ref{Fig13} for the 4-point vertex function. It is understood that the pairs of incoming and outgoing reggeized gluons are in color singlet states. The result is symbolically depicted in Fig.~\ref{Fig14} 
which contains, in the last line, two independent 6-point functions with four reggeized gluons and two elementary gluons. We therefore need to consider also the flow equation for these 6 point functions.

%%%%%%%%%%%%%%%%%%%%%%%%%%%%%%%%
\subsection{Flow equation for the 5-point and 6-point functions}
\begin{figure}[H]
\begin{center}
\epsfig{file=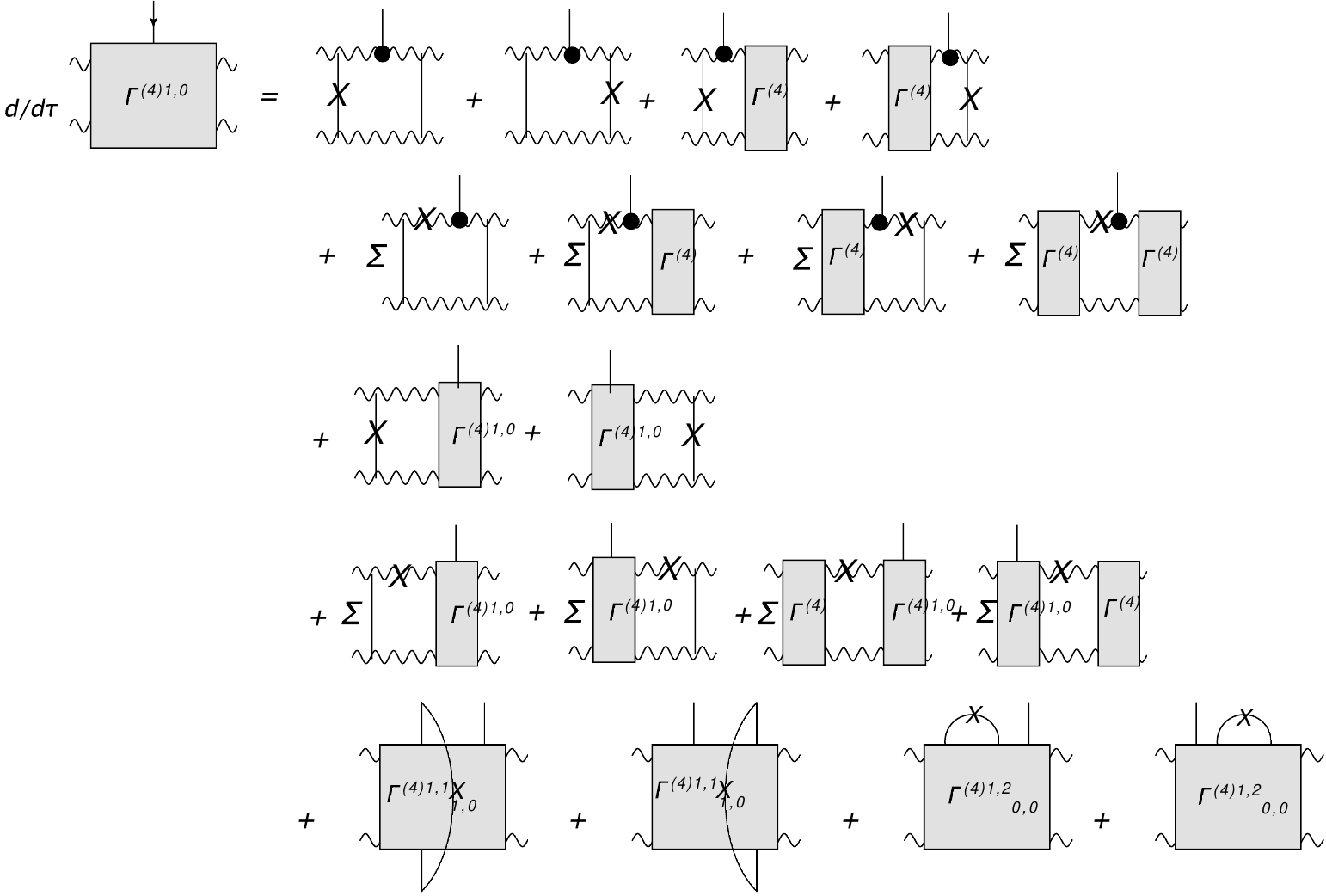,width=16cm,height=7.5cm}
\caption{flow equation for the 5-point function
\label{Fig15}} 
\end{center}
\end{figure}

%%%%%%%%%%%%%%%%%%%%%%%%%%%%%%%%
For simplicity, we start with the 5-point function $\Gamma^{(4)1,0}_{\hspace{0.8cm}0,0}$.   
We proceed in the same way as for 4-point function, and start from the trace illustrated in Fig.~\ref{Fig13}. The flow is given in Fig.~\ref{Fig15} where in the last line we have, once more, disregarded several terms which after a closer inspection are zero and kept only the non vanishing configurations.   

It is not difficult to generalize to the 6-point function and we do not present here the long expression.
From these examples we recognize the general pattern. In the MRK the flow equation of the 4-point function contains the 6-point function, the 6-point function needs the 5-point, 7-point and 8-point functions, and so on. 
In general, the n-point function encounters the (n+2) function, 
and these flow equations form an infinite tower of coupled differential equations.
We shall show that these correspond to the coupled equations obtained by substituting the truncation given in Eq.~\eqref{truncation} into the exact RG flow equation of Eq.~\eqref{eq:exactflow1}.

%%%%%%%%%%%%%%%%%%%%%%%%%%%%%%%%
\subsection{Introducing the running coupling}
%%%%%%%%%%%%%%%%%%%%%%%%%%%%%%%%
Before we derive, from the flow equations, a new nonlinear form of the equations in the MRK,
let us come back to our discussion of the running coupling which we have started at the end of section 2.  
The inclusion of the QCD running coupling effects in the Regge limit is a delicate issue when considering a full resummation.
Strictly speaking this effect goes beyond the Leading Log contribution in the MRK, since one has to take into account emissions of at least two real gluons close in rapidity,
which start from the region called quasi multi regge kinematics. It is also well known that the BFKL Pomeron in NLL accuracy 
has a spectrum which must be cured in the collinear regions with subleading term, 
and several approaches have been proposed (cf. our discussion in the beginning of section 3).  

There is a consensus that a good understanding of the pure running coupling effects can be nevertheless obtained by directly improving 
the picture obtained from the leading logarithmic approximation, that is by simply replacing the fixed coupling by  a running coupling, even if this approach is not unique.
We shall take this attitude and consider in our field theory, for any cubic vertex involving reggeized and real gluons, the following prescription:\\
(i) in a vertex with just one reggeized gluon $A$ or $A^\dagger$ with momentum $\bq$ we make the replacement $g_s\to g_s(\bq^2)$\\
(ii) in a vertex with two reggeized gluons $A$ and $A^\dagger$ with momentum $\bq$ and $\bq'$ we make the replacement $g_s\to \sqrt{g_s(\bq^2) g_s(\bq'{}^2)}$.\\
This means that in the trajectory function $\omega_g(\bq^2)$ we simply put
\be
\alpha_s \to \alpha(\bq^2)\,.
\label{traj-running}
\ee
The real kernel $K_{\text{BFKL}}(\bq,\bq')$ in the forward direction is modified by the substitution
\be
\label{forward-kernel-running}
\alpha_s \to \sqrt{\alpha(\bq^2)\alpha({\bq'}^2)}.
\ee
In the nonforward direction the kernel $K_{\text{BFKL}}(\bq_1,\bq_2;\bq'_1,\bq'_2)$ will be multiplied by
\be
\label{kernel-running}
\alpha_s \to \left(\alpha(\bq_1^2)\alpha(\bq_2^2)\alpha(\bq'_1{}^2)\alpha(\bq'_2{}^2)\right)^{1/4}.
\ee

We consider this as a first approximate attempt to include the running coupling. When going to the color octet channel, the choice (\ref{kernel-running}) does not satisfy the bootstrap condition. A way to 
implement this important self consistency condition has been decribed in  \cite{Braun:1996tc}. In a later step we will implement this method into our investigation.  

In our truncation for the effective average action the running coupling is chosen to be strictly independent of the IR regulator. As mentioned before, in our effective field theory at this level
there is no renormalization of the vertices: the running coupling, therefore, can be considered as an 'external' 
momentum dependent function  which does not change in course of the flow. As a consequence,     
all the results of our analysis performed at fixed coupling can easily be modified by the replacement:
fixed coupling $\to$ running coupling.

%%%%%%%%%%%%%%%%%%%%%%%%%%%%%%%%
\section{A nonlinear equation for the $\tau$-derivative of 4-point function}
%%%%%%%%%%%%%%%%%%%%%%%%%%%%%%%%
 
Starting from this tower of flow equations we now make use of the special features of the higher order 
vertex functions, illustrated in Figs.~\ref{Fig8}-\ref{Fig10}, and show that, starting from the flow equations, 
we can derive the differential equation for the $\tau$ derivative of the 4-point function, presented in Section~\ref{sect3}. 
This means that the truncation relevant for the multi-regge kinematics convert the one loop structure of the flow for the effective average action encoding an infinite set of 1PI vertices into a two loop structure for the flow of the 4-point vertex, whose knowledge is sufficient to define also the higher order vertex functions.

\begin{figure}[h]
 \begin{center}
\epsfig{file=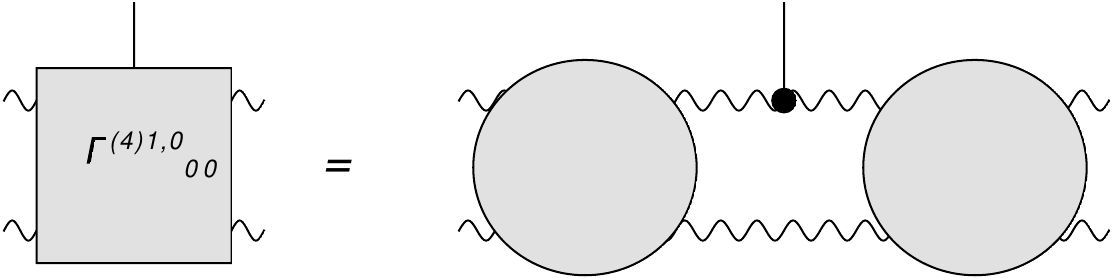,width=8cm,height=2cm}
\caption{ identity for the 5-point function
\label{Fig16}} 
\end{center}
\end{figure}

Let us begin wih the 5-point function. Returning to Fig.~\ref{Fig8} we note that our 1PI 5-point function can be expressed in terms of the 4-point function. For simplicity we use the full 4-point function rather than the 1PI 4-point function, see Fig.~\ref{Fig16}.

\begin{figure}[h]
\begin{center}
\epsfig{file=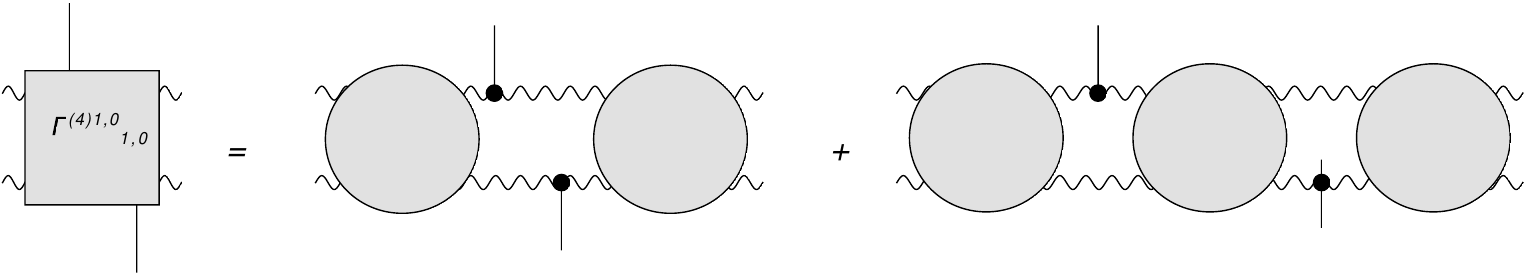,width=14cm,height=3cm}
\caption{identity for the 6-point function
\label{Fig17}} 
\end{center}
\end{figure}

\noindent
Similarly from Fig.~\ref{Fig9} for the 6-point vertex we derive the expression depicted in Fig.~\ref{Fig17}.

\noindent
It is straightforward to generalize these identities to other higher order vertex functions $\Gamma^{(4)m,n}_{\hspace{0.8cm}0,0}$. 

\begin{figure}[H]
\begin{center}
\epsfig{file=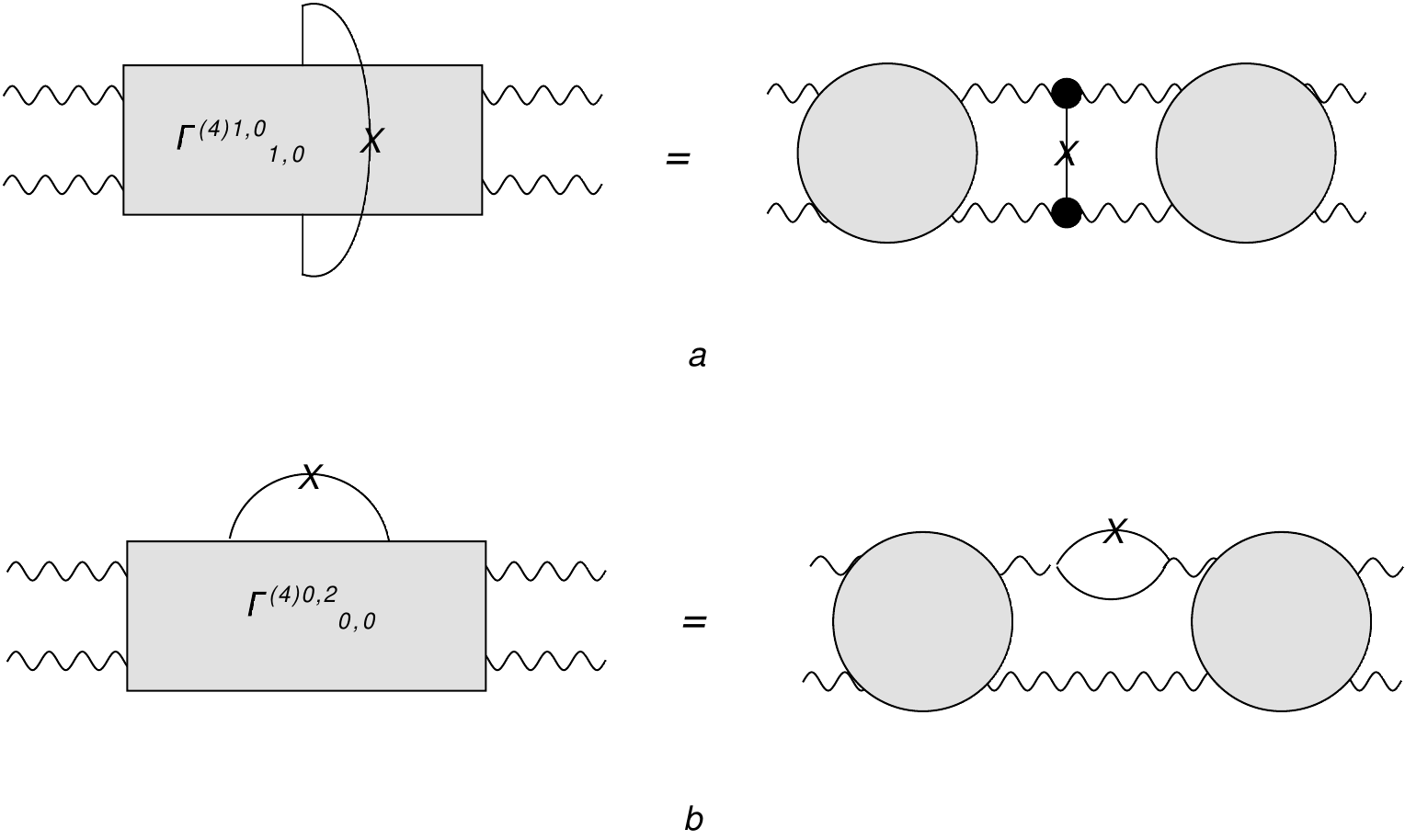,width=12cm,height=6cm}
\caption{closing the 6-point function  
\label{Fig18}} 
\end{center} 
\end{figure}

With these identities we can rewrite the infinite set of the flow equations. We begin with the 4-point function in Fig.~\ref{Fig14}, and rewrite the last line. As we have said before, the two reggeon states are in 
color singlet. For the first term we use Fig.~\ref{Fig17} and arrive at Fig.~\ref{Fig18}a

The second term in the last line of Fig.~\ref{Fig14} is derived from Fig.~\ref{Fig10} and illustrated in 
Fig.~\ref{Fig18}b.
Here we have made use of the fact that the scalar lines do not propagate in rapidity: Therefore, the emission and absorption of the scalar gluon from the upper and lower reggeized gluons happen at the same rapidity and the second term in Fig.~\ref{Fig17} does not 
contribute.     
Next we combine in Fig.~\ref{Fig14} the four terms in the first line, similarly the four terms of the second line.
In this way we arrive at:
\begin{figure}[H]
\begin{center}
\epsfig{file=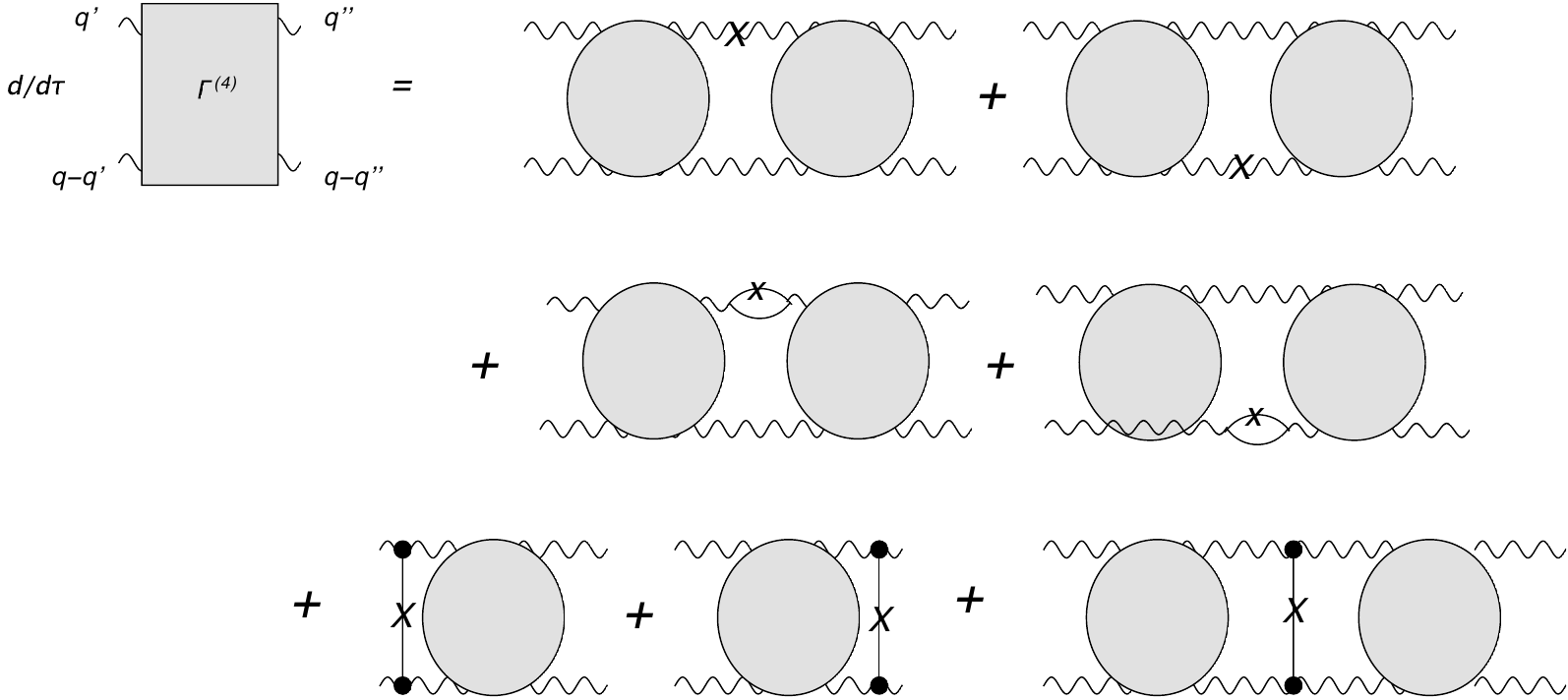,width=11cm,height=4.5cm}
\caption{modification of flow equation in Fig.~\ref{Fig14}  
\label{Fig19}} 
\end{center} 
\end{figure}
\vskip -0.5cm
We note that the rhs  agrees with Fig.~\ref{Fig3}: using the double cross notation of Fig.~\ref{Fig3}, 
the terms on the rhs of Fig.~\ref{Fig19} in the first and second lines can be combined to double crosses. 

\begin{figure}[H]
\begin{center}
\epsfig{file=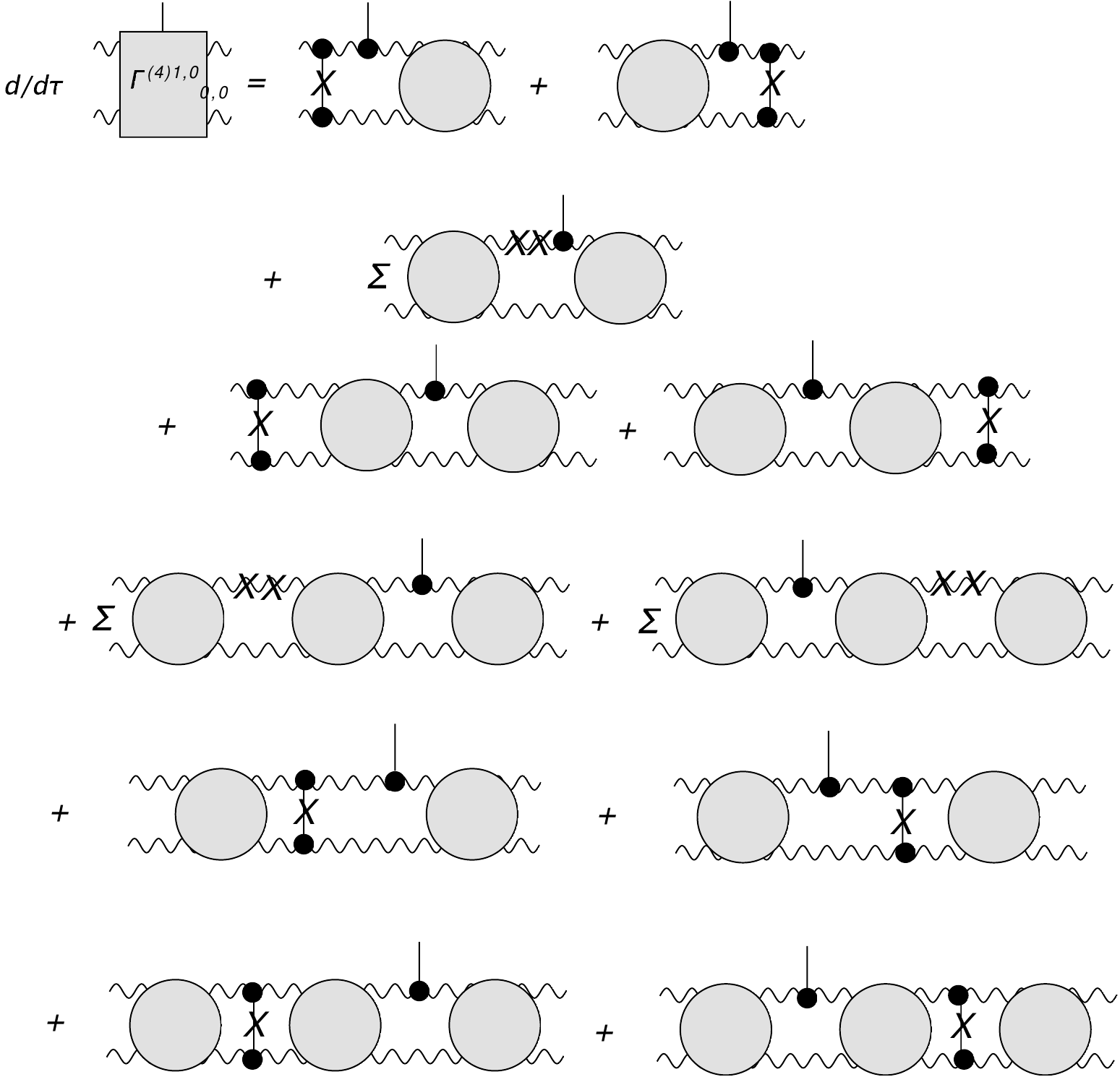,width=11cm,height=10cm}
\caption{Modification of the flow equation in Fig.~\ref{Fig15}. For the terms in the second and fourth lines it is understood that, for the derivative marked by cross we have to sum over all locations on the inner loop (upper and lower reggeon lines, left and right hand side of the production vertex). 
\label{Fig20}} 
\end{center} 
\end{figure}

It is straightforward (although somewhat tedious) to show that also for the higher order vertex functions the flow equations lead to simpler equations. As the simplest example, let us consider the 1PI 5-point vertex with its flow equation in Fig.~\ref{Fig15}.  
First we combine the first and the third lines on the rhs of Fig.~\ref{Fig15} by expressing the 1PI 4-point vertex by the full 4-point function. This leads to the first line of Fig.~\ref{Fig20}.
\vskip -0.2cm

\begin{figure}[H]
\begin{center}
\epsfig{file=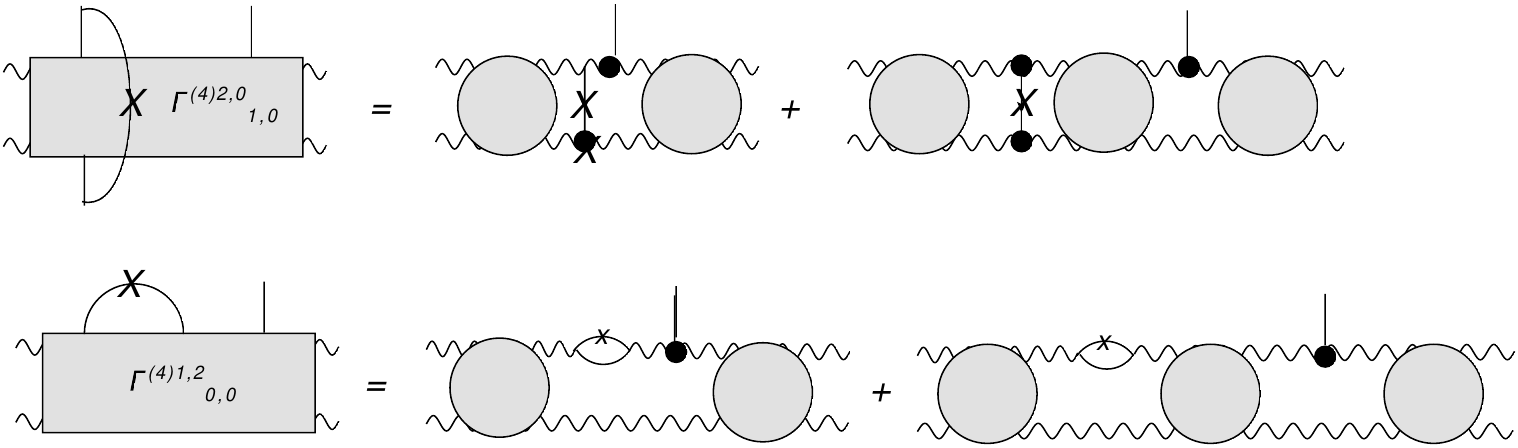,width=13cm,height=4.5cm}
\vspace{0.5cm}
\caption{closing the 7-point function
\label{Fig21}} 
\end{center}  
\end{figure}

Similarly we combine the terms in the second line and a part of the last two terms in the last line (we shall comment more on it later on)
of Fig.~\ref{Fig15} to arrive at the second line of Fig.~\ref{Fig20}.
For the third line of Fig.~\ref{Fig15} we use the identity of Fig.~\ref{Fig16} and obtain the third line of Fig.~\ref{Fig20}. 
In the same way, line 4 together with the other part coming from the last two terms in the last line of Fig.~\ref{Fig15} can be written as shown in the fourth line of Fig.~\ref{Fig20}. 
In the last line of Fig.~\ref{Fig15} the first two terms, by means of the first identity of  Fig.~\ref{Fig21}, lead to the last two lines of Fig.~\ref{Fig20}. 
Finally let us comment on the last two terms in the last line of Fig.~\ref{Fig15}. 
Using the second identity of Fig.~\ref{Fig21} (and a few more terms obtained by symmetry arguments) 
it is evident that they have to be combined with the second and forth lines of Fig.~\ref{Fig15}; they lead to the double crosses in the second and fourth lines of Fig.~\ref{Fig20}.   

Let us compare this equation with the defining identity in Fig.~\ref{Fig16}: we take the derivative with respect to 
$\tau$, and for the derivatives of the 4-point functions we insert Fig.~\ref{Fig19} (or Fig.~\ref{Fig3}). 
The result coincides with Fig.~\ref{Fig20}. With this we conclude that, similarly to the 4-point function, the $\tau$-derivative of the 5-point function again can be expressed in terms of the four point function, this time at third order. 
We expect that an analogous result holds also for higher order vertex functions, with increasing powers of the four point function. This implies that, instead of the infinite coupled set or flow equations, we have an infinite set of decoupled equations.
For our analysis of the BFKL Pomeron we only need the lowest nonlinear equation which is exact without any further truncations.

%%%%%%%%%%%%%%%%%%%%%%%%%%%%%%%%
\section{Numerical studies (1): Higgs-mass regulator}  
%%%%%%%%%%%%%%%%%%%%%%%%%%%%%%%%
In this and the next sections we shall perform a numerical analysis of the IR modified BFKL kernel
(first introducing a simple mass regulator, then  a Wilsonian optimized regulator). We return  to the notations introduced in section 2, in particular to the non-amputated BFKL kernel.     
We begin with the fixed QCD coupling, and in a second step  we also consider a running gauge coupling. Our main focus is on the spectrum of the integral kernel: eigenvalues, eigenfunctions, and $\bq^2$ slopes of the 
Regge poles. It is for a discrete spectrum that one can find new effective degrees of freedom and make a link at large distances with the local Pomeron fields of a RFT. 
Therefore we shall look for evidence of such a case.

The numerical analysis proceeds in two steps. For comparison we first study the BFKL Pomeron with the 
Higg's mass regulator\footnote{We use the notion 'Higgs mass' for the mass parameter of the spontaneously broken $SU(3)$.
It should not be confused with the mass of the physical Higgs particle.}.   This problem has been addressed before
\cite{Levin:2014bwa,Levin:2015noa,Levin:2016enb}, and partly we follow this paper. What is new in our analysis are the 
$\bq^2$-slopes of the discrete Pomeron states. After this we turn to the Wilsonian IR regulator and, again, compute eigenvalues, eigenfunctions, and $\bq^2$ slopes.   

%%%%%%%%%%%%%%%%%%%%%%%%%%%%%%%%
\subsection{Eigenvalues and eigenfunctions in the forward direction}
%%%%%%%%%%%%%%%%%%%%%%%%%%%%%%%%

We begin with the Higgs'mass regulated BFKL Pomeron with fixed coupling. 
First we define:
\be
\label{mom-param}
\bq_1=\frac{\bq}{2} +\bk,\,\, \bq_2=\frac{\bq}{2}-\bk,\,\,\bq'_1=\frac{\bq}{2}+\bk,'\,\, \bq'_2=\frac{\bq}{2}-\bk'\,.
\ee
The symmetrized kernels are
\ba
\frac{2\pi}{ \bar{\alpha}_s} K(\bq,\bk,\bk')= \sqrt{\frac{\bq_1^2+m^2}{\bq_2^2+m^2}} \frac{1}{(\bk-\bk')^2+m^2} \sqrt{\frac{\bq_2'{}^2+m^2}{\bq_1'{}^2+m^2}}\nonumber\\
+\sqrt{\frac{\bq_2^2+m^2}{\bq_1^2+m^2}} \frac{1}{(\bk-\bk')^2+m^2} \sqrt{\frac{\bq_1'{}^2+m^2}{\bq_2'{}^2+m^2}}\nonumber\\
-\frac{\bq^2+\frac{N_c^2+1}{N_c^2} m^2}{\sqrt{(\bq_1^2+m^2)(\bq_2^2+m^2)(\bq_1'{}^2+m^2)(\bq_2'{}^2+m^2)}}
\ea
and 
\be
\tilde{K}(\bq,\bk,\bk')= K(\bq,\bk,\bk')+\delta^{(2)}(\bk-\bk')  \left( \omega_g(\bq_1^2) +\omega_g(\bq_2^2) \right)\,,
\ee
whereas the trajectory function has the form:
\ba
\omega_g(\bk^2)=-\frac{\bar{\alpha}_s}{4\pi} \int d^2 k' \frac{\bk^2+m^2}{({\bk'}^2+m^2)((\bk-\bk')^2+m^2)}\nonumber\\
= -\frac{\bar{\alpha}_s}{2\pi}\int d^2 k' \frac{\bk^2+m^2}{({\bk'}^2+m^2)({\bk'}^2+(\bk-\bk')^2+2m^2)}\,.
\ea
We first consider the forward direction $\bq^2=0$ where the kernel simplifies:
\ba
\label{Higgs-forwardkernel}
\frac{2\pi}{ \bar{\alpha}_s} K^{(0)}(\bk,\bk')= \frac{2}{(\bk-\bk')^2+m^2} - \frac{\frac{N_c^2+1}{N_c^2} m^2}{(\bk^2+m^2)({\bk'}^2+m^2) }
\ea 
and 
\be
\tilde{K}^{(0)}(\bk,\bk')= K^{(0)}(\bk,\bk')+2 \delta^{(2)}(\bk-\bk') \omega_g(\bk^2)\,.
\ee

For the numerical calculation we combine terms which contain the potentially singular denominator $1/(\bk-\bk')^2$ and 
rewrite the eigenvalue equation
\be
\tilde{K}^{(0)} f_\omega = \omega f_\omega
\ee
in the following form \cite{Siddikov}:
\ba
\omega f(\bk)& =&\frac{\bar{\alpha}_s}{2\pi} \int d^2k' \Big[ 
\frac{2 f(\bk')({\bk'}^2+m^2)- 2f(\bk) (\bk^2+m^2)}{({\bk'}^2+m^2)((\bk-\bk')^2+m^2)} - \frac{\frac{N_c^2+1}{N_c^2} m^2}{(\bk^2+m^2)({\bk'}^2+m^2) }f(\bk')\Big] \nonumber\\
 &&+\frac{\bar{\alpha}_s}{2\pi} \int d^2k'\frac{ f(\bk) (\bk^2+m^2)}{({\bk'}^2+m^2)({\bk'}^2+(\bk-\bk')^2+2m^2)}
 \Big].
\ea 
By restricting ourselves to spherically symmetric eigenfunction, the angular integrations can be done by using the formula
\be
\frac{1}{2\pi} \int_0^{2\pi} d \varphi \frac{1}{a+b \cos \varphi} = \frac{1}{\sqrt{a^2-b^2}}.
\ee 
For the remaining  integration in ${\bk'}^2$ we change to logarithmic variables $t'=\ln \frac{{\bk'}^2}{m^2}$ with 
$d{\bk'}^2 =dt' {\bk'}^2$ and introduce a lattice in the new variables $t'$. Introducing the limits $k^2_{min}=10^{-40},
t_{min}=\ln \frac{k^2_{min}}{m^2}$ and $k^2_{max}=10^{80}, t_{max}= \ln \frac{k^2_{max}}{m^2}$
and dividing the interval $\Big[t_{min}, t_{max}\Big]$
into $N_{step}$ equal steps, we define the lattice points 
\be
t_i=t_{min}+i \frac{t_{max}-t_{min}}{N_{step}},\,\, k_i^2=m^2 e^{t_i},\,\,i=0,...,N_{step}
\ee
and arrive at the discrete vector $f_i=f(k_i)$ and matrix $K_{ij}=K(k_i,k_j)$:
\be
\int d{\bk'}^2 K(\bk,\bk') f(\bk') = \int dt' K(t,t') {k'}^2  f(k') \rightarrow \sum_{j} K_{ij} {k'_j}^2 f_j\,.
\ee
For the eigenfunctions we symmetrize, in the eigenvalue equation, the kernel and write instead 
\be 
\int dt' \left(k K(t,t') k'\right)   \left( k' f(k')\right) \rightarrow \sum_{j} \left( k_i K_{ij} k'_j \right) \left( k_j f_j \right)
\ee
with the new eigenvectors $\left( k_i f_i\right) $, orthogonal among each other. 
For $N_{step}=600$, when results numerically stabilize the first eigenvalues are listed in Table~\ref{Table1}. 
%mod9e,mod10%
\begin{table}[H]
\begin{center}
\begin{tabular}{cll} \\
n&energy&slope \\\hline
1&-0.5295&0.00000\\
2&-0.5268&0.00001\\
3& -0.5224&0.00003\\
4& -0.5162&0.00005\\
5&-0.5085&0.00007\\
6&-0.4992&0.00010\\
7&-0.4885&0.00014\\
8&-0.4765&0.00017\\
9&-0.4633&0.00021\\
10&-0.4491&0.00025\\
11&-0.4339&0.00029
\end{tabular} 
\end{center}
\caption{Numerical results for the eigenvalues and slope with fixed coupling constant and Higgs-mass regulator\label{Table1}}
\end{table}
As expected, the spectrum of eigenvalues starts near $E_{BFKL}= -\omega_{BFKL} = - 4 \ln 2 \bar{\alpha}_s=-0.5295$, with $\bar{\alpha}_s$ given in Eq.~\eqref{alphasb} for $\alpha_s=0.2$,
and the slopes are very small. The discreteness of the eigenvalues as well as the nonvanishing slopes consequence of the lattization: from analytic calculations it is known that the characteristic function leads to a continuous cut in the energy plane beginning at $ - 4 \ln 2 \bar{\alpha}_s=-0.5295$ and extending to $+\infty$.

%%%%%%%%%%%%%%%%%%%%%%%%%%%%%%%%
\subsection{$q^2$ slopes of the eigenvalues}
%%%%%%%%%%%%%%%%%%%%%%%%%%%%%%%%

Next we leave the forward direction and consider the $\bq^2$ dependence of the eigenvalues.
The $q^2$ slopes of the eigenvalues are obtained from
\be
\omega_n(\bq^2)=\omega_n^{(0)}+\bq^2 \frac{\int d^2 \bk \int d^2 \bk' f_n({\bk'}^2)\Big[ K^{(1)}(\bk,\bk')+2\delta^{(2)}(\bk-\bk')\omega_g^{(1)}(\bk^2)\Big]  f_n(\bk^2)}{\int d^2\bk |f_n(\bk^2)|^2}\,,
\ee
where $\omega_n^{(0)}$ are the eigenvalues of the forward kernel $K^{(0)}$, $f_n(\bk^2)$ the corresponding (rotationally invariant) eigenfunctions, and $K^{(1)}$, $\omega^{(1)}$ the corrections
of the order $q^2$ to the forward BFKL kernel and the gluon trajectory, resp. 
We find it convenient to introduce
\be
E_n=-\omega_n\,.
\label{slope}
\ee

In order to find $K^{(1)}(\bk,\bk')$ we expand the kernel in the small $\bq^2$ region to first order in $\bq^2$:
\be
\label{kernelexp}
K_(\bq,\bk,\bk')=K^{(0)}(\bk,\bk')+\bq^2 K^{(1)}(\bk,\bk').
\ee
With the shorthand notations
\ba
D=\bk^2+m^2,\,\, D'={\bk'}^2+m^2,\,\,D_0=(\bk-\bk')^2+m^2
\ea
we find:
\ba 
\label{kernelexp}
K (\bq,\bk,\bk')= \frac{ \bar{\alpha}_s}{2\pi}\Big[  \frac{2}{D_0} \left(1-\frac{(2\bq\bk)(2\bq \bk')}{4D D'}+\frac{(2\bq\bk)^2}{8D^2}+\frac{(2\bq\bk')^2}{8{D'}^2}\right)\nonumber\\
-\frac{m^2 \frac{N_c^2+1}{N_c^2}}{DD'}  
\left(1+\frac{1}{2} (\frac{ \bq \bk}{D})^2 +\frac{1}{2} (\frac{ \bq \bk'}{D'})^2
-\frac{\bq^2}{4}(\frac{1}{D}+\frac{1}{D'})\right)-\bq^2 \frac{1}{DD'}\Big]\,.
\ea 
Since in (\ref{slope}) the eigenfunctions are rotational invariant, we can average over the angles of 
$\bk$ and $\bk'$ (keeping the angle $\psi$ between $\bk$ and $\bk'$ fixed); this eliminates terms 
proportional to the scalar products $\bq \bk$ or  $\bq \bk'$ (which in (\ref{kernelexp}) have already been dismissed), Moreover, we use
\ba
\frac{1}{2\pi} \int_0^{2\pi}  d\phi  \int_0^{2\pi}  d\phi' 4(\bq \bk)^2 &=& \int_0^{2\pi} d\psi \,2 \bq^2 \bk^2 \nonumber\\
\frac{1}{2\pi} \int_0^{2\pi}  d\phi  \int_0^{2\pi}  d\phi' 4(\bq \bk')^2 &=& \int_0^{2\pi} d\psi \,2\bq^2 \bk^2 \nonumber\\
\frac{1}{2\pi} \int_0^{2\pi}  d\phi  \int_0^{2\pi}  d\phi' 4(\bq \bk)(\bq\bk') &=& \int_0^{2\pi} d\psi \,2\bq^2 \sqrt{\bk^2 \bk'{}^2} \cos \psi\nonumber\\
&=& \int_0^{2\pi} d\psi\, \bq^2\left(-(\bk-\bk')^2 +\bk^2+{\bk'}^2 \right)\,,
\ea
where $\psi$ is the angle between $\bk$ and $\bk'$. With these identities the bracket in the first line in (\ref{kernelexp}) can be written as 
\be
1+\frac{\bq^2}{4}\Big[ \frac{(\bk-\bk')^2}{DD'} -m^2\left(\frac{1}{D}-\frac{1}{D'}\right)^2\Big]
\ee   
and 
\ba
&&K^{(1)}(\bk,\bk')=\\
&&\frac{\bar{\alpha}_s}{2\pi} \int_0^{2\pi} d\psi 
\Big[\frac{1}{2D_0} \left(\frac{(\bk-\bk')^2}{DD'} -m^2\left(\frac{1}{D}-\frac{1}{D'}\right)^2 \right)
+\frac{m^4 \frac{N_c^2+1}{N_c^2}}{4DD'}   \left( \frac{1}{D^2}+ \frac{1}{{D'}^2} \right)- \frac{1}{DD'}\Big]\,. \nonumber
\ea
We do the remaining angular integral over $\psi$ and obtain: 
\ba
&&K^{(1)}(k,k')=\nonumber\\
&&\bar{\alpha}_s
\Big[ \frac{1}{2DD'} \left(1-\frac{m^2}{S_0} \right) -\frac{m^2}{2S_0} \left( \frac{1}{D}-\frac{1}{D'} \right)^2 
+\frac{m^4 \frac{N_c^2+1}{N_c^2}}{4DD'}   \left( \frac{1}{D^2}+ \frac{1}{{D'}^2} \right)- \frac{1}{DD'}\Big]\,,
\ea
where
\be
S_0=\sqrt{(\bk^2-{\bk'}^2)^2+2m^2(\bk^2+{\bk'}^2)+m^4}.
\ee

For the trajectory function we put:
\be
\label{trajexp}
\omega_g( (\frac{\bq}{2}+\bk)^2 )= \omega_g(\bk^2) +\omega^{(1/2)}+\bq^2\omega^{(1)}(\bk^2)\,,
\ee
where $\omega^{(1/2)}$ is of the order $\cal{O}$$(\bq)$. From the $\bq^2$ expansion of the trajectory function
\ba
\label{traj-q2}
\omega_g( (\frac{\bq}{2}+\bk)^2 )&=& \omega_g(\bk^2) 
-\frac{\bar{\alpha}_s}{2\pi} \int d^2 \bk''  \left(\frac{2\bq\bk}{2D'' D_1}(1-\frac{D}{D_1})
+\frac{2\bq \bk'' D}{2D'' D_1}\right. \\ 
&&\left.+\frac{\bq^2}{4} \frac{1}{D'' D_1}(1-\frac{D}{D_1})+(2\bq(\bk-\bk''))^2  \frac{D}{4 D'' D_1^3}-
(2\bq \bk)(2\bq(\bk-\bk'')) \frac{1}{4 D'' D_1^2} \right) \nonumber
\ea
we derive
\be
\omega^{(1/2)}=-\frac{\bar{\alpha}_s}{2\pi} \int d^2 \bk''  \left(\frac{2\bq\bk}{2D'' D_1}(1-\frac{D}{D_1})
+\frac{2\bq \bk'' D}{2D'' D_1}\right)\,.
\ee
After angular integration this contribution vanishes. For the second term we find (after angular integrations): 
\ba  
\label{delta-omega_higgs}
\bq^2 \omega_g^{(1)}(\bk^2) &=& -
\frac{\bq^2}{4} \frac{\bar{\alpha}_s}{2\pi} \int d^2\bk'' \Big[ 2\left( \frac{1}{D_1^2} - 
\frac{D}{D_1^3}\right) +\frac{m^2}{D''}\left( \frac{1}{D_1^2} - 2 \frac{D}{D_1^3}\right) \Big].\nonumber\\
&=&-\bq^2 \frac{\bar{\alpha}_s}{8}  \int d{\bk''}^2 
\Big[ \frac{a}{S_1^3} \left(2+\frac{m^2}{D''} \right)-\frac{D(2a^2+b^2)}{S_1^5} \left(1+\frac{m^2}{D''}\right)\Big]\,,
\ea
where we have used the short hand notation
\be
D''= {\bk''}^2+m^2,\,\,D_1={\bk''}^2+(\bk-\bk'')^2+2m^2
\ee
and
\ba
&&a= \bk^2+2({\bk''}^2+m^2),\,\,b=-2\sqrt{\bk^2 \bk''{}^2},\nonumber\\
&&S_1=\sqrt{(\bk^2-{\bk''}^2)^2+2({\bk''}^2+2m^2)(\bk^2+{\bk''}^2) +({\bk''}^2+2m^2)^2}\,.
\ea

With these results we rewrite (\ref{slope}):
\be
\label{slope-final}
E_n(q^2)=E_n^{(0)}-\frac{\bq^2}{2} \frac{\int d \bk^2  \int d{\bk'}^2 f_n({\bk'}^2) \Big[K^{(1)}(\bk,\bk') 
+ 4 \delta(\bk^2-{\bk'}^2) \omega_g^{(1)}(\bk^2)\Big]
f_n(\bk^2)}{\int d \bk^2 |f_n(\bk^2)|^2}
\ee
and insert the expressions derived above.

Numerical results for the slopes are also listed in Table~\ref{Table1}. 

%%%%%%%%%%%%%%%%%%%%%%%%%%%%%%%%
\subsection{Running coupling}
%%%%%%%%%%%%%%%%%%%%%%%%%%%%%%%%

Let  us now turn to the physical case of the running coupling. As a first step we 
simply replace the fixed coupling $\alpha_s$ by
\be 
\label{runningalpha}
\alpha_s(\bq^2) = \frac{3.41}{\beta_0 \ln (\bq^2+R^2_0)}
\ee
and
\be
\bar{\alpha}_s(\bq^2) = \alpha_s(\bq^2) \frac{N_c}{\pi}
\ee
with $\beta_0=(11 Nc - 2 Nf)/12$, $N_f=3$. Its normalization is chosen to match the measured value at the $Z$ mass scale.
$R_0$ defines the scale below which the running coupling is 'frozen'. Both $q^2$ and $R^2_0$ are in units of $\Lambda^2_{QCD}$, and $R_0$  has to be well above  
$ \Lambda_{QCD}^2=0.15^2$ GeV$^2$. In our calculations we use $R_0=0.54$ Gev. More accurate models allowing for different number of flavors  can be easily considered.

In our numerical computations with the Higgs regulator we actually find it convenient to 
follow the conventions used in \cite{Levin:2014bwa,Levin:2016enb}: we 
define momenta and $R_0$ in units of the regulator mass $m=m_h=0.54 GeV$.
This leads to the modification of (\ref{runningalpha}):
\be 
\label{runningalpha-pract}
\alpha_s(\bq^2) = \frac{3.41}{\beta_0 \Big[ \ln (\bq^2+R^2_0)+\ln \frac{m_h^2}{\Lambda^2_{QCD}}\Big]}
\ee
with $R_0=1$.  With this convention in all our previous expressions the mass $m=m_h$ will be replaced by unity.  

As discussed before, for the forward direction the eigenvalue equations will be modified in the following way:
\be
 \alpha_s({\bk'}^2) K(\bk',\bk'') \rightarrow  \sqrt{\alpha_s({\bk'}^2)} K(\bk',\bk'')\sqrt{ \alpha_s({\bk''}^2)},
\ee
and the trajectory functions will be simply multiplied by  $\alpha_s({\bk}^2)$.

For the slopes we have to leave the forward direction. In addition to the $\bq^2$ expansions of the kernel and of the trajectory function described in section 7.2, we also need the expansion of the running couplings in (\ref{traj-running}) and (\ref{kernel-running}). With the parametrization (\ref{mom-param}) and the short-hand notation 
\be
L=\bk^2+R_0^2,\,\, L'={\bk'}^2+R_0^2,\,\,s= \frac{m_h^2}{\Lambda^2_{QCD}}
\ee
(\ref{runningalpha-pract}) can be written as:
\be
\alpha_s(\bk^2)=\frac{A}{\ln (sL)},\,\, A=\frac{3.41}{\beta_0}.
\ee  
We put 
\be
\alpha_s((\frac{\bq}{2}+\bk)^2)=\alpha_s(\bk^2)\Big[1 + \alpha^{(1/2)}+\bq^2 \alpha^{(1)}(\bk^2) \Big]\,,
\ee
where $ \alpha^{(1/2)}$ is of the order $\cal{O}$$(\bq)$.
From the expansion
\be
\label{coupling-q2}
\alpha_s((\frac{\bq}{2}+\bk)^2)=\alpha_s(\bk^2) \Big[ 1-\frac{\bk \bq}{L \ln (sL)}- \bq^2 \frac{1}{4 L \ln (sL)}+
\frac{(\bk \bq)^2}{L^2} (\frac{1}{\ln^2 (sL)}+ \frac{1}{2\ln (sL)}) \Big]
\ee
we deduce
\be
\alpha^{(1/2)}=-\frac{\bk \bq}{L \ln (sL)}
\ee
and 
\ba
\bq^2 \alpha^{(1)}(\bk^2)&=&- \bq^2 \frac{1}{4 L \ln (sL)}+
\frac{(\bk \bq)^2}{L^2} (\frac{1}{\ln^2 (sL)}+ \frac{1}{2\ln (sL)})\nonumber\\
&=&\bq^2 \Big[ -\frac{1}{4 L \ln(sL)} +\frac{\bk^2}{2L^2} \left(\frac{1}{\ln^2(sL)}+ \frac{1}{2 \ln (sL)}\right)
\Big]\,.
\ea

For the coupling in front of the kernel (cf.(\ref{kernel-running})) we encounter the product
\ba
\alpha_s(\bq_1^2)\alpha_s(\bq_2^2)&=&\alpha_s(\bk^2)^2\Big[1+2 \bq^2 \alpha^{(1)}(\bk^2)- ({\alpha^{(1/2)}})^2\Big]\nonumber\\
&=&\alpha_s(\bk^2)^2\Big[1+2\bq^2 \tilde{\alpha}(\bk^2)\Big]\,,
\ea
where
\be
\tilde{\alpha}(\bk^2)= -\frac{1}{4 L \ln(sL)} +\frac{\bk^2}{4L^2} \left(\frac{1}{\ln^2(sL)}+ \frac{1}{ \ln (sL)}\right)
\ee
This leads to the following factor in front of the kernel: 
\be
\left( \alpha_s(\bq_1^2)\alpha_s(\bq_2^2)\alpha_s({\bq'_1}^2)\alpha_s({\bq'_2}^2) \right)^{\frac{1}{4}}
=\sqrt{\alpha(\bk^2) \alpha({\bk'}^2)} \Big[1+\bq^2 \frac{\tilde{\alpha}(\b\bk^2)+\tilde{\alpha}({\bk'}^2)}{2}\Big]\,.
\ee
Note that in this product the terms proportional to $\bq \bk$ have cancelled, and we can simply use the expansion  (\ref{kernelexp}),
without considering such terms.  

For the trajectory the situation is a bit more complicated, and both in the expansion (\ref{trajexp})
and (\ref{coupling-q2}) terms linear in $\bq$ have to be kept. In the product of the two expansions 
we find:
\ba
&&\Big[1 + \alpha^{(1/2)}+\bq^2 \alpha^{(1)}(k^2) \Big] \Big[  \omega_g(\bk^2) +\omega^{(1/2)}+\bq^2\omega^{(1)}(k^2)\Big]\nonumber\\
&&=\omega_g(\bk^2) +  \alpha^{(1/2)}\omega^{(1/2)}+ \bq^2 \alpha^{(1)}(\bk^2) \omega_g(\bk^2)+\bq^2
\omega^{(1)}(\bk^2)\,.
\ea
In the second term we average of the azimuthal directions and obtain:
\ba
<\alpha^{(1/2)}\omega^{(1/2)}> &=& \int_0^{2\pi} \frac{d \phi}{2\pi}\left(\alpha^{(1/2)}\omega^{(1/2)}\right)\nonumber\\
&=&
\bq^2 \frac{\alpha(\bk^2)}{8 \pi L \ln(sL)} \int d^2\bk'' \frac{1}{D'' D_1} \Big[D \left( 1-\frac{D}{D_1}+2 
\frac{D''}{D_1} \right) -m^2 \left( 2- \frac{D}{D_1}\right) \Big]\nonumber\\
&=&\bq^2 \frac{\alpha(\bk^2)}{8  L \ln(sL)} \int d{\bk''}^2 \Big[ \frac{D-2m^2}{D'' S_1} + \frac{D(4{D''}^2-(k^2)^2)}{D'' S_1^3}\Big]\,.
\ea

Putting these expressions together we obtain for the terms proportional to $\bq^2$:
\ba
\label{slope-running}
&&\sqrt{\alpha(\bk^2) \alpha({\bk'}^2)} K^{(1)}(\bk,k\b') + 2 \delta^{(2)}(\bk-\bk') \omega^{(1)}(\bk^2)
+ 2 \delta^{(2)}(\bk-\bk') <\alpha^{(1/2)}\omega^{(1/2)}>\nonumber\\
&&+\Big [\sqrt{\alpha(\bk^2) \alpha({\bk'}^2)}K^{(0)}(\bk,\bk') + 2 \delta^{(2)}(\bk-\bk') \omega_g(\bk^2)\Big]
\frac{\tilde{\alpha}(\bk^2)+\tilde{\alpha}({\bk'}^2)}{2} \,.
\ea
The first two terms are analogous to those of the fixed coupling case, the remaining ones are due to the running coupling.    

We are now ready to present numerical results for the eigenvalues and for the slopes.
Table~\ref{Table2} contains our results for the leading states (up to $n=8$):
%mod10
\begin{table}[H]
\begin{center}
\begin{tabular}{ccll} \\
n&energy&slope&radius [GeV]
\\\hline
1&-0.4384&0.0464&3.92\\
2&-0.2253&0.0144&9.07$\times 10^1$\\
3&-0.1508&0.0070&2.32$\times 10^3$\\
4&-0.1131&0.0041&6.18$\times 10^4$\\
5&-0.0904&0.0027&1.68$\times10^6$\\
6&-0.0753&0.0019&4.59$\times10^7$\\
7&-0.0644&0.0014&1.26$\times10^9$\\
8&-0.0563&0.0011&3.48$\times10^{10}$
\end{tabular}
\end{center}
\caption{Numerical results for the eigenvalues, slopes and  radii  with running coupling constant and Higgs-mass regulator 
\label{Table2}}
\end{table}
The wavefunctions for $n=1,2,5$ are shown in Fig.22:
\begin{figure}[H]
\begin{center}
\epsfig{file=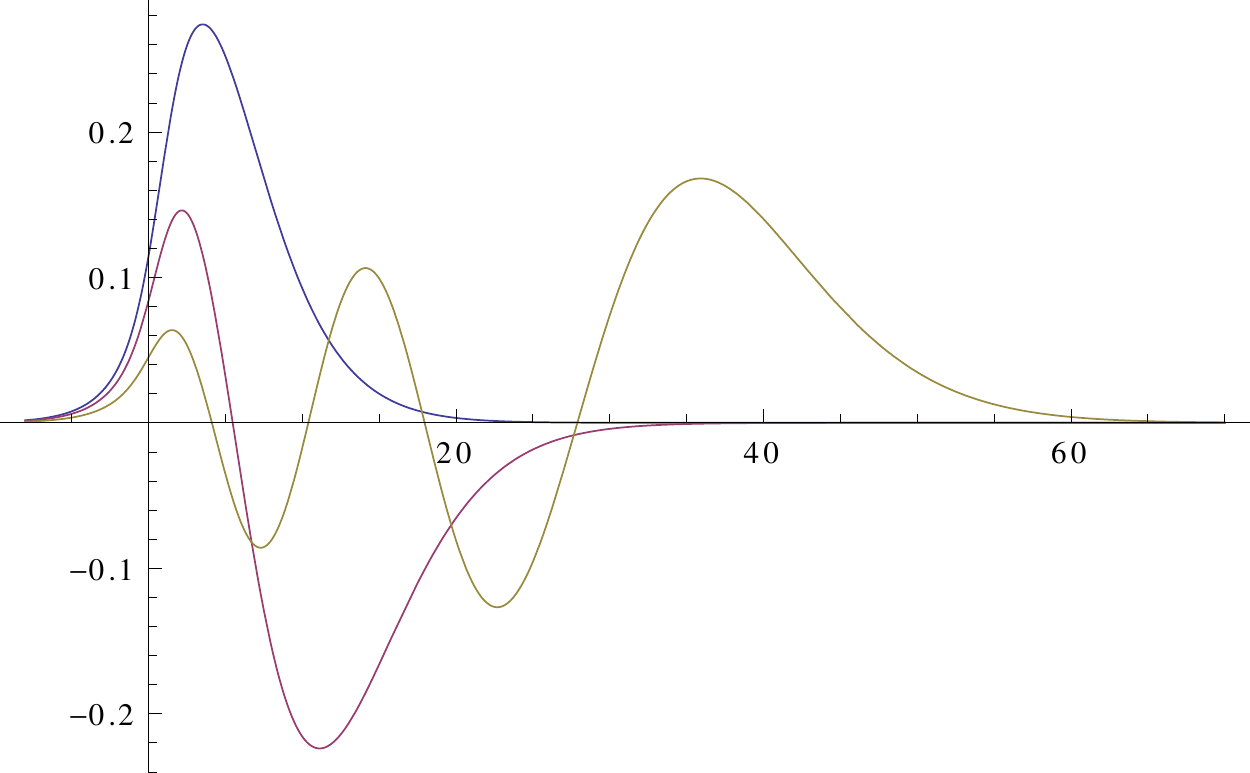,width=12cm,height=5cm}\\
\caption{three leading wavefunctions (No 1,2,5)  
as a function of $\ln q^2$.
\label{Fig22}} 
\end{center}
\end{figure}

Describing the results of our numerical analysis in more detail, we begin with the Regge poles in the negative energy region 
(we remind that we are using 'reggeon energy' $E = - \omega= -(j-1)$: the most negative ($n=1$) 
one has the leading intercept). Moving to larger $n$, the distance between neighbouring eigenvalues decreases. 
At some point the eigenvalues become positive and increase further up to some maximal values. Because of our lattization the spectrum is completely discrete.

In order to analyse these results, it may be useful to first summarize the 
general feature of the spectrum of the continuum BFKL equation with running coupling and an infrared cutoff, 
as discussed, for example, in  \cite{Kowalski:2015paa,Levin:2016enb}. The spectrum consists of 
a discrete part, located at $E<0$, and a continuum part extending from zero to infinity $0< E<\infty$. 
The discrete part starts with the most negative (leading) eigenvalue and has an  
infinite number of eigenvalues (Pomeron poles) which accumulate at zero. In our lattice analysis,
also the continuos part becomes discrete, and ends at some maximal value. 
Both the spacing and the maximal value depend upon the lattice size. 
On the negative side, the leading eigenvalues should come out correctly, 
since - at least for the lattice being large enough - the spacing between the discrete eigenvalues is much larger than the discretization due to lattice effects. In fact,   
in the spectrum of our eigenvalues we recognize two different scales of spacing: whereas the leading eigenvalues show a spacing of the order 0.1, 
the very subleading ones have a much smaller spacing (order $10^{-3}$ or smaller). 
Near the accumulation point at $E=0$ it is difficult so distinguish between discretization due to lattice effects and genuine discrete eigenvalues, 
and we have to decide up to which value $n$ we can believe our results for the discrete part of the BFKL spectrum.   

We use several criteria. First we look at the $n$-dependence of the energy eigenvalues: up to 
$n\approx 20$ the inverse eigenvalues, $1/E_n$, lie on a straight line. A linear fit leads to:
\be
E_n \approx \frac{1}{0.0878 - 2.235 \,n} \,.
\ee
Above $n=20$ the values of $1/E_n$  start to show a clear deviation from the straight line. 
We mention that such a $n$-dependence of the energy eigenvalues has already been predicted  
in \cite{Lipatov:1985uk} and confirmed in \cite{Kowalski:2017umu}.

Similarly, the inverse of the slopes, when plotted as function of $n^2$, lie on a straight line. This holds up to $n=18$:
\be
\alpha'_n \approx \frac{1}{24.54 + 13.91 n^2}\,.
\ee
In order to have a measure for the extension of the wave functions into the region of large momenta
we define a logarithmic radius:
\be
r_n= < \ln \bq^2> = \frac{\int d\bk^2 |f_n(\bk)|^2 \ln \bk^2}{\int d\bk^2 |f_n(\bk)|^2}\,,
\ee
where momenta are in units of $m_h=0.45$GeV.
Our analysis shows that, again for $n<18$,  $r_n$ grows linearly with $n$:
\be
r_n \approx  -2.7 + 6.6 n.
\ee  
By exponentiating this logarithmic radius we translate to the linear scale (in units of GeV) and we find the values listed in the third column of Table 2. 
Clearly, with increasing $n$ the support of the wave functions quickly becomes dominated by the UV region, only $n=1$ is concentrated in the 'soft' region.  The radii are related to the oscillatory structure
of the wave functions. As an example, the first maximum of the $n=1$ wave function lies at  
$|\bq|=2.7$ GeV, whereas the radius in Table 2 is $3.9$ GeV. For $n=8$ the largest maximum lies at
$6.8 \times 10^{12}$ GeV, whereas the radius was found to be $3.5\times 10^{10}$ GeV. For comparison we just mention that these features of the wave functions are very different from the  fixed coupling case:
here the radii are almost constant with $n$ ($r_n\approx 92$) , i.e. even for small $n$ oscillations are    
in the UV region, i.e. all states are much 'harder'.

Summarizing these observation, we use as a first estimate $n=18$ as the critical value up to which we interpret our eigenvalues and eigenfunctions as being genuine parts of the discrete BFKL spectrum.

%%%%%%%%%%%%%%%%%%%%%%%%%%%%%%%%
\section{Numerical studies (2): Wilsonian optimized IR regulator}
%%%%%%%%%%%%%%%%%%%%%%%%%%%%%%%%

We now turn to the regulator introduced in Section~\ref{sect3} and repeat the analysis described above. We put
\ba
\label{denom}
D(\bq)&=&\bq^2+R_k(\bq^2)\nonumber\\
&=&k^2 \Theta(k^2-\bq^2) +\bq^2 \Theta(\bq^2-k^2).
\ea
After symmetrization the kernel becomes:
\ba
\label{reg-BFKL-kernel}
K_{BFKL}(\bq,\bk',\bk'')&=& \frac{\bar{\alpha}_s}{2\pi} \Big[
-\bq^2\frac{1}{\sqrt{D(\bq_1)D(\bq_2)}} \frac{(\bk'-\bk'')^2}{D(\bk'-\bk'')}\frac{1}{\sqrt{D(\bq'_1)D(\bq_2')}}\nonumber\\
&&+\frac{\bq_1^2}{\sqrt{D(\bq_1)D(\bq_2)}} \frac{1}{D(\bk'-\bk'')}\frac{{\bq'}_2^2}{\sqrt{D(\bq'_1)D(\bq_2')}}\nonumber\\
&&+\frac{\bq_2^2}{\sqrt{D(\bq_1)D(\bq_2)}} \frac{1}{D(\bk'-\bk'')}\frac{{\bq'}_1^2}{\sqrt{D(\bq'_1)D(\bq_2')}}\Big].
\ea

We begin with forward direction. The kernel simplifies and takes the form:
 \be
\label{reg-forw-BFKL-kernel}
K_{\text{BFKL}}(0,\bk',\bk'') = \frac{\bar{\alpha}_s}{2\pi} \
\frac{{\bk'}^2}{D(\bk')} \frac{2}{D(\bk'-\bk'')}  \frac{{\bk''}^2}{D(\bk'')}.
\ee
For the numerical integration over the angle $\varphi$ between the vectors $\bk'$ and $\bk''$ we need some care. Because of the $\Theta$ functions in $D(\bk'-\bk'')$ (cf.(\ref{denom})) the limits of the $\varphi$ integral will depend upon $|\bk'|$ and  $|\bk''|$. 
With $(\bk'-\bk'')^2=(|\bk'|-|\bk''|)^2+4 |\bk'| |\bk''|  \sin^2 \frac{\varphi}{2}$ we define the function $R_{crit}$:
\be   
R_{crit}(\bk',\bk'')=\frac{\bk^2-(|\bk'|-|\bk''|)^2}{4 |\bk'| |\bk''|}
\ee
and the angle
\be
\varphi_{crit}= 2 \arcsin \sqrt{R_{crit}}, \hspace{0.5cm} \text{if}\,\, 0<R_{crit}<1.
\ee
There are several  different regions: 
\begin{center}
\begin{tabular}{lccl}
$R_{crit}<0$&&$k^2-(\bk'-\bk'')^2<0$&$D(\bk'-\bk'')=(\bk'-\bk'')^2$ \\
$R_{crit}>1$&&$k^2-(\bk'-\bk'')^2>0$&$D(\bk'-\bk'')=k^2 $\\
$0<R_{crit}<1$&$0<\varphi<\varphi_{crit}$&$k^2-(\bk'-\bk'')^2>0$&$D(\bk'-\bk'')=k^2$ \\
&$2\pi-\varphi_{crit}<\varphi<2\pi $&$k^2-(\bk'-\bk'')^2>0$&$D(\bk'-\bk'')=k^2$ \\
&$\varphi_{crit}<\varphi<\pi$&$k^2-(\bk'-\bk'')^2<0$&$D(\bk'-\bk'')=(\bk'-\bk'')^2$\\
&$\pi<\varphi<2\pi- \varphi_{crit}i$&$k^2-(\bk'-\bk'')^2<0$&$D(\bk'-\bk'')=(\bk'-\bk'')^2$.
\end{tabular}
\end{center}

For the $\bq^2$-slope we need the expansion in powers of $\bq^2$. Expanding first the denominator $D(\bq+\bk')$
we find:
\ba
D(\bq+\bk') &=& D(\bk')+ 2(\bq \bk')\Theta({\bk'}^2-k^2)+ \bq^2 ({\bk'}^2-\bk^2)\delta({\bk'}^2-k^2)
\\ 
&&+2(\bq \bk')^2 ({\bk'}^2-k^2) \delta'({\bk'}^2-k^2)+ 4 (\bq \bk')^2\delta({\bk'}^2-k^2)
+\bq^2  \Theta({\bk'}^2-k^2). \nonumber
\ea
Anticipating the integration over ${\bk'}^2$, the last time in the first line vanishes, and the first term in the second line (after partial integration) can be replaced by  $-2 (\bq \bk')^2  \delta({\bk'}^2-k^2)$. We thus arrive at:
\ba
D(\frac{\bq}{2}+\bk') = D(\bk')+ 2(\frac{\bq}{2} \bk')\Theta({\bk'}^2-k^2) +2 (\frac{\bq}{2} \bk')^2 \delta({\bk'}^2-k^2)
+\frac{\bq^2}{4}  \Theta({\bk'}^2-k^2).
\ea
With this result we expand the kernel in (\ref{reg-BFKL-kernel}) and do the angular integrals, observing that the wave functions only depend upon the absolute values ${\bk'}^2$ and ${\bk''}^2$. For the term proportional to $\bq^2$ we find:
\ba
\label{kernel-theta}
K^{(1)}_{BFKL}&=&- \bq^2 \frac{\bar{\alpha}_s}{4\pi} \frac{1}{D(\bk')D(\bk'')D(\bk'-\bk'')}\nonumber\\ &&\cdot \Big[\frac{{\bk'}^4{\bk''}^2}{D(\bk')} \delta({\bk'}^2-k^2)+
\frac{{\bk''}^4 {\bk'}^2}{D(\bk'')} \delta({\bk''}^2-k^2)+(\bk'-\bk'')^2\Big].
\ea

Next we expand the trajectory function $\omega_{g,k}((\frac{\bq}{2}+\bk)^2)$ given in Eq.~\eqref{reg-traj} and write:
\be
\omega_{g,k}( (\frac{\bq}{2}+\bk')^2 )= \omega_{g,k}({\bk'}^2) +\omega^{(1/2)}+\bq^2\omega^{(1)}({\bk'}^2).
\ee
We start from 
\be
\omega_{g,k}((\frac{\bq}{2}+\bk')^2)
=-\frac{\bar{\alpha}_s}{2 \pi} \int d^2 \bk'' \frac{(\frac{\bq}{2}+\bk')^2}
{D(\bk'') \Big[D(\frac{\bq}{2}+\bk'-\bk'')+D(\bk'')\Big]}
\ee
and expand:
\ba
&&\omega_{g,k}((\frac{\bq}{2}+\bk')^2)=\omega_{g,k}({\bk'}^2)\\
&&-\frac{\bar{\alpha}_s}{2 \pi} \int \frac{d^2 \bk''}{D_2D''} \Big[ \bq \bk' -\frac{{\bk'}^2\bq (\bk' -\bk'') }{D_2} \,\Theta +\frac{\bq^2}{4}\nonumber\\
&&+\left( -\frac{(\bq \bk') (\bq(\bk'-\bk''))}{D_2} +{\bk'}^2 \left(-\frac{\bq^2}{4 D_2} +
\frac{(\bq (\bk'-\bk''))^2}{D_2^2} \right) \right) \Theta -\frac{{\bk'}^2 (\bq (\bk'-\bk''))^2}{2D_2}\, \delta\Big].
\nonumber
\ea
Here we have used
\be
D''= D({\bk''}),\,\,\,D_2=D(\bk'-\bk'') +D(\bk''),\,\,\Theta=\Theta \big[(\bk'-\bk'')^2-k^2\big],\,\,
\delta=\delta\big[(\bk'-\bk'')^2-k^2\big].
\ee

From this expansion we derive
\be
\omega^{(1/2)}=-\frac{\bar{\alpha}_s}{2 \pi} \int \frac{d^2 \bk''}{D_2D''} \Big[ \bq \bk' -\frac{{\bk'}^2\bq (\bk' -\bk'') }{D_2} \,\Theta\Big]
\ee
and (after angular integration)
\be
\label{traj-theta}
\bq^2\omega^{(1)}({\bk'}^2)=-\bq^2 \frac{\bar{\alpha}_s}{2 \pi} \int \frac{d^2 \bk''}{4 D_2^2 D''} \Big[
D'' +k^2+({\bk''}^2-k^2-\frac{2{\bk'}^2 D''}{D_2})\, \Theta - k^2 {\bk'}^2 \,\delta \Big].
\ee
After these remarks we perform a numerical analysis for the fixed coupling case:
%mod11a%
\begin{table}[H]
\begin{center}
\begin{tabular}{cll} \\
n&energy&slope \\\hline
1&-0.5295&0.00002\\
2&-0.5269&0.00090\\
3& -0.5225&0.00019\\
4& -0.5165&0.00033\\
5&-0.5088&0.00050\\
6&-0.4997&0.00068\\
7&-0.4892&0.00088\\
8&-0.4774&0.0011\\
9&-0.4644& 0.0013\\
10&-0.4503& 0.0015\\
11&-0.4354&0.0017
 \end{tabular}
\end{center} 
\caption{Numerical values for eigenvalues and  slopes, with fixed coupling, using the Wilsonian   IR regulator 
\label{Table3}}
\end{table}
Comparison with the results for the Higgs mass regulator listed in Table~\ref{Table1} shows very small differences in the spectrum. 
Slopes are still small, but differ from those of the Higgs mass regulator. 

For the running case we use the same form of the coupling as described in  (\ref{runningalpha}), and the infrared cutoff will be  given in units of $m_h$. 
In particular, for $k=m_h$ we expect the results to be close to the results obtained for the Higgs mass regulator. We use our results in (\ref{kernel-theta})
and (\ref{traj-theta}) and insert them into (\ref{slope-running}). For $<\alpha^{(1/2)}\omega^{(1/2)}>$
we find:
\be
<\omega^{1/2} \alpha^{1/2}>=\bq^2 \alpha_s({\bk'}^2) \frac{{\bk'}^2}{8\pi L' \ln (sL')} \int \frac{d^2\bk''}{D_2 D''}
\Big[2-\Theta -\frac{{\bk'}^2-{\bk''}^2-D''}{D_2}\, \Theta \Big].
\ee

\begin{figure}[H]
\begin{center}
\epsfig{file=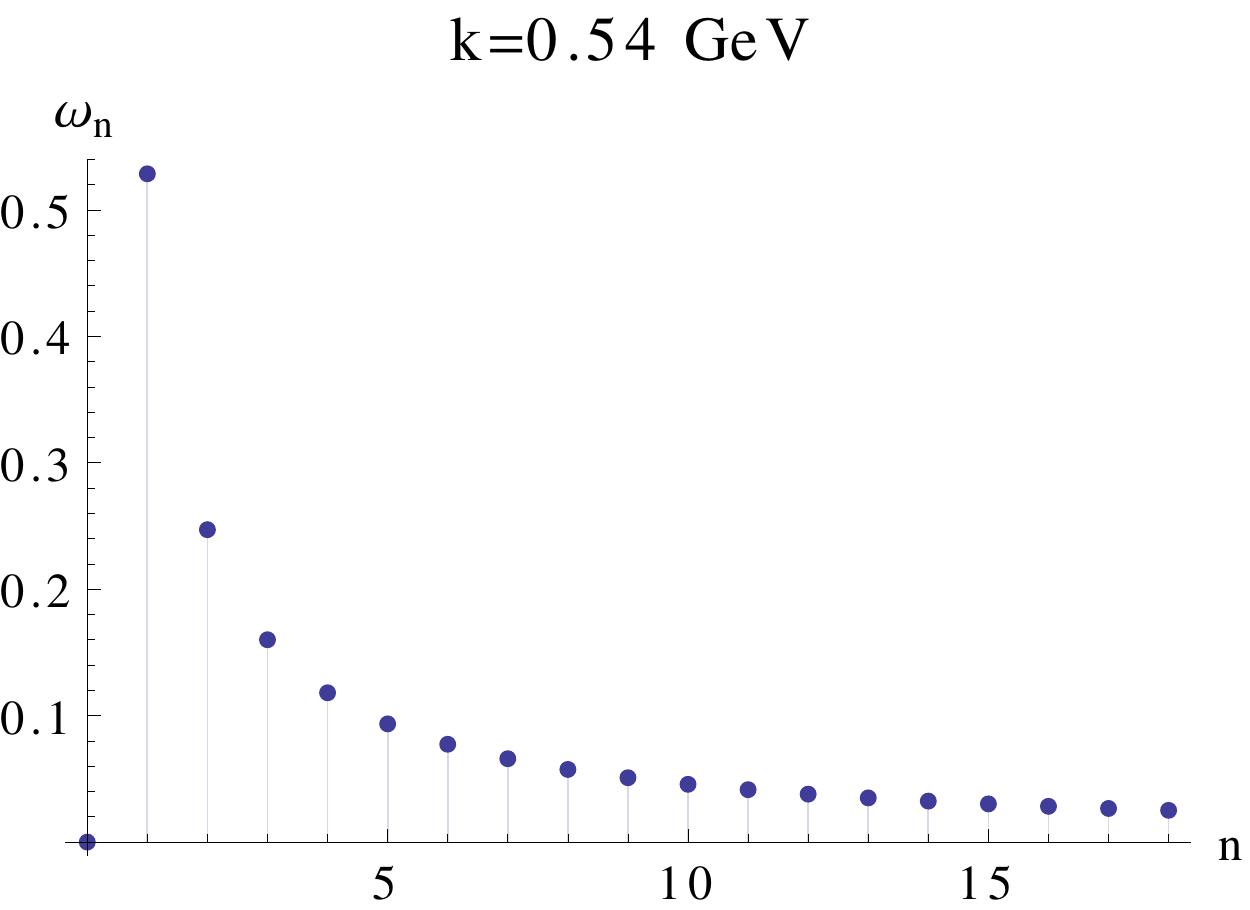,width=4.5cm,height=3cm} 
\epsfig{file=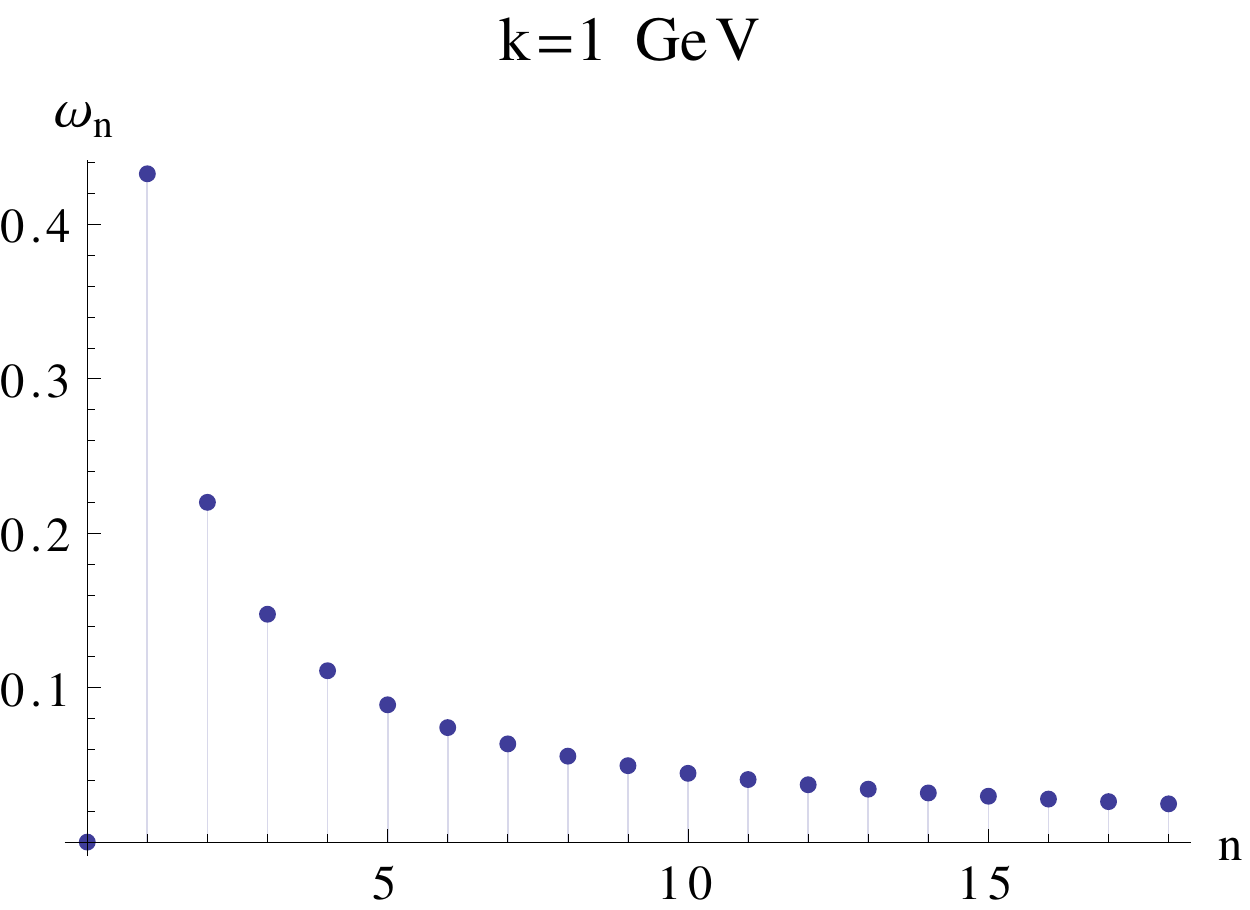,width=4.5cm,height=3cm}
\epsfig{file=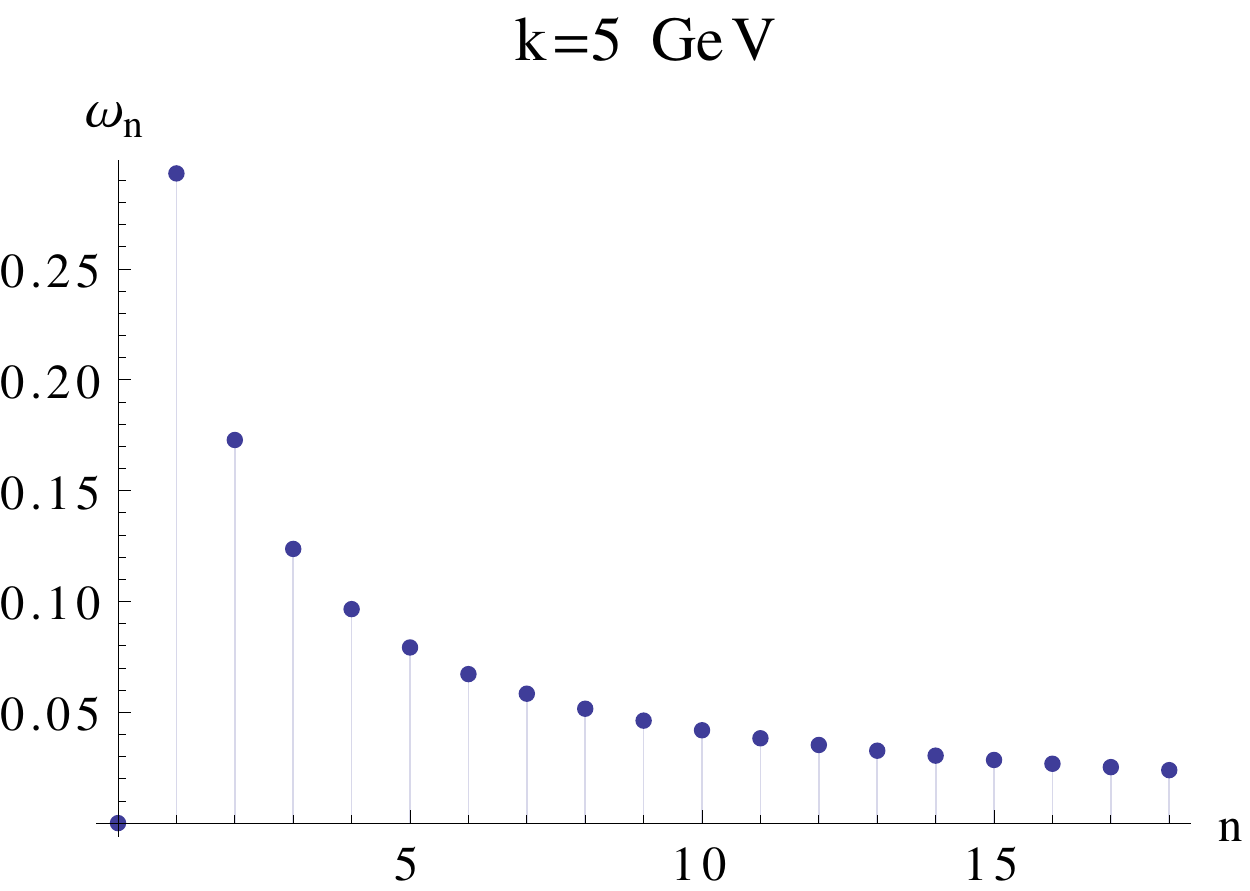,width=4.5cm,height=3cm}\\ 
\vspace{1cm}
\epsfig{file=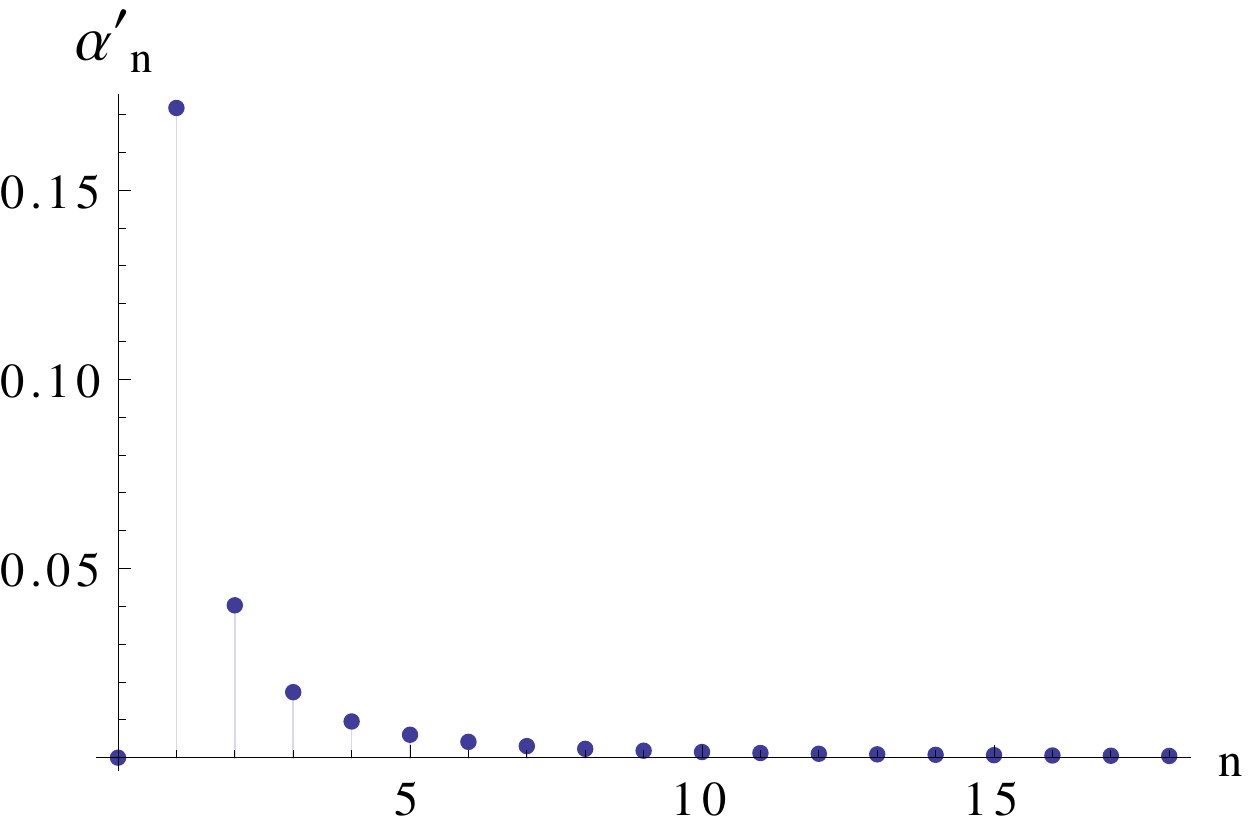,width=4.5cm,height=3cm} 
\epsfig{file=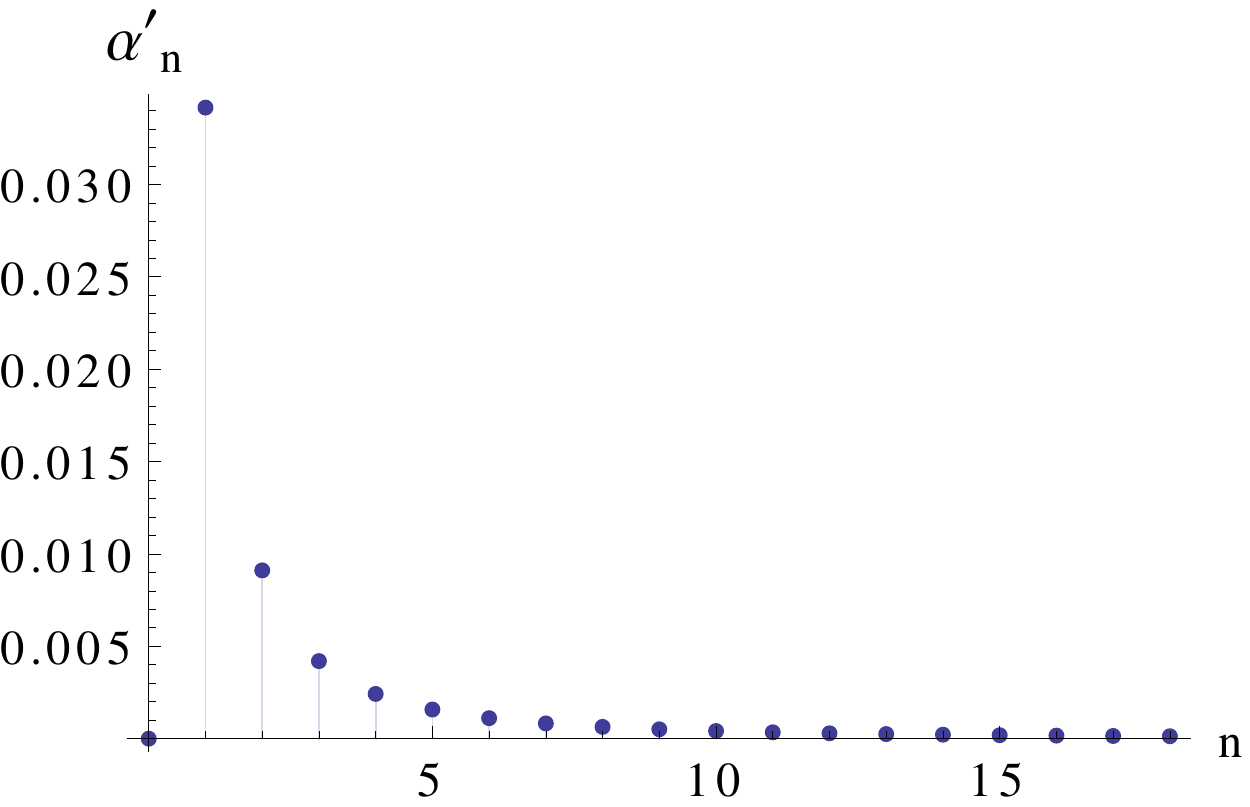,width=4.5cm,height=3cm}
\epsfig{file=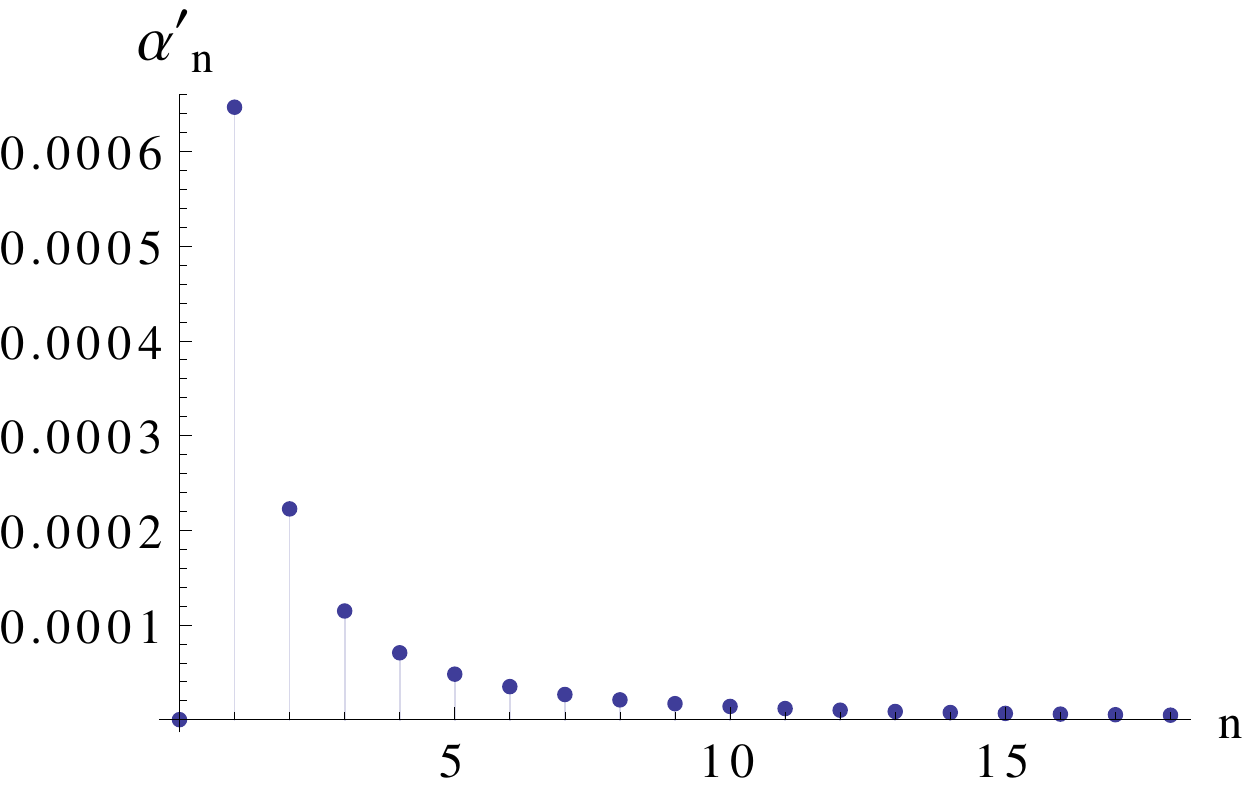,width=4.5cm,height=3cm}\\
\vspace{1cm}
\epsfig{file=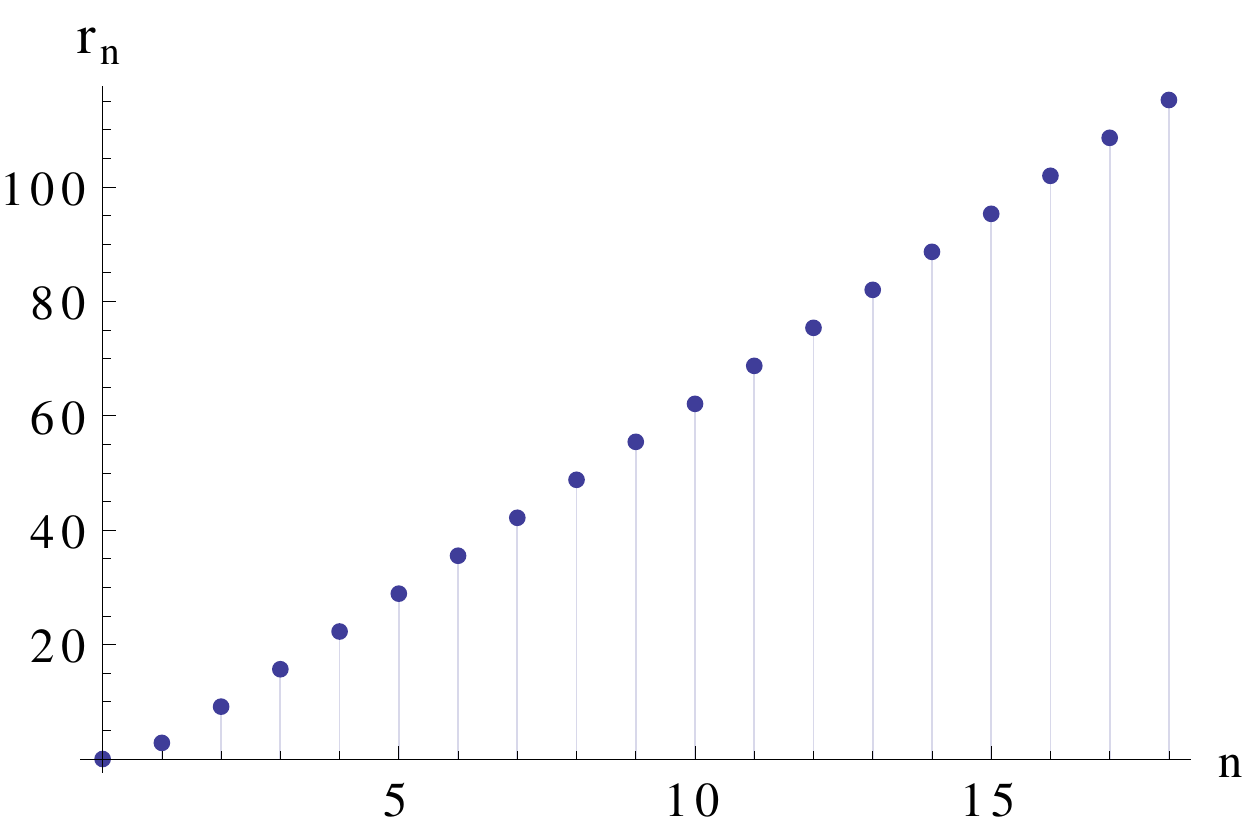,width=4.5cm,height=3cm} 
\epsfig{file=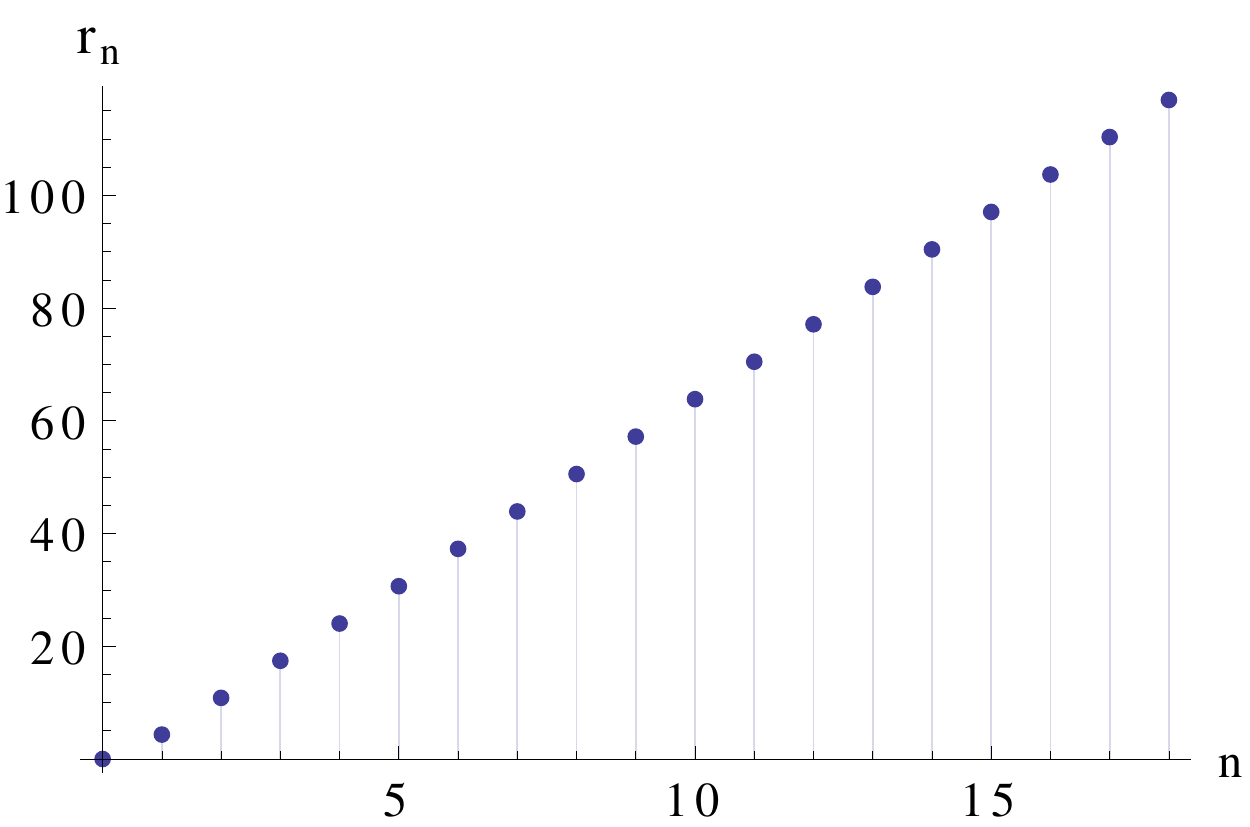,width=4.5cm,height=3cm}
\epsfig{file=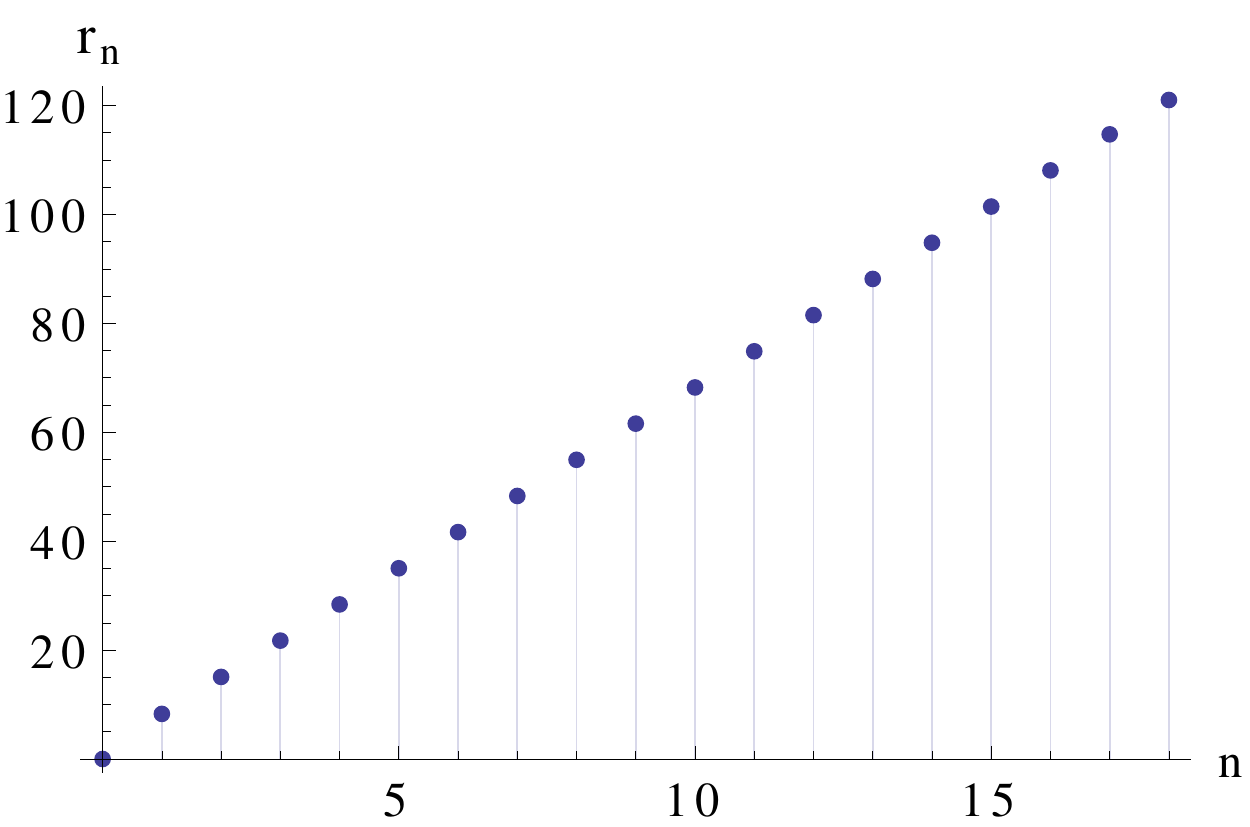,width=4.5cm,height=3cm}\\
\caption{Graphical illustration of Table 4. First row: eigenvalues $\omega_n$ vs.$n$. Second row: slopes $\alpha'_n$ vs.$n$.Third row: radii $r_n$  vs.$n$. 
\label{Fig23}} 
\end{center} 
\end{figure}

With these results we now compute eigenvalues, eigenfunctions, slopes, and radii.
We consider different values of the infrared cutoff (in GeV):
(1) $k=0.54$ GeV; (2) $k = 1$GeV; (3) $k= 5$ GeV.
We collect the results for the energy eigenvalues, for $\bq^2$-slopes and for the radii  in Table~\ref{Table4}, up to $n=8$:
\begin{table}[H]
\begin{center}
\begin{tabular}{clll|lll|lll} \\
&&energy&&&slope  &&&radius &\\
n& k=0.54 &k=1&k=5&k=0.54&k=1&k=5&k=0.54&k=1&k=5\\
\hline
1&-0.53& -0.43 &-0.29  &0.17 &0.034    &0.00065    &2.2&4.7&3.4$\times10^1$\\
2&-0.25& -0.22&-0.17   &0.040 &0.0091   &0.00022   &5.2$\times10^1$&1.2$\times10^2$&1.0$\times10^3$\\
3&-0.16& -0.15& -0.12  &0.017 &0.0042    & 0.00012  &1.4$\times10^3$&3.3$\times10^3$& 2.8$\times10^4$\\
4&-0.12& -0.11& -0.097  &0.0097 &0.0024   & 0.000070   & 3.7$\times10^4$& 8.9$\times10^4$&7.8$\times10^5$\\
5&-0.094&-0.089& -0.079&0.0061 & 0.0016     &0.000048      &1.0$\times10^6$      &2.5$\times10^6$&2.2$\times10^7$\\
6&-0.077&-0.074& -0.067 & 0.0042 & 0.0011     &0.000035    &2.8$\times10^7$      &6.7$\times10^7$&6.0$\times10^8$\\
7&-0.066&-0.064&-0.058 &0.0040 &0.00083   &0.000026    &7.8$\times10^8$&1.9$\times10^9$&
1.7$\times10^{10}$\\
8&-0.058& -0.056&-0.052 &0.0023&0.00064    &0.000021       
        &2.2$\times10^{10}$&5.2$\times10^{10}$&4.6$\times10^{11}$
        \end{tabular}
\end{center}
\caption{Numerical values for eigenvalues, slopes and radii, with running coupling and for different values of the IR cutoff $k$ 
\label{Table4}}
\end{table}
For illustration we also represent of our results of Table~\ref{Table4} graphically. In Fig.~\ref{Fig23} we show the energy eigenvalues for the first 18 eigenvalues (first row), slopes (second row) and radii (third row); note that in Table 4 we have listed the linear radii obtained by exponentiating the logarithmic radii $r_n$,
whereas in Fig.~\ref{Fig23} we plot the logarithmic radii $r_n$.  

Our discussion given above for the results of the Higgs mass regulator qualitatively also  applies to  Table 4.
New is the dependence on the cutoff parameter $k$:  with increasing cutoff parameter $k$ the intercepts as well as the slopes become smaller, whereas the radii grow. For the energy eigenvalues we again observe that, up to $n=20$, the inverse values $1/E_n$ lie on straight lines. Similarly, the inverse of the slope increase proportional to $n^2$. Numerical fits yield: 
\ba
k=0.54GeV:&\hspace{1cm}E_n \approx \frac{1}{0.452 - 2.24 \, n} &\hspace{1cm} \alpha'_n \approx \frac{1}{-2.183 + 6.731 \, n^2}   \\
k=1.00 GeV:& \hspace{1cm}E_n\approx \frac{1}{-0.052 - 2.24 \, n} &\hspace{1cm} \alpha'_n\approx \frac{1}{153.11 + 23.342 \,  n^2}\\
k=5.00 GeV:&\hspace{1cm}E_n\approx \frac{1}{-1.308 - 2.25 \,  n}&\hspace{1cm} \alpha'_n\approx \frac{1}{4733.15 + 647.164 \,  n^2}
\ea
It is interesting to note that the coefficient of $n$ is fairly independent of $k$, and it is nearly the same 
for the two definitions of the infrared regulator.  Interesting enough, it is also not far from the value 
given in \cite{Kowalski:2017umu}: $1/0.52223\approx 1.91$.

\begin{figure}[H]
\begin{center}
\epsfig{file=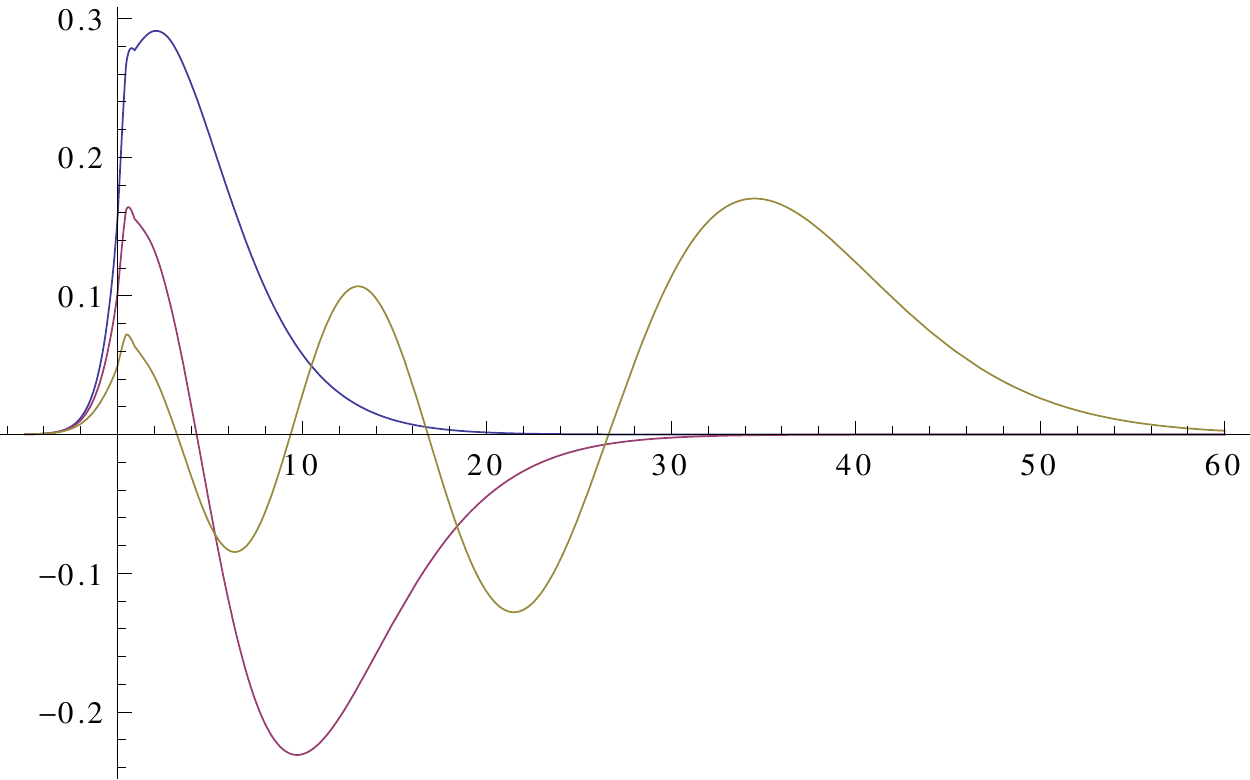,width=12cm,height=4.5cm}\\
\caption{three leading wavefunctions (No 1,2,5) for the scale $k=0.54$GeV, 
as a function of $\ln q^2$.
\label{Fig24}} 
\end{center}
\end{figure}

\begin{figure}[H]
\begin{center}
\epsfig{file=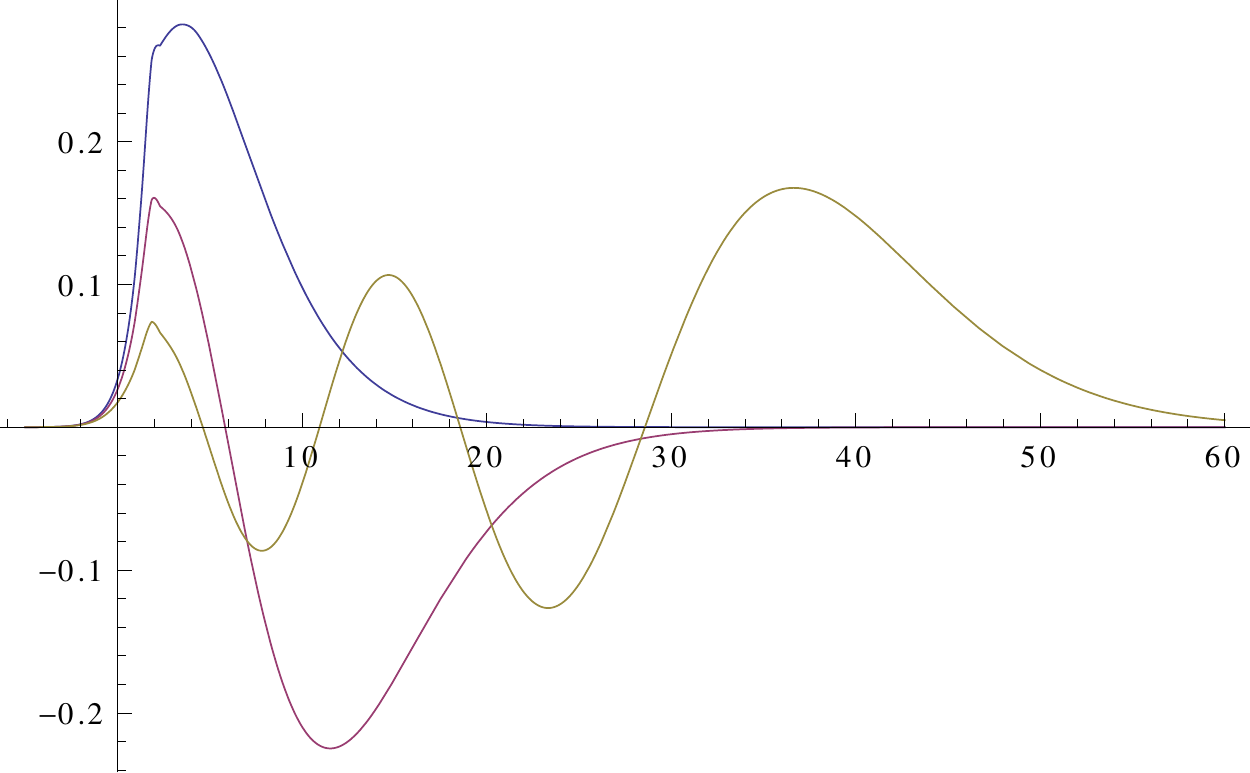,width=12cm,height=4.5cm}\\
\caption{three leading wavefunctions (No 1,2,5) for the scale $k=1$GeV, 
as a function of $\ln q^2$  
\label{Fig25}} 
\end{center}
\end{figure}

\begin{figure}[H]
\begin{center}
\epsfig{file=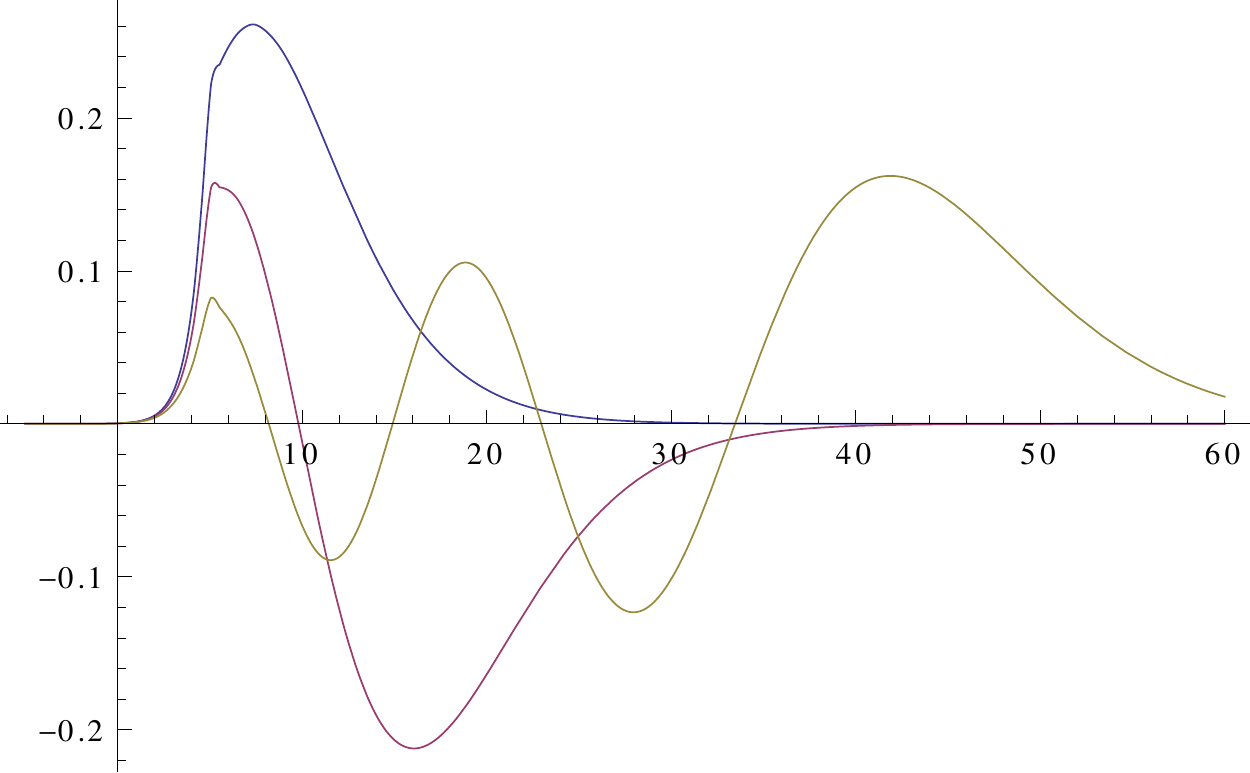,width=12cm,height=4.5cm}\\
\caption{three leading wavefunctions (No 1,2,5) for the scale $k=5$GeV, 
as a function of $\ln q^2$  
\label{Fig26}} 
\end{center}
\end{figure}

In Figs.~\ref{Fig24}-\ref{Fig26} we show the shape of the wavefunctions (at $\bq^2=0$). For large values $\bq^2$ the eigenfunction go to zero as 
$1/ \sqrt{\bq^2}$; we therefore plot the products: eigenfunction$\times  \sqrt{\bq^2}$.   
For each scale we observe that the nonleading eigenfunctions No 2 and 5 have their support at larger momenta, in agreement with he behavior of the radii listed in table 4: 
for example, in the first case the leading eigenfunction (zero node) has its maximum at $\log \bq^2 \approx  3$, whereas the 
fifth eigenfunction (4 nodes) has its largest extremum near $\log \bq^2 \approx 34$. For the third case the leading eigenfunction peaks near $ \log \bq^2 \approx 8$ whereas the fifth eigenfunction reaches out to $\log \bq^2 \approx 42$. This illustrates that the 
nonleading eigenfunctions are more 'ultraviolet' than the leading one and should be less important in the description of the large transverse distance IR region. 
A similar observation was made in \cite{Kowalski:2017umu}.

%%%%%%%%%%%%%%%%%%%%%%%%%%%%%%%%
\section{Summary and Outlook}
%%%%%%%%%%%%%%%%%%%%%%%%%%%%%%%%

In this paper we have started to address the UV part of our renormalization group program which aims at finding an interpolation between the perturbative BFKL pomeron in QCD and the soft Pomeron which describes the high energy scattering of hadrons. 

Our first goal was to find a formulation of the BFKL Pomeron with an infrared cutoff. This regulator was constructed in such a way that the BFKL Pomeron becomes part of the exact renormalization group equations in the Multi Regge Kinematics. 
As a first step we have derived a partial differential equation of the BFKL Green's function (corresponding to a perturbative Bethe-Salpeter resummation) with respect to the infrared cutoff parameter. 
This equation has the nonlinear form of the infrared evolution equations first found by Lipatov and Kirschner.

In a second step we have defined, starting from the high energy effective action of Lipatov, an effective field theory which describes the LO BFKL equation. It contains the reggeized gluon as well as elementary gluons and, at least for fixed coupling, it takes care of the bootstrap property of the reggeized gluon induced by s-channel unitarity. For this field theory we then have defined a set of RG equations which have the form of the Wetterich equations for the proper vertices in the MRK. By making use of very special features of this effective field theory we have derived, from the flow equations, the same nonlinear evolution equation obtained before. 
This means that the BFKL resummation can be obtained from an IR regulated Wilsonian flow for a suitable infinite truncation in 1PI vertices of the effective average action. 

We have also discussed how to move to another dynamically equivalent effective description in terms of Pomeron Fields, which is directly suitable for the transition to a RFT.
Conceptually the Pomeron field appears as a bound state of two reggeized gluons. However, since the BFKL Green's function, once an infrared regulator has been introduced, contains an infinite number of discrete poles in the angular momentum plane, we have to define a sequence of Pomeron fields which in the next step serve as input to an RFT analysis in the infrared region.
We therefore come to the third part of this paper where we have performed a numerical analysis of the BFKL equation with our infrared regulator.
As  a first step we have investigated the leading poles in the angular momentum plane: for different  numerical values of the 
infrared cutoff we have computed the positions of the leading poles, the $\bq^2$ slopes of the trajectory functions and the bound state wave functions. Our results show several features which will play an important role in the next step of our analysis: only the wavefunction of the leading eigenstate is 'soft' and has its support in the region of small momenta, whereas the nonleading states extend more and more into the UV region. Moreover, the $\bq^2$ slopes of the trajectory functions are largest for the 
leading eigenvalue and then go to zero for nonleading states.     

These numerical results define the Pomeron fields  which will serve as an input for the next step of our program, the RGE analysis of the interactions of these fields in the infrared region.
For this we shall need to consider the interactions among the leading Pomeron states, 
which can be obtained by projecting the $2\to4$ reggeized gluon vertex
onto the various states of the Pomeron spectrum one wants to take into account.                                 

One can also extend our formal analysis of the first and second part to include the QCD odderon states~\cite{Bartels:1999yt,Janik:1998xj}~\footnote{The known QCD Odderon states have different intercept and coupling properties to external particles~\cite{Bartels:2001hw}.}, 
analyzing in the generalized MRK the propagation of 3 reggeized gluons in a color singlet state (C and P odd) defined by the symmetric color tensor $d_{abc}$. 
This would require to introduce the BKP kernel~\cite{Bartels:1980pe,Kwiecinski:1980wb} and the corresponding non local 6 reggeizing gluon vertex in the truncation, together with the tower of the higher order inelastic vertex functions with $\chi$ and $a$ field emissions.
Similarly a numerical analysis on the discretized spectrum of the odderon states should be performed together with a subsequent evaluation of the interacting vertices.
The LL $2\to 6$ reggeized gluon vertex with the necessary quantum numbers is indeed known~\cite{Bartels:1999aw, Bartels:2004hb}. 

Another line of research would be to push all this analysis to the next-to-leading accuracy, i.e. work with the Quasi-Multi-regge-Kinematics accuracy.\\ 

\noindent
{\bf Acknowledgements:} J.B expresses his gratitude for support and hospitality of the Departamento de Fisica, Universidad Tecnica Federico Santa Maria, Valparaiso, Chile and for the support of the INFN and the hospitality of the Bologna University. C. C. thanks for the financial support from the grant FONDECYT 1180118, Chile. 

%%%%%%%%%%%%%%%%%%%%%%%%%%%%%%%%
\appendix
%%%%%%%%%%%%%%%%%%%%%%%%%%%%%%%%

%%%%%%%%%%%%%%%%%%%%%%%%%%%%%%%%
\section{Rapidity-dependence of propagators} 
%%%%%%%%%%%%%%%%%%%%%%%%%%%%%%%%

As discussed before, in order to reproduce the BFKL kernel, we introduce separate fields for the reggeized gluons, $A_i(t,x),\,\Adag_i(t,x)$ 
and for elementary gluons, $a_i(t,x),\,\adag_i(t,x)$ (i=1,2) and $\chi,\chidag$.  
Following the derivation of Lipatov 's effective action, the elementary gluons do not propagate in time $t$ (rapidity), i.e. their Green's function are proportional to delta functions:
\be
\delta(t'-t) = \frac{1}{2\pi i} \left(\frac{1}{t'-t-i \epsilon}-\frac{1}{t'-t+i \epsilon}\right).
\ee
As usual, the infinitesimal regulator $\epsilon$ will be removed at the of the calculations. In contrast, the reggeized gluon propagates with rapidity, and the Green's function is proportional to $\theta(t'-t)$. Here we have to make sure that
the zero-time propagation of the elementary gluon and the finite-time propagation of the 
reggeized gluon do not overlap. We therefore introduce an infinitesimal cutoff, $\delta>0$:
\be
\theta(t'-t) \to \theta(t'-t-\delta),
\ee       
which will be removed at the end of the calculations.  
With these modifications the Green's functions become
\be
G_a(\omega,\bq^2) \sim \frac{e^{-\omega \epsilon} \theta(\omega)+ e^{\omega \epsilon} \theta(-\omega)}{\bq^2+R_k(\bq^2)}
\ee
and 
\be
G_A(\omega,\bq^2) \sim\frac{e^{\pm i\omega \eta}}{\mp Z_A i \omega +(\bq^2+R_k(\bq^2)) 
I_k(\bq^2)} \cdot \frac{1}{\bq^2+R_k(\bq^2)}.
\ee  
With our results for the gluon trajectory function we have
\be
\left(\bq^2+R_k(\bq^2)\right)I_k(\bq^2) = \Theta(k^2-\bq^2) \Big[ \frac{g^2}{4\pi^2} + \frac{\bq^2}{k^2} \frac{g^2}{16\pi^2}+...\Big] + 
\Theta(\bq^2-k^2) \Big[ \frac{\bq^2}{k^2}\frac{g^2}{4\pi^2} +
\left( \frac{\bq^2}{k^2}\right)^2 \frac{g^2}{16\pi^2}+...\Big].
\ee
Keeping only the leading terms:
\be
(\bq^2+R_k(\bq^2)) I_k(\bq^2) \approx \Theta(k^2-\bq^2) \frac{g^2}{4\pi^2}  + \Theta(\bq^2-k^2) \frac{\bq^2}{k^2}\frac{g^2}{4\pi^2} 
\ee
we have, in the language of RFT, a reggeon with zero mass and slope 
$\alpha'=\frac{g^2}{4\pi^2}$.

With these notations it is easy to see that the propagation of the two fields does not overlap. As an example, we consider the 'selfenergy'
diagram shown in Fig.~\ref{Fig27}:
\begin{figure}[H]
\begin{center}
\epsfig{file=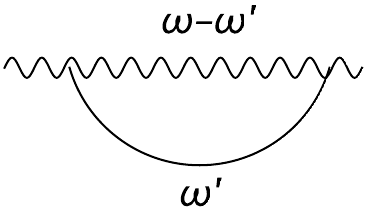,width=5cm,height=2cm}\\
\caption{example of a vanishing diagram. The wavy line denotes the reggeized gluon, the straight line the elementary gluon.
\label{Fig27}}
\end{center}
\end{figure}
If we denote the time coordinates of the left and right vertices by $t_1$ and $t_2$, resp., 
the Green's function of the elementary gluon is nonzero only for $t_1=t_2$, whereas the 
reggeized gluon is nonzero only for $t_2-t_1>\eta>0$. In fact, the energy integral
\ba
&&\int d\omega' G_a(\omega') G_A(\omega-\omega')\\
&=&\int d\omega' \frac{e^{-i\eta (\omega-\omega')}}{i(\omega-\omega')-\omega_g((\bq-\bq')^2)} \Big[e^{-\omega'\epsilon} \Theta(\omega') +e^{\omega'\epsilon} \Theta(-\omega') \Big]\nonumber\\
&=e^{-i\omega \eta} & \int_0^{\infty} d\omega' e^{-i\omega' \epsilon} \Big[ \frac{e^{i\omega' \eta}}{i(\omega-\omega') -\omega_g((\bq-\bq')^2)} +\frac{e^{-i\omega' \eta}}{i(\omega+\omega') -\omega_g((\bq-\bq')^2)} \Big]\nonumber
\ea     
vanishes as long as $\eta>0$. It is easy to see that, while $\epsilon$ could have been set to zero from the beginning, the parameter $\eta$ has to be kept positive and nonzero until the end of the calculations.  
It is straightforward to generalize this proof of equivalence also to the higher order vertex functions. 
%For the 6-point function we present this proof to the appendix.    

%%%%%%%%%%%%%%%%%%%%%%%%%%%%%%%%
 
\end{document}